\documentclass[journal]{IEEEtran}

\usepackage{xcolor,soul,framed} %

\colorlet{shadecolor}{yellow}
\usepackage[pdftex]{graphicx}
\graphicspath{{../pdf/}{../jpeg/}}
\DeclareGraphicsExtensions{.pdf,.jpeg,.png}

\usepackage[cmex10]{amsmath}
\usepackage{array}
\usepackage{mdwmath}
\usepackage{mdwtab}
\usepackage{eqparbox}
\usepackage{url}
\usepackage{multirow}
\usepackage{amssymb}
\usepackage{subcaption}
\usepackage{listings}
\usepackage{xcolor}
\usepackage{graphicx}
\usepackage{caption}
\usepackage{booktabs}
\usepackage{makecell}

\usepackage[english]{babel}

\usepackage{bm}
\usepackage[colorlinks=true, allcolors=blue]{hyperref}

\begin{document}

\tableofcontents
\newpage

\title{Foundation Models for Music: A Survey}
\author{Yinghao Ma, Anders Øland, 
Anton Ragni,  
Bleiz MacSen Del Sette, 
Charalampos Saitis, 
Chris Donahue, 
Chenghua Lin,   
Christos Plachouras, 
Emmanouil Benetos,    
Elona Shatri, 
Fabio Morreale, 
Ge Zhang, 
Gy\"orgy Fazekas,   
Gus Xia,  
Huan Zhang, Ilaria Manco, Jiawen Huang, Julien Guinot, Liwei Lin, Luca Marinelli, 
Max W. Y. Lam,  
Megha Sharma, 
Qiuqiang Kong, 
Roger B. Dannenberg,   
Ruibin Yuan, Shangda Wu, Shih-Lun Wu, Shuqi Dai, 
Shun Lei, 
Shiyin Kang,  
Simon Dixon,  
Wenhu Chen, %
Wenhao Huang, %
Xingjian Du, Xingwei Qu, 
Xu Tan, 
Yizhi Li, Zeyue Tian, 
Zhiyong Wu, 
Zhizheng Wu, 
Ziyang Ma,
and Ziyu Wang
\thanks{Yinghao Ma, Bleiz MacSen Del Sette, Charalampos Saitis,~\IEEEmembership{} Christos Plachouras, Emmanouil Benetos, Elona Shatri, Gy\"orgy Fazekas,~\IEEEmembership{} 
Huan Zhang, Ilaria Manco, Jiawen Huang, Julien Guinot, Luca Marinelli, and Simon Dixon, are with the Centre for Digital Music, Queen Mary University of London, London E1 4NS, U.K., email: \{yinghao.ma, emmanouil.benetos\}@qmul.ac.uk}
\thanks{Anders Øland, Chris Donahue,~\IEEEmembership{}
Roger B. Dannenberg, Shih-Lun Wu, and Shuqi Dai are with the School of Computer Science, Carnegie Mellon University. 15213, PA, U.S.}
\thanks{Anton Ragni,~\IEEEmembership{} University of Sheffield, Western Bank, Sheffield, S10 2TN, U.K.}
\thanks{Chenghua Lin,~\IEEEmembership{}
Xingwei Qu, and Yizhi Li, The University of Manchester, Oxford Rd, Manchester, M13 9PL, U.K.}
\thanks{Fabio Morreale,~\IEEEmembership{}  University of Auckland}
\thanks{Ge Zhang, Wenhao Huang, are with 01.AI, Beijing 100089, CN}
\thanks{Gus Xia,~\IEEEmembership{} Liwei Lin, and Ziyu Wang, are with Music X Lab, Mohamed bin Zayed University of Artificial Intelligence \& New York University Shanghai}
\thanks{Max W. Y. Lam,~\IEEEmembership{}  independent researcher}
\thanks{Megha Sharma, University of Tokyo}
\thanks{Qiuqiang Kong, is with the Chinese University of Hong Kong}
\thanks{Ruibin Yuan, Zeyue Tian, are with the Division of Emerging Interdisciplinary Areas, The Hong Kong University of Science and Technology, Hong Kong SAR}
\thanks{Shangda Wu, is with the Central Conservatory of Music, CN}
\thanks{Shun Lei, is with the Tsinghua Shenzhen International Graduate School, Tsinghua University}
\thanks{Shiyin Kang, is with the Skywork AI PTE. LTD., Beijing}
\thanks{Wenhu Chen, is with the Department of Computer Science,  University of Waterloo, N2L 3G1, CA \& Vector Institute, CA}
\thanks{Xingjian Du~\IEEEmembership{} is with University of Rochester}
\thanks{Xu Tan, is with Microsoft}
\thanks{Zhiyong Wu, is with the Tsinghua Shenzhen International Graduate School, Tsinghua University}
\thanks{Zhizheng Wu, is with the Chinese University of Hong Kong, Shenzhen. %
} %
\thanks{Ziyang Ma, is with Shanghai Jiao Tong University}}

\maketitle

\begin{abstract}
In recent years, foundation models (FMs) such as large language models (LLMs) and latent diffusion models (LDMs) have profoundly impacted diverse sectors, including music. 
This comprehensive review examines state-of-the-art (SOTA) pre-trained models and foundation models in music, spanning from representation learning, generative learning and multimodal learning. 
We first contextualise the significance of music in various industries and trace the evolution of AI in music. By delineating the modalities targeted by foundation models, we discover many of the music representations are underexplored in FM development. 
Then, emphasis is placed on the lack of versatility of previous methods on diverse music applications, along with the potential of FMs in music understanding, generation and medical application. 
By comprehensively exploring the details of the model pre-training paradigm, architectural choices, tokenisation, finetuning methodologies and controllability, we emphasise the important topics that should have been well explored, like instruction tuning and in-context learning, scaling law and emergent ability, as well as long-sequence modelling, etc. 
A dedicated section presents insights into music agents, accompanied by a thorough analysis of datasets and evaluations essential for pre-training and downstream tasks. 
Finally, by underscoring the vital importance of ethical considerations, we advocate that following research on FM for music should focus more on such issues as interpretability, transparency, human responsibility, and copyright issues. 
The paper offers insights into future challenges and trends on FMs for music, aiming to shape the trajectory of human-AI collaboration in the music realm.
\end{abstract}

\begin{IEEEkeywords}
Self-Supervised Learning, Foundation Model, Music Information Retrieval, Music Instruction Following, Music Generation
\end{IEEEkeywords}
\IEEEpeerreviewmaketitle

\section{Introduction}
Music is an important part of human culture, universal in its cross-cultural presence,
yet taking many different forms across cultures. Its functions include emotion regulation, communication, and promoting social cohesion; it appears in art, entertainment, worship, and advertising; and represents a large industry contributing to the global economy. It presents opportunities to benefit both human society culturally and music industries economically, as well as unique technical challenges when combined with AI.

The field of computer music is at the intersection of music, computer science, electrical engineering, and artificial intelligence, drawing upon fields such as philosophy (aesthetics), psychology (perception, cognition, and production), and physics (acoustics). 
Computational approaches to music often employ signal processing and other techniques to extract features from audio signals, and then apply machine learning algorithms for music information retrieval (MIR) tasks or music composition. 

Although natural language processing, computer vision, and speech processing have widely used foundation models (FMs), we are still only scratching the surface of AI for art, of which music is an essential component. One challenge specific to music is polyphonic signal modelling. Unlike speech and language signals, music usually has several simultaneous ``speakers'', and the ``meaning'' of what they ``say'' is not grounded in real-world objects or events. The occurrences of different note events are not independent, making it a challenging modelling task to capture the ``language(s)'' of music.
Moreover, music typically has a much longer duration with a much higher sample rate compared to speech or general audio, making it harder to model the whole musical piece.

Recent advances in pre-trained language models (PLMs) significantly outperform traditional algorithms on a range of music-related computational tasks, demonstrating the potential of modern machine learning techniques to understand and process music on an unprecedented scale \cite{li2023mert}. However, a critical bottleneck has emerged in terms of dataset size and quality. For algorithms to be reliable, especially those deployed in complex, realistic scenarios, they need to be trained on diverse and representative datasets. The performance of these algorithms is deeply dependent on the size of the annotated dataset and the quality of its annotations, which justifies the need for large quantities of high-quality data. Unfortunately, music datasets are often size-constrained due to the limited availability of copyright-free public domain data and the high costs associated with labelling and annotation.

FMs address this problem by employing self-supervised learning (SSL) approaches for pre-training on a large amount of unlabelled music data.  
SSL enables the model to learn meaningful representations without the need for explicit labelling, by exploiting intrinsic structures within the data. 
This approach is similar to the natural human learning process. For example, when children hear different instruments played, they learn the characteristics of each unknown instrument and are able to identify the instruments in new pieces of music without necessarily knowing their names. Similarly, SSL enables machine learning models to derive general knowledge from large unlabelled datasets, thereby improving their performance on downstream tasks that lack large amounts of labelled data. As has proven successful across other domains, models trained through such approaches show promising results for music understanding and generation.

\subsection{What is a Foundation Model?}
The term \emph{foundation model} was coined to describe a multi-purpose machine learning model that, rather than being trained for a single specific task, serves as the basis of multiple derived models that are able to perform a wide range of tasks \cite{bommasani2021opportunities}.
This terminology reflects the shift from the traditional emphasis on the specifics of architectures or tasks to a focus on broadly applicable models whose emergent abilities and generalisation are unlocked by significantly increasing the number of model parameters \cite{wei2021finetuned, chowdhery2022palm}.
Contrary to terms such as \emph{large language models} or \emph{self-supervised learning}, which emphasise narrow aspects of AI development, \emph{foundation model} captures the essence of the generality of these models.

The rise of foundation models has come about due to advances in computational hardware, architectural innovations in neural networks (e.g., the Transformer architecture), and an enhanced focus on minimally supervised training paradigms.
Foundation models employ deep neural network architectures that are typically trained on large-scale unlabelled datasets with SSL. Following this \emph{pre-training} phase, foundation models can be adapted for various downstream tasks via a relatively lightweight finetuning or in-context learning stage, for example, using a labelled dataset that is orders of magnitude smaller than the pre-training data. %

Beginning with language models such as Google's BERT (Bidirectional Encoder Representations from Transformers \cite{devlin2018bert}) and OpenAI's GPT (Generative Pre-trained Transformer \cite{brown2020language}) series, foundation models have demonstrated the power of SSL for training on extensive web-sourced datasets, freeing researchers from reliance on labelled data, which does not scale economically to web-scale data sizes.
In addition to text analysis and text generation, such PLMs have also demonstrated their utility across various modalities, including image processing with CLIP \cite{radford2021learning}, DALL-E \cite{ramesh2021zero}, and Flamingo \cite{alayrac2022flamingo}; speech and audio generation with Audiobox \cite{vyas2023audiobox}; music generation with Jukebox \cite{dhariwal2020jukebox}, MusicLM \cite{agostinelli2023musiclm} and MusicGen \cite{copet2024simple}; and robotic control with RT-2 \cite{brohan2023rt}.

The release of Stable Diffusion\footnote{\url{https://github.com/CompVis/stable-diffusion}} and ChatGPT\footnote{\url{https://chatgpt.com/}} in 2022 marked a significant turning point for foundation models, in terms of public impact, as well as industrial and academic interest in the creation of AI-generated content (AIGC).
This significant progress is primarily due to the ability to follow language instruction,  emergent ability in algorithmic advances when scaling up to large language models (LLMs), and the authentic quality of Latent Diffusion Models (LDMs) \cite{ldm}. These methods suggest a paradigm shift in AI, as generalised frameworks can support multiple applications across disparate domains. Although developing AI with universal capability for multiple and unseen tasks has been the goal of AI researchers since its earliest days \cite{newell1959gps}, most AI research in the ensuing decades has focused on a single or a limited number of pre-defined task(s). In addition, access to advanced problem-solving capabilities through natural language interaction facilitates uptake by non-specialists.
Although the development of foundation models demands substantial financial and computational investments plus significant human effort, the adaptation of pre-existing models for tailored needs is more cost-effective, and the release of open-source foundation models like Stable Diffusion, Llama \cite{touvron2023llama}, Mistral \cite{jiang2023mistral}, and MAP-NEO \cite{zhang2024map} gives access to users, developers and researchers alike to explore the possibilities of the models.

In this paper, we will discuss two types of self-supervisedly pre-trained foundation models which can perform multiple downstream tasks. The first is the single-modality pre-trained model in the waveform or symbolic domain that requires finetuning on downstream tasks. This can be some variant of PLM for music understanding such as MERT\cite{li2023mert} or music generation such as Jukebox\cite{dhariwal2020jukebox}. The second is multimodal pre-trained models that can take both natural language and music as input and have the potential to solve downstream tasks with in-context learning. This includes Latent Diffusion Models (LDMs) with multiple text inputs such as MusicLDM\cite{musicldm}, a music encoder prepended to an LLM such as Mu-llama\cite{liu2024music} or an LLM with multimodal tokenisers such as AnyGPT\cite{zhan2024anygpt}, Gemini 1.5\cite{reid2024gemini} and GPT-4o.

\subsection{Why Foundation Models for Music?}
\begin{figure*}
    \centering
    \includegraphics[width=0.75\linewidth]{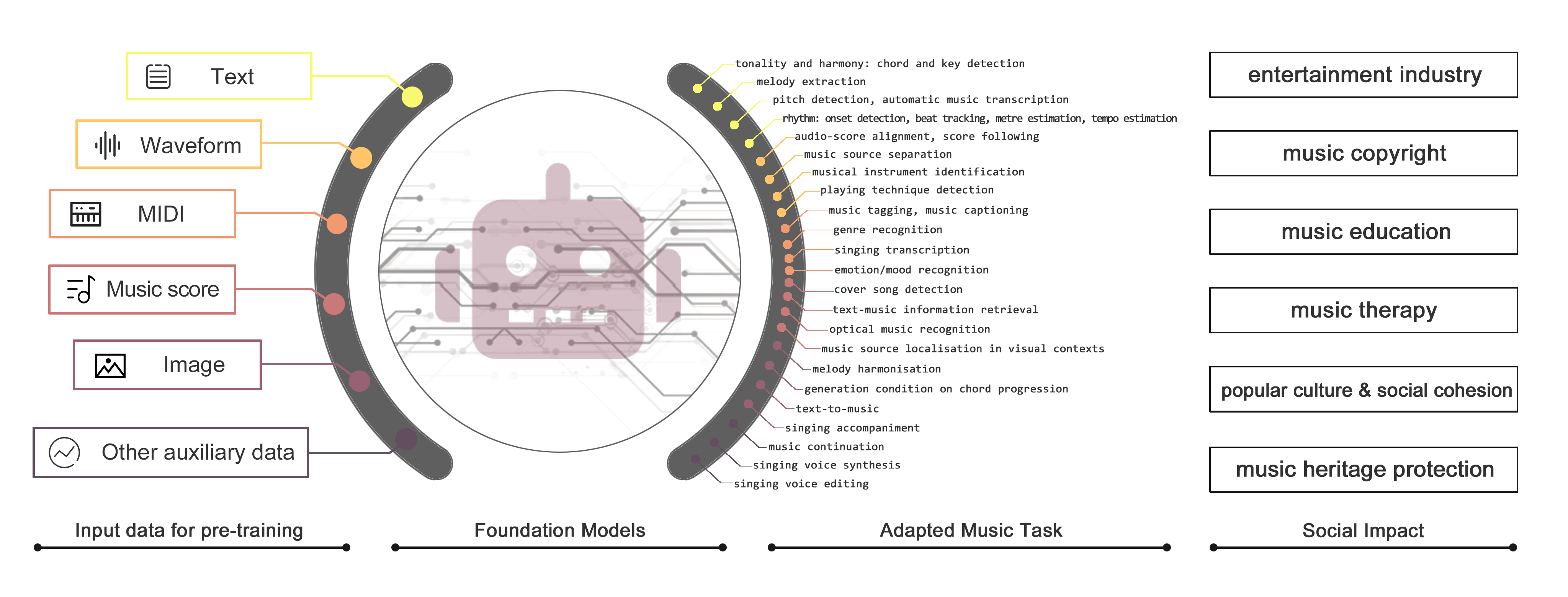}
    \caption{The input modalities, downstream applications and social impacts of foundation models for music}
    \label{fig:fm1}
\end{figure*}
FMs for music not only address data scarcity and reduce annotation costs, but also enhance generalisation in music information retrieval and creation. By pre-training on large music datasets, these models provide a better understanding capability of unseen structures, genres, or instruments. These algorithms can also contribute to the protection of the cultural heritage of music through world music analysis, music education, and new forms of artistic expression.

\subsubsection{Impact on Industry}
Foundation models have, or will potentially have, more robust and commercially viable applications for music than previous methods, including in creative processes, music understanding, and approaches within the entertainment industry.

In the area of \textbf{Creative Applications}, AIGC is perhaps the most obvious application of foundation models, including music, such as personalised music generation and collaborative composition with musicians.
Foundation models enable the generation of music based on user-specified preferences as input such as genre, mood, tempo, and instruments. Following recent progress in LLMs and LDMs in music, many music generation startups with proven commercial impact such as SunoAI\footnote{\url{https://suno.com/}}, TiangongAI\footnote{\url{https://music.tiangong.cn/}} and Udio\footnote{\url{https://www.udio.com/}}, have emerged. Musicians and producers can manipulate aforementioned parameters to steer the composition processes, assisting in the ideation process.
Such music generation applications can enable new forms of interaction with users and musicians. Music can change based on the listener's feedback or input prompts, potentially creating more immersive and personalised listening experiences.
Additionally, FMs show potential in collaborating with musicians or music editors by following their instructions more professionally and robustly.

Foundation models address several aspects of \textbf{Music Understanding}.
By analysing listening habits and understanding musical preferences, FMs can offer listeners more personalised recommendations, improving the user experience on streaming platforms.
FMs may also be used to better detect cover songs and identify  copyright infringement, helping artists and companies protect their intellectual property more efficiently.
They may also provide analyses of musical pieces to aid musicologists in understanding music structure, characteristics, and innovation.

For \textbf{Entertainment and Media},
foundation models can create adaptive soundtracks that correspond to the narratives of visual media for musicians and music editors, enhancing the impact and immersion of movies and video games.

\subsubsection{Social Impact}

Foundation models for music, with their capacity to understand, generate, and process music, can provide profound implications for culture and society. As FM has the potential to better resolve all kinds of music-related tasks, most main applications of MIR can be regarded FM territory, and therefore FMs have the potential to change the way we interact with, preserve, and understand music, raising significant ethical and heritage considerations. 

Concerning \textbf{Cultural Preservation and Diversity},
foundation models can play a role in preserving world cultural and musical traditions that are at risk of being lost. By analysing diverse musical datasets, these models can identify unique characteristics of styles, compositions, and performances from cultures around the world, much similar to current LLMs' capability of understanding minor languages. Moreover, FMs can promote cultural awareness by facilitating exploration of music from different parts of the world.

For the field of \textbf{Music Anthropology},
foundation models may serve as a tool for studying the evolution of music across different nations and eras. By analysing vast amounts of music data, FMs can uncover music patterns and cultural influences. By relating this analysis to social and historical data, FMs could potentially provide insights into music's role in human societies.

Foundation models can improve access to \textbf{Music Education} by creating personalised learning experiences adapted to the learner's pace and style; For example as a virtual tutor that provides theoretical and practical knowledge, feedback, virtual accompaniment and simulated ensemble playing.
This could make music education more accessible, regardless of access to traditional music education resources, encouraging a more inclusive culture of music learning, and removing barriers that have historically limited participation in music-making.

In \textbf{Music Therapy}, 
FMs could be tailored to produce music for therapeutic purposes, aligning with individual therapeutic goals or emotional needs, potentially offering mental health support.
Likewise, in non-clinical settings, by generating music that reflects or counteracts listeners' emotional states, foundation models can play a role in mood regulation and wellness practices.

The ability of foundation models to generate music that mimics human compositions raises important \textbf{Ethical Considerations}. The fact that the models benefit from the intellectual property of the millions of musicians and artists who created the training data leads to legal challenges and debates about the lawful use of data. Ethical discussions focus on issues of copyright, originality, and the role of AI in creative processes, ideally with interpretability and transparency. As these models become more prevalent, society must navigate between leveraging technology for innovation in music creation and respecting the rights and contributions of human artists.

The impact of foundation models for music is likely to be profound, offering new tools for the generation, analysis and interaction of music, as well as for music education and therapy. As these models are developed, it is essential to consider their ethical implications thoughtfully, ensuring that they serve to enrich human culture and promote a more fair and inclusive global society. For more information about the ethical issues of FMs in music, please refer to Section \ref{sec:ethic}.

\subsection{Aims of the Survey}
The purpose of this survey is to provide a thorough and comprehensive overview of foundation models as they relate to the domain of music, including LLMs and LDMs. 
While some previous overview papers have addressed FMs~\cite{bommasani2021opportunities} or LLMs~\cite{zhao2023survey, he2024llms} in general and as they apply to specific areas such as vision~\cite{zhang2024vision},  speech~\cite{zhou2023comprehensive, mehrish2023review, latif2023survey} and audio~\cite{wu2024towards, mei2022automated, latif2023sparks, triantafyllopoulos2024computer}, they do not comprehensively cover music-related applications of FMs. 
Besides, previous music surveys also exhibit limitations in providing a comprehensive overview of FMs. For example, \cite{ji2020comprehensive} fails to incorporate new advancements post-2021, particularly in LLMs and audio LDMs. Similarly, \cite{hayes2024review} focuses predominantly on digital signal processing methods, neglecting the integration of FMs into music synthesis and understanding. \cite{hernandez2022survey} briefly mentions LLMs and LDMs but lacks an in-depth exploration of their applications in music understanding as well as multimodality. \cite{zhu2023survey} provides a limited discussion on music generation models, primarily focusing on commercial scenarios and overlooking critical technical details and ethical considerations.

Our survey aims to bridge this gap by reviewing the wide array of FM applications, from music understanding to generation, therapy, and the ethical implications associated with these technologies. By doing so, we seek to highlight the unique challenges and opportunities that musical data presents for FMs, including aspects such as modelling long-term temporal dependencies, and the evaluation of artistic outputs. Additionally, this survey endeavours to update the literature with recent advances in LLMs and audio LDMs that are not covered in existing surveys.

The survey provides a detailed exploration of FMs in music. Section 2 examines music modalities and representations, including psychoacoustics, audio representations, symbolic music representations, and their integration with other modalities. We then turn to the diverse applications of FMs in music in section 3, spanning understanding, generation, and medical application.
Section 4 covers FMs' technical aspects, focusing on pre-training strategies, (instructional) finetuning, model architecture, audio tokenisation, applying LLM foundation models, music agents, scaling laws and emergent abilities, along with future work. The discussion in Section 5 extends to datasets and evaluation approaches, highlighting the challenges and solutions in both acoustic and symbolic domains of music understanding and generation tasks.
The last sections critically assess the ethical and social impacts along with copyright concerns of utilising FMs in music. They address potential cultural issues, including transparency and interoperability of algorithms, effect on humans, people's responsibility, and copyright issues. 
We suggest researchers in general ML focus on sections 2 and 3, and computer music researchers focus on FM methodology in section 4. For quick start, please refer to the GitHub repository\footnote{\url{https://github.com/nicolaus625/FM4Music}}.%

\section{Representations of Music}\label{sec:representations}

Music representations can be used to describe, encode, and communicate musical information mentally and computationally. These representations can capture different aspects of music, such as pitch, rhythm, harmony, dynamics, expression, timbre or structure. %
Humans perceive acoustic music signals in both the temporal and frequency domains, at multiple time and frequency scales, and abstract this rich sensory information to perceptual representations (experiences) of musical concepts such as pitch, rhythm, dynamics, and timbre, among others. Such a process inspires people to design a broad range of music representations with often very different focuses and characteristics. 
Many representations have been adopted to or exclusively devised as computer representations of music, and yet more are developed for or used as input representations to neural networks. Studying music representations in the context of foundational music models is essential because the choice of representation affects the model's effectiveness, efficiency and performance. Appropriate representations ensure that the data is relevant and informative, enhancing accuracy and reducing computational load. They may also help extract meaningful features and aid in the generalisation of new data. A suitable representation may also incorporate domain-specific knowledge, better align with the task requirements and facilitate interpretability, ultimately leading to more robust, transparent and explainable models. 

In this section, we discuss manually designed computer representations of musical notation and audio content at various levels of abstraction, starting from a very brief overview of how sound and musical concepts are experienced by humans. 
Additionally, the representations of foundation modelsF(Ms) can also be fully learnable audio tokens other than manually designed representations; please refer to Section~\ref{sec:audio_tokenizers} for more information. 
Finally, In the last subsection of this part, we discuss multimodal representations for music. 

We can tell from the following paragraphs that the current FMs for music are developed on limited modalities, including waveform, MIDI or ABC notations. And FMs with more comprehensive input modalities remain underexplored.

\subsection{Music Perception \& Notation}

Humans process musical information through a complex interplay of perceptual and cognitive processes. 
A sound wave transmits a pattern of oscillations, created by an excitation force (e.g., bowing) that stimulates a vibrating object (e.g., a viola's string), through a physical medium such as air \cite{caetano2019audio}. On the way from the source (e.g., bowed viola string) to our ears, the sound waves carry information about the vibrating object, the driving force, and possibly other interactions through the physical medium (e.g., the walls of a room). When the waves reach the ears, these oscillations are translated into a pattern of neural pulses that the brain processes to make sense of musical elements such as melody, harmony, rhythm, and timbre.
The ear acts like a set of overlapping bandpass frequency filters \cite{lyon2017human}. Sound energy within a frequency range is integrated across frequency and time (the temporal pattern of neural pulses can also encode frequency). 
The notion of the \emph{critical band} (the range of frequencies that are integrated) is important psychoacoustically because it provides a basis for explanations of the degree of masking and loudness summation, but also of frequency discrimination.

Pitch may be defined as the human perceptual correlate of acoustic frequency. Specifically, the perceptual representation of pitch involves at least two dimensions: a circular dimension of \textit{pitch chroma} (the \textit{pitch class}, sometimes called \textit{tonality}) and a vertical dimension of \textit{pitch height} \cite{shepard1982geometrical}. A pitch class is a set of all heights of pitches that are perceived as repeating once per octave (e.g., all C notes). Both pitch chroma and pitch height dimensions generally allow ordering pitches on a scale from low to high, but not always unambiguously (e.g., Shepard tones \cite{shepard1964circularity}). 
Furthermore, the relationship between pitch and frequency is not linear: we can better detect differences in lower frequencies than higher frequencies \cite{stevens1937scale}. For example, we can tell the difference between 500 and 1500 Hz, but probably struggle to discriminate between 9 and 10 kHz, although the frequency distance remains the same.
This is because the width of the critical band increases at higher frequencies (frequencies are smeared) and decreases at lower frequencies (frequencies can be resolved).

Musically, chroma is the most important part of the pitch and forms the basis of melody (distinct pitches played in sequence) and harmony (distinct pitches coinciding with one another), but also of rhythm. 
Rhythm is what orders the movement of melodic and harmonic patterns in time, both in the sense of how listeners perceive and how musicians perform music, consisting of \textit{beats} (perceived pulses that mark equally spaced points in time), \textit{tempo} (their frequency), and \textit{meter} (perceived cyclical groupings of beats into units marked by accents; the \textit{time signature}). 
Staff notation is a \textit{symbolic} visual representation of music that allows musicians to read the pitch and rhythm of notes they are supposed to play, and forms the basis of digital systems to encode musical documents in a machine-readable format. 
The pitch (class) and octave (height) of a note are indicated by the vertical position of the note within, below, or above the staff. Notes have different durations or note lengths, represented by whole notes, half notes, quarter notes, eighth notes etc. Each note length has a specific duration relative to the beat defined by the meter. Roughly speaking, rhythms and melodies are made up of notes with different durations. 

Timbre, or \textit{tone colour}, is a ``catch-all'' term, as it has been called, that refers to everything on what a sound or musical piece ``sounds like,'' except for pitch information, loudness, and rhythm-related features, although changes in pitch, dynamics, and timing also produce changes in timbre \cite{siedenburg2019timbre}. 
Timbre is primarily determined by the relative amplitudes of frequency components. Phase can also play a role, especially when certain changes in the relative phases of harmonics within a critical band can change the resulting envelope modulation  \cite{lyon2017human}.
Temporal characteristics of the signal (e.g., attack time, decay) and vibrato, are key contributors to timbre too \cite{aucouturier2005way}. However, changes in harmonic amplitudes affect timbre in a  much more direct way, such that timbral (i.e., perceptual) distances between sounds correlate well with spectral differences, such as measured in terms of the log power outputs of a one-third-octave filterbank \cite{plomp1976aspects}.
Such an approach to timbre is not accurate or complete \cite{lyon2017human}, but captures a reasonable part of the way a sound or musical piece ``sounds'' and, together with non-linear frequency scales, is the basis for computer representations of musical spectra.

\subsection{Computer Representations of Music}\label{subsec:symbolic-acoustic}
Inspired by the process through which humans perceive music, as mentioned in the previous subsection, we will introduce the various computational methods for music representation. 
Echoing how the ear transforms sound waves into neural signals, we begin by detailing the conversion of the waveform in the time domain into spectral representations using band-pass frequency filters.
Subsequently, we delve into symbolic music representations at both performance-level and note-level that abstract neural signals in terms of pitch, timbre and time through our brain. 
Though MIDI (Musical Instrument Digital Interface) in performance-level and ABC notations in note-level are the most widely-used presentations for symbolic music LLMs, we introduce other music representations which have been used with traditional deep learning methods in this subsection as they have potential to be trained for foundation models. 
We then explore more abstract note-level representations which prioritise relational and structural aspects of music notation. An example is the ABC notation system, which organises music by relative beats in staff notation rather than specific timestamps, offering a more human-understandable representation suited to computational models. 
At last, we will discuss combined representations like note-tuples and the potential of learnable note-level tokenisers. These advanced topics pave the way for integrating these representations into foundation models, enhancing our ability to process and generate music through AI.

\subsubsection{Acoustic-level Representations of Music}

\textbf{The log Mel spectrogram} is a key audio representation technique that synthesises signal processing with psychoacoustic principles to closely mirror human auditory perception. It starts with converting audio signals from the time domain to the frequency domain using the Fast Fourier Transform (FFT). The Short-Time Fourier Transform (STFT) further analyses these signals over time, segmenting the audio into overlapping frames and applying FFT to capture temporal dynamics. Transitioning to the Mel scale involves applying Mel band-pass filters to the STFT output. These filters, aligned with the non-linear nature of human pitch perception, group frequencies into bins that are linearly spaced at lower frequencies and logarithmically at higher ones. The logarithmic transformation of this Mel-scaled output adjusts for the human ear's non-linear response to loudness and pitch, compressing the spectrogram's dynamic range and emphasising perceptually significant changes in sound energy. The log Mel spectrogram has been directly applied in tasks of music generation \cite{Hawthorne2022Multi-instrumentDiffusion, Forsgren_Martiros_2022} due to its perceptual relevance, enabling accurate modelling of musical characteristics. In the foundation model paradigm, however, log Mel spectrograms usually serve as input into audio representation models, which learns an efficient latent space \cite{gongSSASTSelfSupervisedAudio2022a, feiAJEPAJointEmbeddingPredictive2024, huangMaskedAutoencodersThat2022} in which generation operates. 

\textbf{The Mel-Frequency Cepstral Coefficients (MFCCs)} are a crucial audio feature extraction technique that encapsulates the characteristics of human speech perception. This process begins with converting audio signals from the time domain to the frequency domain via FFT. It then applies the STFT to segment the audio into overlapping frames and analyse temporal changes. The Mel scale is employed through Mel filters that mimic the human ear's logarithmic frequency perception, focusing on discriminative lower frequencies. The computation of MFCCs involves taking the logarithm of the energies in each Mel filter, followed by a Discrete Cosine Transform (DCT). Due to their effectiveness in capturing vocal tract configurations, MFCCs are widely used in speech recognition and speaker identification applications. AV-HuBERT \cite{shi_learning_2022}, Hubert \cite{hsu2021hubert}, and MusicHubert \cite{ma2023effectiveness} use MFCC features as re-prediction targets.

\textbf{The Constant-Q Transform (CQT)} enhances audio processing in music analysis by providing a log-frequency spectrogram aligned with the musical scale. Unlike the linear Fourier Transform, the CQT uses a logarithmic scale that reflects the exponential nature of musical pitch, simplifying the extraction of note frequencies. Its key feature is that the centre frequency to bandwidth ratio remains constant (denoted by Q), allowing variable filter lengths that optimise performance. Despite less popularity compared to the log-Mel spectrogram, CQT has proven useful for tasks like musical timbre transfer  \cite{Huang2019TimbretronTransfer} and music representation learning \cite{li2023mert}. 

\textbf{Chroma features} focus on the twelve pitch classes of Western music, capturing the harmonic core of a piece regardless of timbre or instrumentation. These features, displayed in chromagrams, condense music into pitch class profiles that highlight the harmonic and melodic structure. Chroma features are especially effective in identifying chords, key signatures, and modulations due to their alignment with the equal-tempered scale. The most recent application of chromagram is melody-guided music generation in MusicGen \cite{copet2024simple}.

\subsubsection{Symbolic Music Format \& Their Content} 

In this subsection, we introduce the low-level symbolic music formats and their informational content at both performance-level and note-level. Symbolic music formats are essential for representing musical notation and facilitating digital music processing. This subsection will cover key formats, highlighting their structures, capabilities, and limitations, especially in the context of training computer music foundation models.

\paragraph{Performance-level symbolic music repsentation}
\textbf{Musical Instrument Digital Interface (MIDI) protocol} is a standard way of saving and transferring MIDI data for playback on different systems. Unlike MusicXML or MEI, MIDI encodes musical performance data, capturing information about note pitches, durations, velocities, and other performance aspects like dynamics, articulations, and control changes. This makes MIDI particularly suited for applications requiring precise control over musical performance and real-time interaction between devices.
In ML, the detailed performance data in MIDI is invaluable for training models focused on tasks such as music generation, transcription \cite{benetos2019transcription}, Optical Music Recognition (OMR) \cite{Shatri_tenor2020} and style transfer \cite{brunnerMidiVaeModelingDynamics2018}. Models can learn from the nuances of human performances captured in MIDI data, enabling the generation of expressive and realistic musical outputs. MIDI's widespread use and standardisation mean there is a vast amount of data available for training purposes. However, the format's focus on performance rather than notation means it lacks explicit information about musical structure, such as key signatures, time signatures, and notation details.

Note-level representations focus on the structural elements of music, such as pitch and rhythm, offering a simplified, human-readable format ideal for sharing musical scores. In contrast, performance-level formats like MIDI encompass additional expressive details of a piece's execution, including dynamics, musical techniques and duration/tempo variations. While note-level symbols capture the composition's essence, MIDI provides a detailed account of its performance, capturing both the notation and its expressive execution. 

\begin{figure*}[htbp]
\centering

\begin{subfigure}[b]{.18\linewidth}
    \centering
    \includegraphics[width=\linewidth]{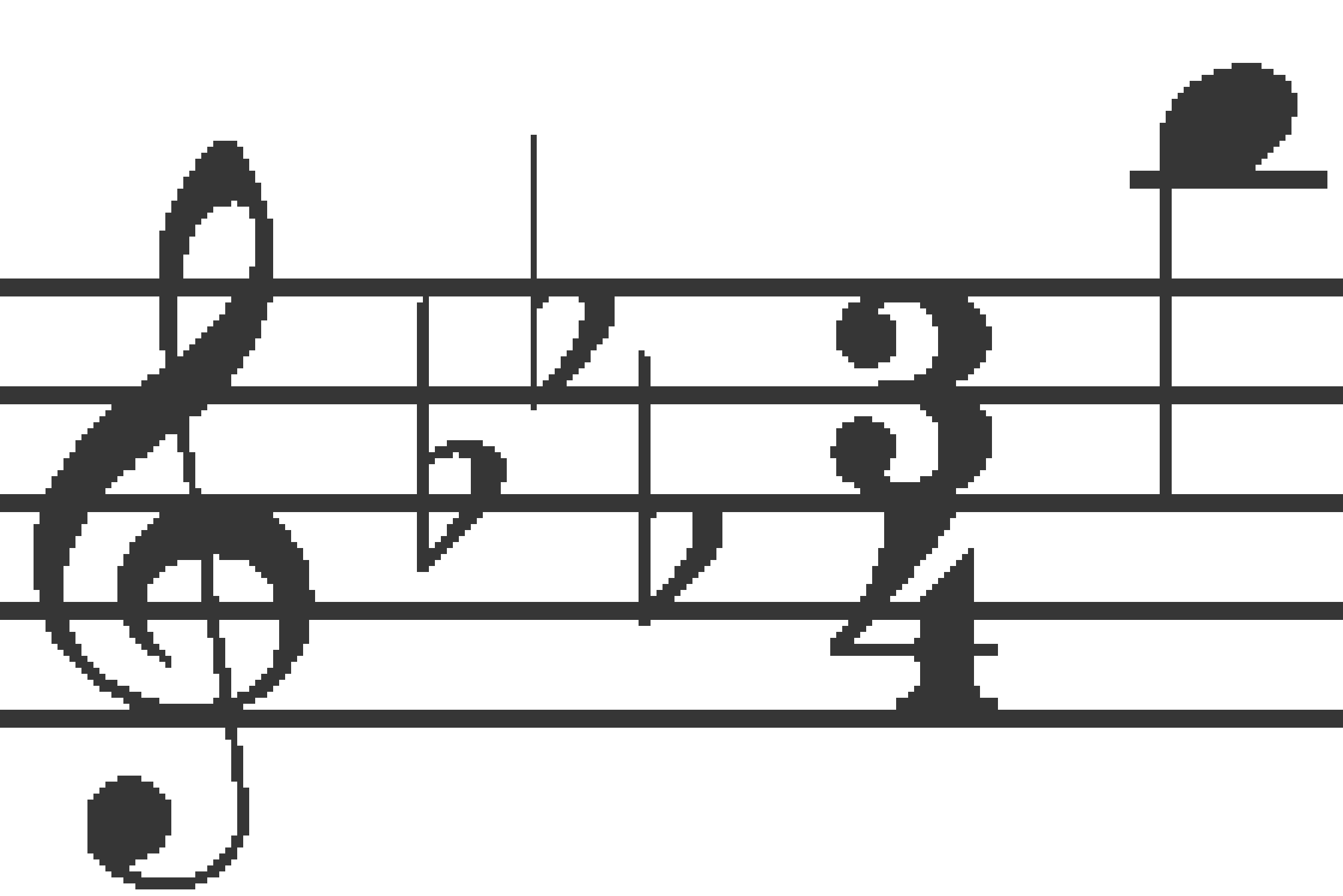}
    \caption{Snippet from a musical score in staff notation.}
    \label{fig:snippet}
\end{subfigure}
\hfill
\begin{subfigure}[b]{.16\linewidth}
    \centering
    \includegraphics[width=\linewidth]{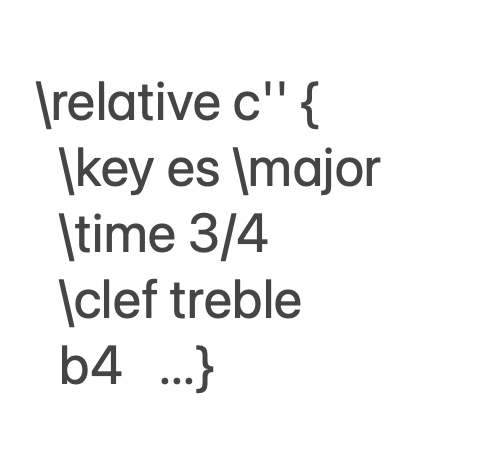}
    \caption{LiyPond}
    \label{fig:lilypond}
\end{subfigure}
\hfill
\begin{subfigure}[b]{.1\linewidth}
    \centering
    \includegraphics[width=\linewidth]{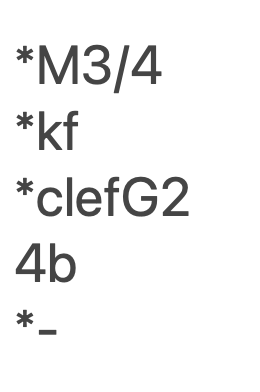}
    \caption{**kern}
    \label{fig:kern}
\end{subfigure}
\hfill
\begin{subfigure}[b]{.1\linewidth}
    \centering
    \includegraphics[width=\linewidth]{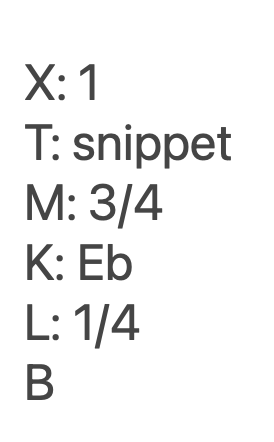}
    \caption{ABC}
    \label{fig:abc}
\end{subfigure}
\hfill
\begin{subfigure}[b]{.16\linewidth}
    \centering
    \includegraphics[width=\linewidth]{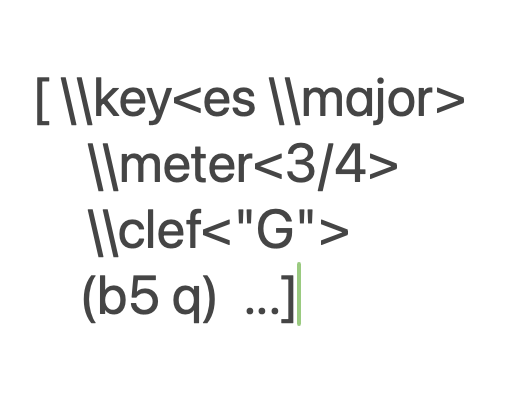}
    \caption{GUIDO}
    \label{fig:guido}
\end{subfigure}
\hfill

\vspace{1cm}
\begin{subfigure}[b]{.47\linewidth}
    \centering
    \includegraphics[width=\linewidth]{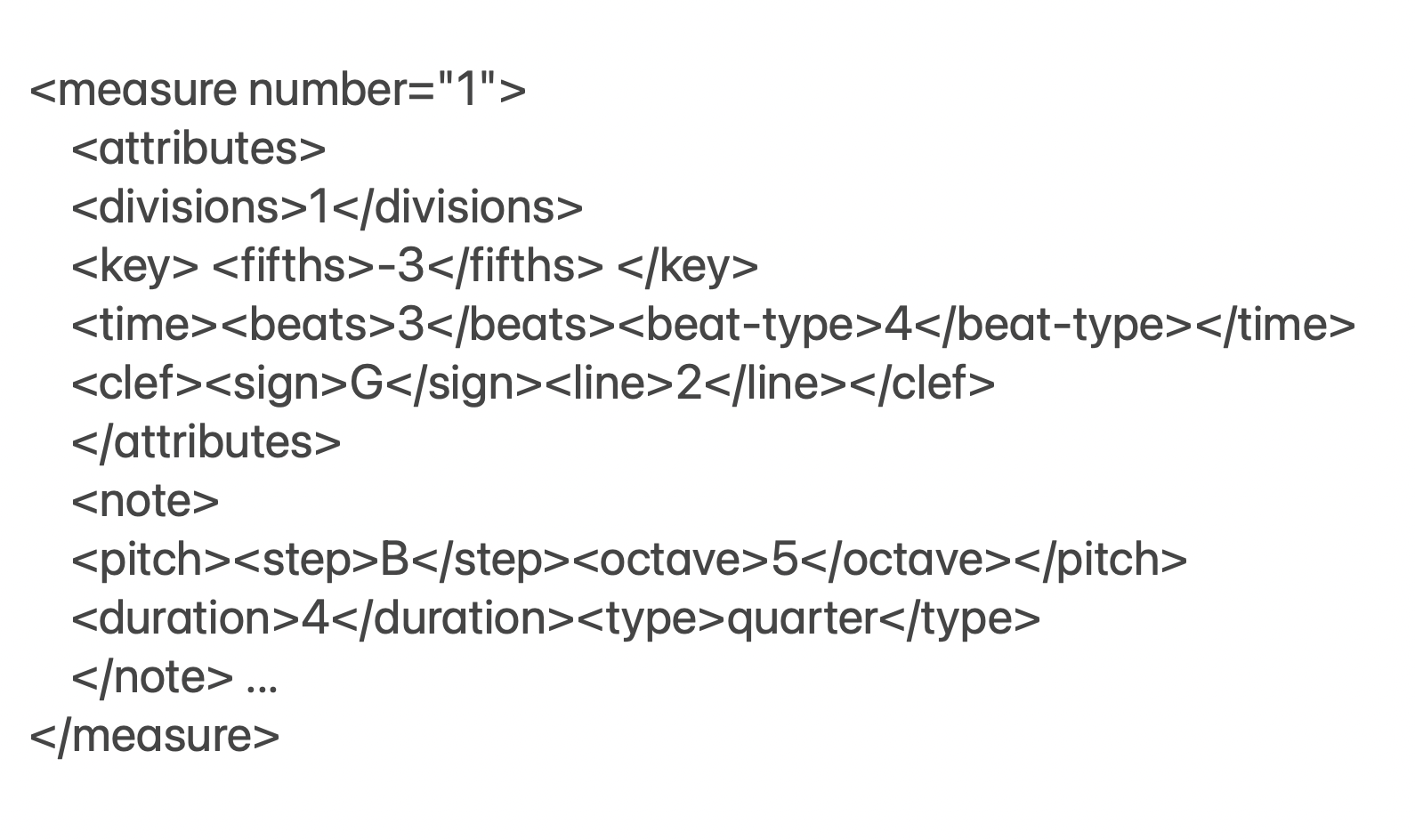}
    \caption{MusicXML}
    \label{fig:msuicxml}
\end{subfigure}
\hfill
\begin{subfigure}[b]{.51\linewidth}
    \centering
    \includegraphics[width=\linewidth]{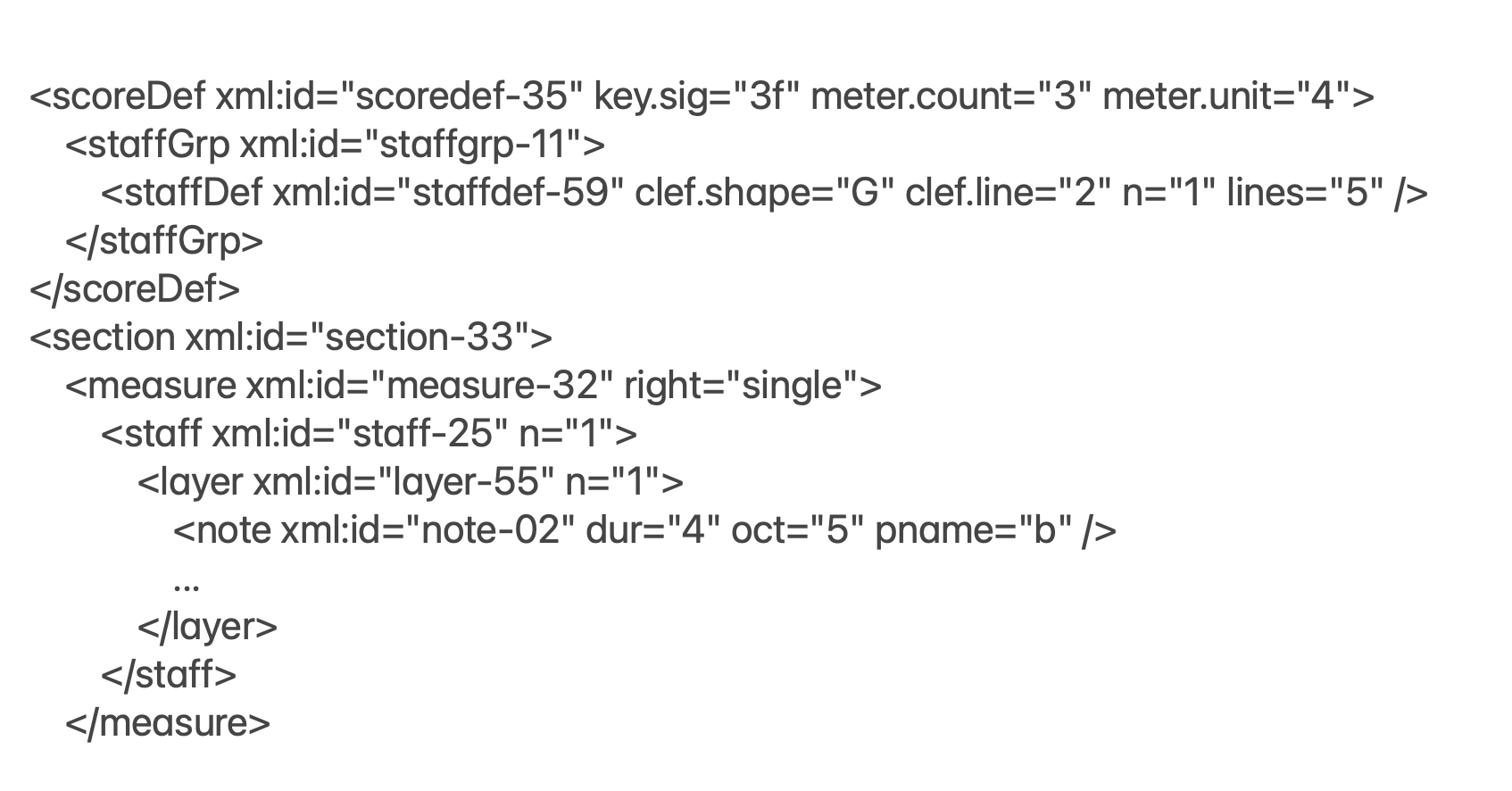}
    \caption{MEI}
    \label{fig:mei}
\end{subfigure}

\caption{Symbolic music representations for the same piece of music}
\label{fig:all_images}
\end{figure*}

\paragraph{Note-level symbolic music representation} The following paragraphs introduce the diverse note representation of symbolic music which are shown in Figure \ref{fig:all_images}.

\textbf{MusicXML} is a significant format in the digital music domain, encapsulating the complexities of musical notation in a universally transferable manner \cite{good2012musicxml}. MusicXML is rooted in the Extensible Markup Language (XML), and offers a textual format for encoding music scores (see Figure \ref{fig:msuicxml}), ensuring human and machine readability. This format enables a detailed representation of musical elements, from the pitch and duration of notes to their graphical representation in sheet music. By design, MusicXML does not encode all document information in a fully-featured music notation system, which limits capabilities for other non-CWMN scores.

Despite MusicXML's strength in accurately representing musical notation and the fact that many music scores are shared and published in MusicXML format, it is not commonly used to train FMs for music in previous works. For one thing, this is due to its symbolic nature, like ABC notation, which, while rich in notational detail, lacks the audio information necessary for models to learn the acoustic properties of music. Foundation models, especially those geared towards music generation and understanding, require data that encompass the timbral, dynamic, and expressive aspects of music as it is heard, not just as it is written. For another, the process of encoding or decoding MusicXML can be cumbersome for AI models, which thrive on larger datasets and longer context lengths for development. The complexity and verbosity of XML-based formats may introduce additional challenges in processing and interpreting data efficiently, but may also provide more benefits on music details compared to other music notations. Therefore, while MusicXML excels in notation accuracy and interoperability between software, its application in training FMs is limited by the challenges and still underexplored.

\textbf{Music Encoding Initiative (MEI)} aims at creating a digital format to represent music notation \cite{roland2002music}. MEI utilises an XML-based schema, see Figure \ref{fig:mei}, providing rules for documenting the intellectual and physical properties of music notation, enabling consistent search, retrieval, display, and exchange of information across platforms.
MEI's modular and extensible structure supports encoding various music notation systems, including common Western notation, mensural (Renaissance), and neume (Medieval) notations. This flexibility ensures an accurate representation by preserving the unique structure and semantics of each notation system rather than merely imitating them visually. One of the primary goals of MEI is to create a semantically rich model for music notation. It enables the encoding of traditional facsimile, critical, and performance editions. Its foundation on open standards and platform independence promotes the development of extensive and international archives of notated music, which serve as essential resources for editions, performances, analyses and research.
In comparison, while MEI and MusicXML encode musical elements such as notes, staves, rests, and clefs using XML, they differ in focus. MusicXML primarily facilitates interchange between notation editors. In contrast, MEI provides a more detailed and systematic encoding of notation information, supporting various notation systems beyond standard common Western notation. Moreover, MEI can document relationships among digital page images, notational features and audio recordings, offering a more comprehensive framework for music representation. Similarly to MusicXML, MEI faces challenges in training computer music FMs. Its complexity, lack of standardisation, and focus on detailed musical notation make it less suitable for ML, which requires more streamlined and uniform data formats. MEI's emphasis on visual representation rather than audio features, large data files, and limited integration with machine learning tools further hinder its effectiveness. Foundational models benefit more from compact, standardised data representations \cite{barate2016advances}.

\textbf{LilyPond} is a syntax similar to \LaTeX, allowing users to write musical notation in plain text files, which are then compiled into engraved scores \cite{nienhuys2003lilypond}. The syntax is comprehensive (see Figure \ref{fig:lilypond}) covering a wide range of musical symbols and formatting options, making it suitable for complex and professional-quality scores. It has been previously used for automatic transcription of organ tablatures \cite{schneider2021automatic} and polyphonic music transcription \cite{carvalho2017towards}. However, its complexity and focus on visual presentation can pose challenges for model training and efficiency.

\textbf{**kern} format is a symbolic music representation system designed for encoding and analysing Western music notation. Developed as part of the Humdrum Toolkit \cite{huron2002music}, **kern aims to provide a flexible and comprehensive way to represent musical scores in plain text. The format encodes musical information such as pitch, rhythm, meter, and articulation using a straightforward syntax, see Figure \ref{fig:kern}, facilitating both human readability and computational processing. **kern is highly extensible, allowing users to include additional musical parameters and metadata as needed. Its standardised text-based structure makes it suitable for training models in symbolic music analysis, pattern recognition, and music generation \cite{rahal2021separated}. Models can leverage **kern to explore musical structures, identify stylistic features, and generate compositions that adhere to specific musical conventions. The flexibility in encoding can lead to variations that require careful handling to ensure consistent and effective model training. Additionally, while **kern captures a broad range of musical elements, it may lack some of the expressive nuances found in performance-level data, potentially limiting its application in models focused on expressive performance analysis.

\textbf{ABC notation} represents a concise and computer-friendly approach to musical notation, employing a simple alphabetic system alongside ASCII characters to depict musical elements. It primarily utilises the letters a–g and A–G to denote notes, with `z' marking rests (see Figure \ref{fig:abc}). Additional notation specifies musical nuances such as sharps, flats, octave shifts, note lengths, keys, and ornamentation. This notation system, which integrates elements from Helmholtz pitch notation, was designed to simplify the sharing of music online and provide a straightforward language for software development, paralleling the simplicity of tablature and solfège. The structure of ABC notation is divided into a header and a body. The header contains metadata such as reference numbers for organising multiple pieces, titles, time signatures, default note lengths, and keys. Conversely, the body focuses on the musical content, encoding each note and rest into tokens that reflect pitch, duration, and rhythmic placement, delineated by bar lines. Like other symbolic music notations, ABC notation is compatible with natural language notations, making it an ideal candidate for developing text-to-symbolic music models, such as chatMusician \cite{yuan2024chatmusician}. This compatibility allows for the seamless integration of ABC notation with AI models designed to convert textual descriptions into musical compositions, offering a bridge between linguistic descriptions and musical notation. By leveraging the simplicity and clarity of ABC notation, developers can create models that not only generate music from text but also understand and manipulate music in its symbolic form. %

\textbf{GUIDO Music Notation (GMN)} is a text-based format designed to represent musical scores in a human-readable and machine-processable way \cite{hoosyguido, hoos2001representing}. It aims to provide a robust and flexible method for encoding music notation, focusing on simplicity and ease of use. The format utilises a straightforward syntax to represent pitches, durations, dynamics, articulations, and other musical elements. Its hierarchical structure allows for the clear organisation of musical information, making it suitable for a wide range of musical styles and notational complexities. GUIDO's design prioritises both readability and computational efficiency, facilitating its use in various digital music applications. The basic concept behind the GUIDO design is to represent simple musical ideas in a simple manner, reserving complex representations for complex notions.
GMN's simplicity and structured syntax (see Figure \ref{fig:guido}) make it an option for training models in symbolic music analysis, composition, and pattern recognition. ML models can utilise GUIDO to analyse musical structures, generate new compositions, and study stylistic patterns across different genres. The human-readable format also aids in debugging and understanding the encoded data, which is advantageous during model development and evaluation. However, the trade-off for GUIDO's simplicity is its potential lack of detail compared to more complex formats like MusicXML or MEI. Although GUIDO captures essential musical information, it may not encompass all the expressive nuances required for detailed performance analysis or highly intricate compositions. Additionally, variations in encoding practices can introduce inconsistencies, necessitating careful preprocessing for effective model training. This format has yet to be explored in the realm of ML/DL.

\textbf{Ontologies for music representation:} In the past two decades, along with the development of Web Ontologies, several music-related ontologies were also developed. However, they either need more expressiveness or are too complex for machine learning models. Ontologies like the Music Theory Ontology (MTO) \cite{rashid2018music} and the Music Score Ontology (MusicOWL) \cite{jones2017musicowl} attempt to model music-theoretical concepts and notation, respectively. MTO is limited in expressiveness for polyphonic and multi-voice music, while MusicOWL focuses on semantic representation and only partially captures engraving rules and visual formatting. The Music Note Ontology \cite{poltronieri2023music} focuses on high-level concepts and musical notes but falls short of fully integrating visual score characteristics. While these ontologies provide a rich semantic framework, their complexity and focus on semantics over visual layout limit them for use in machine learning models that need to align and understand image data (scores) with symbolic data (notation).

\subsubsection{Transformed Symbolic Music Representations for Foundational Models}

\begin{figure*}[htbp]
\centering

\begin{subfigure}[b]{.45\linewidth}
    \centering
    \includegraphics[width=\linewidth]{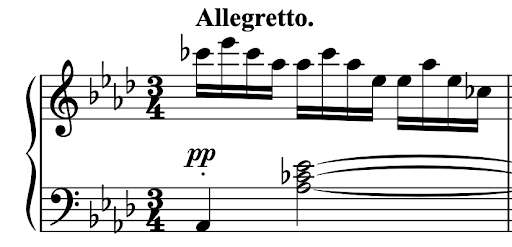}
    \caption{Score Notation}
    \label{fig:symb_figure1}
\end{subfigure}
\hfill
\begin{subfigure}[b]{.45\linewidth}
    \centering
    \includegraphics[width=\linewidth]{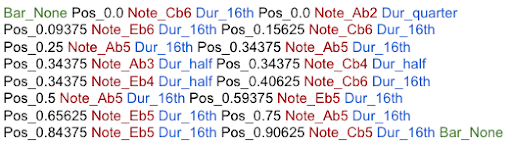}
    \caption{REMI tokenization}
    \label{fig:symb_figure2}
\end{subfigure}
\hfill

\vspace{1cm}
\begin{subfigure}[b]{.3\linewidth}
    \centering
    \includegraphics[width=\linewidth]{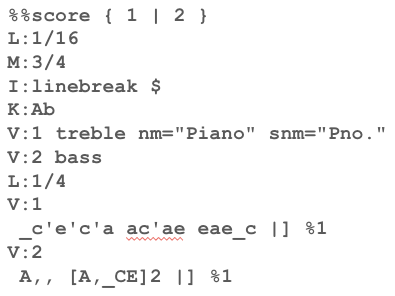}
    \caption{ABC notation}
    \label{fig:symb_figure3}
\end{subfigure}
\hfill
\begin{subfigure}[b]{.3\linewidth}
    \centering
    \includegraphics[width=\linewidth]{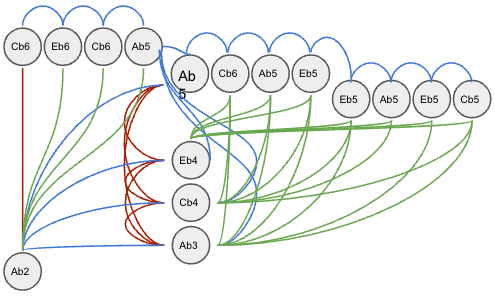}
    \caption{ self-defined graph }
    \label{fig:symb_figure_4}
\end{subfigure}
\hfill
\begin{subfigure}[b]{.3\linewidth}
    \centering
    \includegraphics[width=\linewidth]{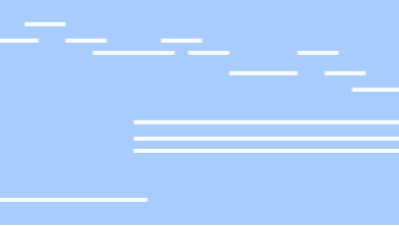}
    \caption{Generic piano-roll matrix}
    \label{fig:symb_figure_5}
\end{subfigure}

\caption{Excerpt of Schubert's \textit{Impromptu Op. 90 No.4} and its input visualisations }
\label{fig:symb_figure}
\end{figure*}

Unlike audio, symbolic music files in the previous section often contain varied informational content and do not provide a straightforward way of processing through deep-learning-based models. In this section, we introduce different mid-level representations that have been applied in literature (a visualisation can be found in Figure~\ref{fig:symb_figure}), with an emphasis on foundation models.

\textbf{Tokenised sequence representation:}  
Tokenised sequence is currently the most popular way of inputting symbolic music. It can be extracted from MIDI or MusicXML, designed with flexible encodings and emphasise specific musical information. 
Early designs like \cite{dong2018musegan} encode music as sequences comprising pitch, duration, chord, and bar position, with chords represented by a 12-dimensional binary vector and bar positions indicating a note's relative location within a bar. Other approaches \cite{Boom2020RhythmNetworks} include representing a lead sheet with sequences of pitch, rhythm, and chord, emphasising monophonic texture, or using dual sequences for pitch and rhythm with unique symbols for note continuation and rests. 
DeepBach \cite{Hadjeres2017DeepBachGeneration} distinguishes itself by using multiple lists to denote Bach's chorales across four parts, incorporating symbols for pitch maintenance and quantising time to accommodate rhythm and fermatas. 
Anticipation-RNN \cite{hadjeres2017interactive} adapts by using real note names or operationalising scores into time series to simplify the learning of polyphonic music with complex rhythms. Some models \cite{Jiang2020RL-DuetLearning, Chen2020MusicRhythm} introduce special symbols or encoding strategies for note continuation, varying from shared ``hold'' symbols to part-specific indications, thereby balancing computational efficiency with the need to capture detailed musical structures. 

\textbf{Note tuples} is a subcategory of note sequences that groups tokens in a more structural way. For instance, in the BachBot project \cite{Liang2017AutomaticLSTM}, each note in a Bach chorale is encoded as a tuple containing pitch and tie information, with each time frame comprising four tuples to represent the four voices. This method organises notes by descending pitch and uses special delimiters and symbols to indicate frame separations, fermatas, and the start and end of scores. Following a similar methodology, BandNet \cite{Zhou2019BandnetMachine} employs a Z-scan for score scanning. Furthermore, NoteTuple in \cite{Hawthrone2018TransformerPerformances} includes additional attributes like velocity, duration, and time offset for piano performance modelling, offering a succinct representation that avoids the interleaving of notes. This tuple-based representation, also used in studies like BachProp \cite{Colombo2018BachPropStyles} and PianoTree VAE \cite{Wang2020PianotreeMusic}, simplifies musical information encoding, making it a preferred choice for many researchers.

More recent MIDI tokenization designs emphasise capturing structures and reducing sequence length: The REMI (REvamped MIDI-derived events) tokenization \cite{Huang2020PopCompositions} enhances MIDI representation by focusing on rhythm modeling and structural clarity. REMI introduces Note Duration events to directly represent rhythms, replacing traditional Noteoff events. It uses a combination of Bar and Position markers to delineate musical structure, providing a clear metric grid that reflects the actual layout of music, which is employed in works like \cite{guoMusIACExtensibleGenerative2022} and \cite{chouBERTlikePretrainingSymbolic2024}. The Compound Word \cite{cpword2021} employs embedding pooling to shorten the sequence length: note tokens (Pitch, Velocity, and Duration) are independently converted to embeddings, then merged into a single one. Similarly, Octuple \cite{zengMusicBERTSymbolicMusic2021} uses embedding pooling to make sure each pooled embedding represents a single note. MuMIDI \cite{popmag2020} is designed for multitrack tasks. It contains a \textit{Track} token that precedes note tokens and includes a “built-in” and learned positional encoding. The python package MIDITok \cite{Fradet2021MiditokTokenization} provides a unified implementation for these methods.

\textbf{ABC tokens}: ABC format, as introduced in the previous section as a file type, is by itself a natural sequential representation that's compatible with language models without additional conversion steps. ChatMusician \cite{yuan2024chatmusician} is among the first to leverage the language models' ability to understand and generate symbolic music, with the text-compatible ABC music as input. \cite{qu2024mupt} further developed a Synchronised Multi-Track ABC Notation (SMT-ABC) that aims to preserve coherence across multiple musical tracks by concatenating elements from different tracks and then enclosing them by newly introduced symbols \texttt{<|>} indicating the barline. Large-scale cross-modal pre-training also demonstrated ABC notation's capability in semantic search and zero-shot classification \cite{wu2023clamp}.

\textbf{Piano roll:} The piano roll serves as a historical symbolic representation of music, dating back to the era of player pianos. These self-playing instruments used piano rolls—continuous paper rolls with punched perforations—to automatically perform music. The perforations on the roll, representing note control data, trigger the playing of notes as they pass over a tracker bar. Player pianos were capable of capturing and reproducing the performances of renowned pianists, encoding not only the pitch and duration of notes but also the dynamics of the performance. In modern music technology, the piano roll has evolved into a geometric visualisation that is used for music analysis and generation. This representation plots time on the horizontal axis and pitch on the vertical axis, with each note depicted as an axis-parallel rectangle that encodes onset time, pitch, and duration. Such a two-dimensional representation is particularly compatible with diffusion-based models, and it has been applied to tasks like transcription \cite{Cheuk2023DiffrollCapability} and generation \cite{Min2023PolyffusionControls, Li2024Diff-BGMGeneration}.

\textbf{Notes graph:} Graphs emerge as a natural representation of symbolic music since music exhibits innate structures like voices and chords that can be formed into a graph with musical heuristics. \cite{Jeong2019GraphPerformance} introduced a novel approach to representing music scores as graphs, where each note forms a node and various musical relationships between notes are depicted as edges. This graph-based representation utilises six primary types of edges—next, rest, set (onset), sustain, voice, and slur—to capture the intricate connections within the score. 
The model distinguishes forward and backward directions, along with a unique self-connection for each note, culminating in a total of 12 edge types for a comprehensive and detailed musical score representation. Similarly, \cite{Karystinaios2023MusicalProblem} also created a heterogeneous graph from score notes and tackles the voice separation problem as graph link prediction in multi-trajectory tracking. \cite{Zhang2023SymbolicEvaluation} further examined the potential of applying score or performance graphs with various edge designs on music understanding problems and compared them with other representations counterparts.

\textbf{Tonnetz matrix} offers a distinctive two-dimensional approach to representing music in computational models, setting it apart from conventional piano roll representations. Inspired by music theory, it organises music into sequences of 2D matrices, where each node corresponds to one of the 12 pitch classes. These nodes are arranged to reflect harmonic relationships, with horizontal lines following the circle-of-fifths and triangles representing major and minor triads. The network's expandable nature allows pitch classes to appear multiple times, facilitating the depiction of complex harmonic structures. According to \cite{Chuan2018ModelingRepresentation}, the use of the Tonnetz in music generation leads to outputs with greater pitch stability and more repetitive patterns, showcasing its potential for innovative music representation and generation in computer models.

\subsection{Multimodal Music Representations}

Music is a multidimensional artistic medium with various facets, each of which gives musical compositions a unique viewpoint. Apart from the auditory and symbolic aspects that are immediately associated with music, two important modalities in the field of multimodal learning for music have been extensively investigated:

\textbf{Text}: Words are essential to creating and comprehending music, including lyrics, metadata, and user comments. They provide a deeper understanding of the themes and storylines found in music by bridging the gap between musical expression and language context.

\textbf{Visual}: Visual components such as album art, frames from music videos, and associated visuals not only enhance the whole listening experience but also have an aesthetic impact.

The subsequent sections will delve into multimodal representations of music, specifically focusing on the interaction between music tokens and their representations in both textual and visual domains.

\subsubsection{Interactions with Textual Modality}
In this section, we will examine music-text representation learning. Traditional NLP methods, such as TF-IDF, will not be covered. Instead, we will briefly introduce contemporary deep learning approaches, including using pre-trained embedding from encoder models like the encoder of T5 \cite{raffel2020exploring}, using pre-trained language models as decoders like Llama2 \cite{DBLP:journals/corr/abs-2307-09288}, and training from scratch.

\paragraph{Pre-trained Embeddings from Natural Language Encoders}
There are three types of pre-trained text encoders used as text conditioning in previous work: pre-trained encoders such as the encoder of T5, instructed-based encoders like FLAN-T5 \cite{chung2024scaling}, and pre-trained text-audio encoders like CLAP \cite{clap}.

Audiogen \cite{kreuk2022audiogen} is an LLM conditioned on T5 embeddings to generate learned quantised audio tokens autoregressively.
Noise2Music \cite{DBLP:journals/corr/abs-2302-03917}, and Moûsai \cite{schneider2023mo} generate high-quality music clips with a Latent Diffusion Model (LDM) conditioned on pre-trained T5 embeddings of text prompts.
ERNIE-Music utilises ERNIE-M \cite{ouyang2021ernie} to encode multi-lingual inputs to produce music waveforms directly from free-form text \cite{DBLP:journals/corr/abs-2302-04456} with a LDM architecture.

MusTango \cite{melechovsky2023mustango} leverages FLAN-T5, showcasing advancements in enhancing music generation controllability and employing a high-fidelity Latent Diffusion Model (LDM) for text-guided universal music generation \cite{DBLP:journals/corr/abs-2308-04729}. Similarly, MusicGen \cite{copet2023musicgen} integrates FLAN-T5 with T5 and CLAP to generate conditioning text embeddings for an audio LLM.

MusicLDM \cite{DBLP:journals/corr/abs-2308-01546} is a diffusion model that has been proven powerful in text-conditioned music generation. It uses the pre-trained model CLAP during training to condition on audio embeddings. At inference time, CLAP is used to extract text embeddings to condition generation. Effective text-to-music generation is made possible by this method, which guarantees varied, high-quality outputs that complement the original content. MusicLM \cite{agostinelli2023musiclm} and Make-an-audio \cite{makeanaudio} also utilise CLAP embeddings as a text condition for music generation.

\paragraph{Pre-trained Embeddings from Natural Language Decoders}
Significant progress has been achieved in LLM-integrated frameworks for multimodal music generation and understanding.
MusiLingo \cite{deng2023musilingo} combines Llama \cite{llama} and a pre-trained acoustic music representation to perform music captioning and instruction following, such as generating music-related Q\&A pairs. Similarly, LLark \cite{DBLP:journals/corr/abs-2310-07160} combines Llama2 \cite{DBLP:journals/corr/abs-2307-09288} and multiple generative audio encoders, providing good performance in music instruction, specifically in zero-shot situations.

M$^2$UGen \cite{DBLP:journals/corr/abs-2311-11255} understands and produces music through the integration of images and videos, utilising the pre-trained Llama 2 model to achieve a comprehensive understanding of music. Using pre-trained audio representation models, Authors of \cite{DBLP:journals/corr/abs-2308-11276} improve text-to-music creation by adding music-related question responses and captioning.

In addition to the models mentioned previously, pre-trained natural language decoders such as GPT-4 can act as agents to make complex AI music tools more accessible to a larger audience by streamlining a range of music-related tasks. For more information, please refer to Section \ref{sec:interpretability}.

\paragraph{Trained jointly from Scratch}
ChatMusician \cite{DBLP:journals/corr/abs-2402-16153} interprets music as a language by adding ABC notations to Llama vocabulary to train a tokeniser and a GPT model based on it. It is proficient at creating well-organised compositions from ABC notation through understanding subtle musical aspects.

SongComposer is trained based on a tokeniser as a tuple of symbolic song notes and lyrics  \cite{DBLP:journals/corr/abs-2402-17645}, which transforms textual descriptions into musical compositions, improving the coherence between melody and lyrics.

By training models from scratch, researchers can directly align the tokenisation and modelling processes with the unique structure and semantics of musical data.

\subsubsection{Interactions with Visual Modality}
\paragraph{Pre-trained Embedding from Visual Encoder}
Pre-trained visual models like the Vision Transformer (ViT) \cite{dosovitskiy2020image} and Video Vision Transformer (ViViT) \cite{arnab2021vivit} have revolutionised tasks that integrate visual and musical elements. Utilised by systems such as M$^2$UGen~\cite{hussain2023m}, these models adapt NLP-focused transformer architectures to process image patches and spatio-temporal visual tokens.

Both V2Meow \cite{DBLP:journals/corr/abs-2305-06594} and VidMuse \cite{tian2024vidmuse} leverage the multimodal CLIP model \cite{clip} as their visual encoders to effectively bridge the gap between visual and audio modalities. In V2Meow, the visual encoder extracts high-level features from video frames, which are then used to condition the auto-regressive music generation model. Similarly, VidMuse employs CLIP's visual encodings to capture both local and global visual cues. By integrating these cues through long- and short-term modelling, VidMuse produces music tracks that are not only acoustically rich but also semantically synchronised with the video content.

\paragraph{Trained jointly from Scratch}
AnyGPT \cite{zhan2024anygpt} uses a unified approach to process multiple modalities, including speech, text, images, and music with specialised tokenisers. For example, music uses the EnCodec model \cite{encodec} to convert audio into discrete tokens, while images use the SEED tokeniser \cite{DBLP:journals/corr/abs-2307-08041} for visual data. These tokens are processed by the language model and reconstructed using de-tokenisers, ensuring the perceptual integrity of the original modalities and enabling effective content generation across diverse inputs and outputs.

MM-Diffusion \cite{DBLP:conf/cvpr/RuanMYH0FYJG23} introduces a joint audio-video generation framework that employs a sequential multimodal U-Net to generate aligned audio-video pairs from Gaussian noise. The integration of audio and video generation in a single framework ensures that the audio and visual elements are harmoniously aligned, resulting in more immersive and coherent multimedia content.

\section{Applications}\label{sec:applications}

In this section, we will introduce the application of foundation models on music, including music understanding, music generation and musical application. Traditional approaches also have such kinds of applications but lack versatility. Foundation models (FMs) like the product of sunoAI and Q-wen-audio have or potentially have better and more generalised performance due to the self-supervised learning paradigm and a larger number of parameters.

\subsection{Music Understanding}\label{sec:music_understanding}

Music information retrieval (MIR) refers to the field of research focusing on retrieving information from music data \cite{downie2003music}. The field's origins stem from the wider discipline of information retrieval (IR) and information science, although its scope and methodologies have shifted such as the community can be said to more broadly focus on Music Information \emph{Research} \cite{serra2013roadmap}. The focal point of the MIR community is the ISMIR\footnote{https://ismir.net/} society and conference (standing for \emph{International Society for Music Information Retrieval}), first established in 2000 as an international symposium before converting into a conference in 2002. The original scope of the research community during the inception of the conference was on retrieving information from symbolic music representations and managing bibliographic music collections (with a large part of stakeholders being music libraries), although over the years audio signals have become the predominant modality used in MIR research. The term MIR is sometimes used interchangeably with \emph{Music Informatics} or \emph{Music Information Processing} \cite{ç}, and there are also strong links between MIR and the signal processing sub-field of \emph{Music Signal Processing} (showcased in conferences such as IEEE ICASSP\footnote{https://ieeexplore.ieee.org/xpl/conhome/1000002/all-proceedings}) and more recently with the sub-field of AI focusing on \emph{Music AI} \cite{miranda2021handbook}.

To the authors' knowledge, there is no systematic and commonly agreed taxonomy of the different problems and tasks that have been of interest to the MIR community over the years, which could subsequently form the basis for developing music foundation models. Certain task categorisations and groupings have been proposed in various MIR survey papers and textbooks \cite{serra2013roadmap, ç, schedl2014music}, and a fairly comprehensive list of popular MIR tasks can be found as part of the MIREX public evaluation challenge\footnote{https://www.music-ir.org/mirex/wiki/MIREX\_HOME} that ran annually between 2005 and 2021. The below paragraphs will provide a brief overview of common MIR tasks organised based on different musical dimensions (e.g. rhythm, harmony, timbre), noting that the majority of MIR research activity uses audio as its primary modality. Increasingly, the MIR community also includes tasks beyond analysis, which include music generation or manipulation; these will be covered in the subsequent subsection \ref{sec:generation}; in addition, tasks related specifically to multimodal MIR and singing voice will also be presented in a separate subsection \ref{subsec:multimodal_understand} and \ref{subsec:vocal_understand} respectively.

\subsubsection{Traditional Music Information Retrieval Tasks}

While coming up with a single MIR taxonomy is challenging and out of scope for this work, one suggested task categorisation could be as follows:
\begin{itemize}
    \item Tasks related to tonality and harmony: mode, chord, and key detection
    \item Tasks related to melody and pitch: melody estimation, pitch \& multi-pitch detection, note tracking, automatic music transcription
    \item Tasks related to rhythm: onset detection, beat \& downbeat tracking, metre estimation, tempo estimation
    \item Temporal alignment tasks: score following, audio-to-score alignment, score alignment
    \item Source separation tasks: musical instrument source separation, harmonic-percussive source separation
    \item Timbre-related tasks: musical instrument identification, playing technique detection
    \item Clip-level classification tasks: music tagging, genre recognition, emotion/mood recognition
    \item Content-based audio retrieval: audio identification, audio matching, cover song detection
    \item Temporal segmentation tasks: music detection, music structure segmentation, time boundary identification
    \item Tasks with visual score input: optical music recognition and subtasks, including staff line identification, music symbol identification
    \item Performance-related understanding: technique identification, performer identification, performance assessment, difficulty estimation
\end{itemize}

Given the extensive task list above, there is no single unified resource for providing task definitions and benchmark methodologies, although there exist textbooks that cover a subset of the above tasks (e.g. \cite{ç, lerch2022introduction}), with ISMIR conference proceedings\footnote{https://ismir.net/conferences/} being useful for providing additional technical details. Review and overview papers do exist for specific tasks (which might or might not be up to date with related literature depending on the publication year), including on onset detection \cite{bello2005tutorial}, music transcription \cite{benetos2019transcription}, chord estimation \cite{mcvicar2014automatic}, melody extraction \cite{salamon2014melody}, score following \cite{orio2003score}, music source separation \cite{cano2018musical}, music classification \cite{fu2010survey}, audio identification \cite{cano2005review}, cover song detection \cite{serra2010audio}, optical music recognition \cite{calvo2020understanding}, plus general roadmaps for the community \cite{serra2013roadmap}.

Attempting to summarise methodological trends, most MIR tasks above focus on audio as the input modality, with a smaller subset of tasks using machine-readable scores or visual representations as input. The vast majority of MIR methods focus on Western tonal music, with a much smaller subset of research focusing on non-Western cultures or folk/traditional music. The varying temporal granularity between different tasks is a key differentiator, with MIR ranging from clip-level classification tasks such as audio tagging (or even higher level tasks such as setlist identification) to tasks which require prediction of musical cues at a fine temporal level (e.g. pitch detection, onset detection). These varying task requirements in terms of temporal granularity, combined with the diversity of music cultures in a global scale, might potentially prove an obstacle for the creation of a general-purpose music foundation model, as will be discussed in the following sections. On the other hand, the clear connections that exist between different tasks (e.g. onset detection is directly linked with beat tracking) can lead to the creation of music representations and MIR models that can potentially address multiple tasks, possibly at the expense of musical diversity.

More recently, large music audio models based on pre-training and self-supervised learning have been proposed to tackle a range of MIR tasks, therefore leading to the emergence of foundation models for music. Technical details of those models will be presented in Section~\ref{sec:technical-details-foundation-models}, and a brief summary for completeness will be provided here. In what might be one of the first attempts towards establishing a common testbed for multiple MIR tasks (thus aiming towards the creation of music understanding foundation models), MARBLE \cite{yuan2024marble} introduced a benchmark with a hierarchical taxonomy for 18 MIR tasks across acoustic, performance, score, and high-level description levels. 

\begin{figure}[t]
\includegraphics[height=2.5in]{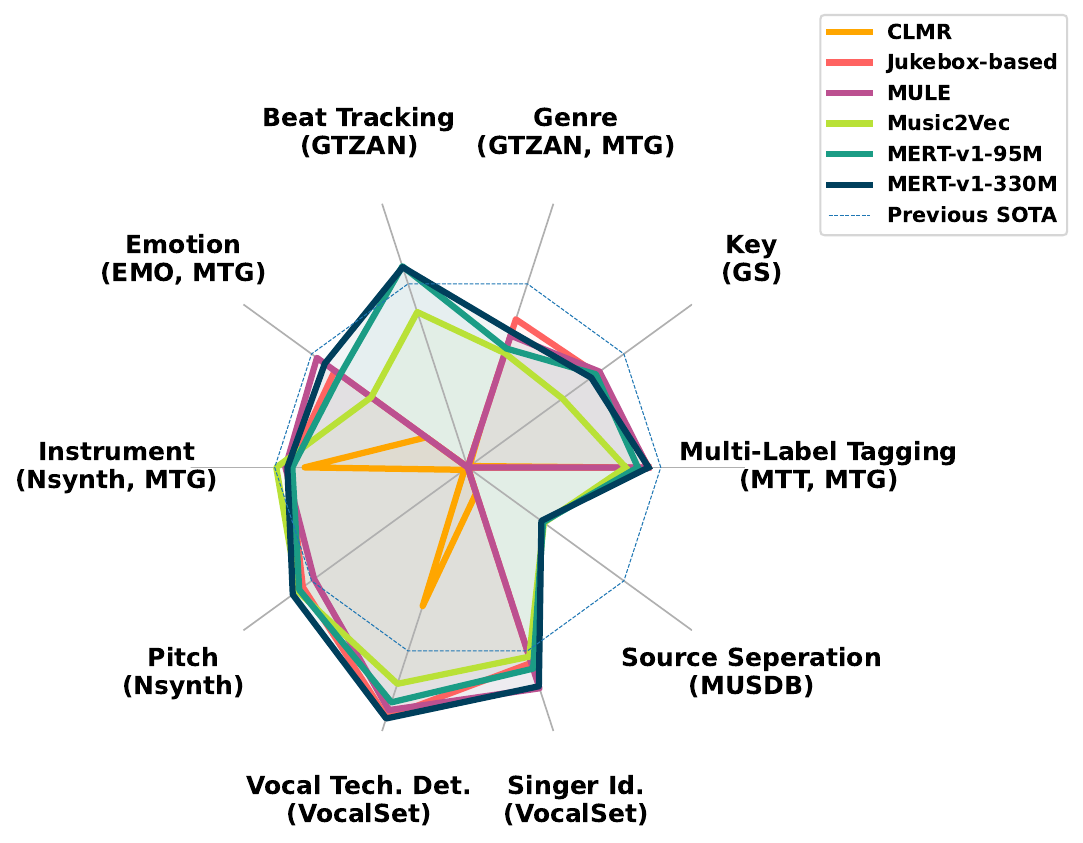}
\caption{Comparison of various music audio self-supervised models evaluated on a range of different tasks, as reported through the MARBLE benchmark \cite{yuan2024marble}. Figure reprinted with permission.}
\label{fig:MARBLE}
\end{figure}

Specific models that might be classed as foundation models for music or large music audio pre-training models include JukeMIR \cite{castellon2021codified}, which explores learnt representations from Jukebox \cite{dhariwal2020jukebox} for four MIR tasks; MULE \cite{mccallum2022supervised}, which is a self-supervised model that combines a SlowFast component with a variant of the well-known ResNet architecture, pre-trained on the MusicNet dataset and applied on music classification tasks; Music2Vec \cite{li2022map}, which is a self-supervised model using a masked prediction strategy with student and teacher models and applied to a variety of MIR tasks not limited to classification; MERT \cite{li2023mert} which combines two teacher models using a masked language modelling acoustic pre-training approach, applied to 14 MIR tasks; and finally MusicFM \cite{MusicFM}, which also applies a self-supervised learning approach taking advantage of random quantisation and applied to 5 MIR tasks. Fig.~\ref{fig:MARBLE} shows results of music audio pre-trained models across a range of MIR tasks, as reported through the MARBLE benchmark \cite{yuan2024marble}.

\subsubsection{Multimodal Music Understanding Tasks}\label{subsec:multimodal_understand}

multimodal information is abundant in the real world, where agents understand and interact through various channels such as vision, language, speech, and touch. 
Therefore, AI systems need to effectively process and integrate data from these diverse sources to improve their ability to comprehend and engage with their environment. 
This section reviews extensive research on multimodal music understanding, which connects the domain of music with other modalities like vision and language.
It highlights key advancements and methodologies from recent studies, showcasing how these developments enhance our understanding and interaction with music through multiple forms of information.
Subsequently, we organise these methodologies into: 1. Audio-Visual Joint Representation Learning and Synthesis; 
2. Audio-Language Joint Representation Learning and Synthesis; %
3. %
multimodal Extensions of Large Language Models. For the multimodal application to music generation, typically text2music generation, please refer to section\ref{sec:generation}.

\paragraph{Audio-Visual Joint Representation Learning}

\textbf{Music source localisation in visual context.}
PixelPlayer\cite{Zhao_2018_ECCV} learns to isolate musical sources and spatially localise them within the visual context from unlabelled video sequences. Through a joint audio-visual learning approach, PixelPlayer can identify objects and their sounds, localise objects in images or video, and separate audio elements emitted by each object. 

\textbf{Video music retrieval \& recommendation}
\cite{li2019query} deploys a dual-stream network for music retrieval from videos. This network learns to match music and videos by understanding their cross-modal distances, and it is trained end-to-end with music-video pairs. Meanwhile, it creates an emotion-based latent space, pre-trained through audio and video emotion tagging.
\cite{DBLP:journals/spm/MullerABDW19} addresses the requirement for cross-modal retrieval in music, allowing users to query one modality (audio, for example) and retrieve related information from other modalities (sheet music, album covers, etc.) . 
 VM-NET employs a specialised loss function to pair videos with music, preserving the unique characteristics of each modality~\cite{DBLP:conf/mir/HongIY18}. 
 Improving upon this, a study enhances retrieval quality by shifting from manual feature extraction to learned feature representations~\cite{DBLP:conf/ijcnn/PretetRP21}. 
 MRCMV, a framework for selecting background music for detailed virtual-content videos, leverages SlowFast and VGGish networks to extract video and audio features. It further employs self-attention and cross-attention modules to understand intra- and inter-modal relationships, with a fusion gate determining the contribution of each feature type~\cite{DBLP:conf/sigir/LiSZLWZYS21}. 
~\cite{DBLP:conf/cvpr/SurisVRS22} capture the extended temporal context within music and video modalities, offering a sophisticated approach to content pairing. 
 ~\cite{DBLP:journals/sensors/GuSL23} utilises a dual-path cross-modal network to integrate content and emotional information to enhance music recommendations for videos. 
 Addressing the challenges of label noise and the inadequate capture of key video segments, the Saliency-based Self-training for Video-Music Retrieval (SSVMR) framework has been proposed~\cite{DBLP:conf/icassp/ChengZLLZ23}. ViML takes a trimodal approach, recommending music by leveraging integrated representations from both video and text inputs, showcasing the potential of combining multiple modalities~\cite{DBLP:conf/cvpr/McKeeSSR23}.

\textbf{Visual \& audio classification}
MAViL~\cite{huang2024mavil} combines Masked Autoencoders (MAE) and contrastive learning to develop a self-supervised learning method for handling the heterogeneous audio and video modalities. MAViL reconstructs heavily masked inputs, learning audio-video representations and aligning them in a unified latent space. Moreover, MAViL introduces a pre-training task that reconstructs unified, context-aware audio-video representations instead of simple single-modality input reconstruction.

\paragraph{Audio-Language Joint Representation Learning}
\textbf{Music captioning}, which involves creating descriptive text for music, represents a unique blend of linguistic expression and musical perception. This emerging field aims to enhance music accessibility and understanding through natural language. 
The curation and augmentation of datasets are crucial in this field. 
To tackle the data scarcity issue in music captioning, \cite{DBLP:conf/ismir/DohCLN23} proposes a method that employs LLMs to generate pseudo captions from tags. 
\cite{DBLP:journals/corr/abs-2311-10057} introduces the Song Describer Dataset (SDD), a collaborative collection featuring more than one thousand natural language descriptions for 706 musical recordings.
To advance music captioning, novel model architectures have been developed. 
\cite{DBLP:conf/ijcnn/MancoBQF21} propose a multimodal encoder that combines a CNN for audio and an LSTM for jointly representing audio and text. This method effectively captures high-level semantics and summarises information across varying levels of input granularity. 
~\cite{DBLP:conf/ismir/ZhangJXD22} integrates a CNN-SA audio encoder with a pre-trained language model to improve song lyrics interpretation.
ALCAP~\cite{DBLP:conf/emnlp/HeHLCLS23} utilises an alignment module through cross-modal contrastive learning between music and lyrics, enhancing the latent representation of music and lyrics. 
\cite{DBLP:conf/hcmir/LeeDJ23} finds that subjective annotator insight is crucial for crafting deep learning models tailored to music captioning. 
Based on a multi-layer cross-attention mechanism,~\cite{du2024joint} employs a joint attention model to improve the interaction between musical and textual information across various semantic layers.

\textbf{Music instruction following}
Mu-llama \cite{DBLP:journals/corr/abs-2308-11276} and MusiLingo \cite{deng2023musilingo} integrated MERT\cite{li2023mert}, a pre-trained music encoder, and Llama, a pre-trained natural language decoder. Mu-llama employs a Llama-adapter on the upper transformer layers for modality alignment \cite{zhang2023llama} while MusiLingo utilises a simple linear projection on the initial transformer layer, yielding enhanced performance. Given the diverse range of tasks within the Music Understanding category, recent works have explored the application of Multimodal Instruction Tuning to enhance the performance of Music Language Models. 
LLark \cite{gardner2023llark} and JMLA~\cite{du2024joint} investigate how this approach can be effectively applied to the training of Music LLMs, aiming to improve their versatility and effectiveness across a wide spectrum of music-related tasks. Both models employ a similar architectural approach, utilising cross-attention modules to connect audio encoders with pre-trained LLMs. This integration allows for the effective processing of both audio and textual information. To create robust instruction-tuning datasets, these studies leverage traditional music tagging datasets, employing data augmentation techniques to expand them into comprehensive datasets suitable for fine-tuning Multimodal LLMs. 
By incorporating this approach, these models demonstrate the potential for more adaptive and comprehensive music understanding capabilities, effectively bridging the gap between audio processing and natural language understanding in the music domain.

\textbf{Text-music information retrieval} is an evolving field that integrates text and visual data to enhance access to and understanding of music. This interdisciplinary approach enriches interactions between sensory modalities, allowing for more nuanced music analysis and retrieval. 
MuLaP proposes a multimodal pre-training framework for MIR that integrates audio and text features in an early fusion manner~\cite{DBLP:conf/icassp/MancoBQF22}. In contrast, other researchers have adopted a dual-tower architecture focusing on contrastive learning. This method aligns textual descriptions with their corresponding musical representations by leveraging a large corpus of music-text pairs for pre-training~\cite{DBLP:conf/ismir/WuY0S23, manco2022contrastive, mulan}. This dual-tower approach allows for distinct processing of the modalities before their interaction, potentially leading to more effective cross-modal alignment.

\textbf{Text-music understanding \& generation}
AudioGPT~\cite{huang2023audiogpt} is a multimodal model that processes and generates audio from text and audio inputs. It follows four steps: It converts speech to text, interprets the task to understand user intent, uses GPT to match tasks with audio models based on features like prosody and timbre, and then the audio models generate a response for the user.
AudioPaLM~\cite{rubenstein2023audiopalm} uses a dual-stream architecture with text~\cite{anil2023palm} and speech~\cite{audiolm} language models. It integrates speech and text into a multimodal generative model and creates a joint vocabulary, enabling the model to process and generate both speech and text.

\paragraph{Audio-Visual-Language Joint Representation Learning}
Several studies leverage Large Language Models (LLMs) like GPT and LLaMA for multimodal applications, extending their capabilities to perform extensive music understanding tasks.

AnyGPT \cite{zhan2024anygpt} converts multimodal input into semantic tokens which are capable of being subsequently converted back to their original forms by de-tokenisers. This method preserves the perceptual integrity of the original modalities while enabling the LLM to handle understanding and generation tasks autoregressively.
M$^2$UGen~\cite{hussain2023m} is a framework designed for multimodal music understanding and generation. M$^2$UGen employs different modality encoders to interpret various inputs: the Vision Transformer (ViT)~\cite{dosovitskiy2020image} for images, ViViT~\cite{arnab2021vivit} for videos, and the MERT~\cite{li2023mert} model for music. After processing the inputs into feature representations, modality-specific adaptors refine them. Subsequently, the LLaMA2~\cite{llama} model interprets these modality signals for various downstream tasks. Specifically, M$^2$UGen then uses AudioLDM2~\cite{audioldm2} for music generation.
NExT-GPT~\cite{wu2023next}, a model constructed on the M$^2$UGen framework, can process and generate content across text, images, videos, and audio. This model employs specific encoders and projection layers to convert inputs from diverse modalities into a unified representation space. It utilises a large language model Vicuna~\cite{chiang2023vicuna} for interpreting these multimodal inputs. NExT-GPT utilises pre-trained diffusion models, such as AudioLDM~\cite{audioldm}, for audio generation. 

\subsubsection{Vocal Music
Understanding}\label{subsec:vocal_understand}

Vocal music understanding is similar to speech understanding and/or speech processing, and many techniques for vocal music or singing voice are inspired by the algorithms for speech processing, such as speaker identification for singer identification and speech recognition for lyrics transcription. The algorithms for speech are typically optimised for more uniform pitch, rhythm, and dynamic articulation. Different from speech signals, the singing voice has a wider pitch range, greater variability in rhythm and pronunciation, and involves complex vocal techniques. Therefore, keeping the differences in mind is essential when applying the algorithms to singing voice.

In the past few years, SSL models for speech processing have proven effective in various vocal understanding tasks. There has also been an exploration into training SSL features specifically for the singing voice. This section introduces representative work in singer identification, lyrics transcription, singing transcription, vocal source separation, and lyrics interpretation.

\paragraph{Singer Identification (Singer ID)} 
This task involves recognising the identity of a singer\footnote{We do not differentiate it from similar tasks such as artist classification}. It is often formulated as a classification problem within a dataset containing a fixed number of singers. Many approaches extract Mel-Frequency Cepstral Coefficients (MFCCs) to capture voice characteristics, paired with Gaussian Mixture Models (GMMs) \cite{DBLP:conf/icmcs/Zhang03, DBLP:conf/icnc/CaiLG11}. Embeddings designed for speaker recognition and verification, such as i-vector \cite{DBLP:conf/interspeech/PorteloART13} and x-vector \cite{DBLP:conf/icassp/SnyderGSPK18}, have also been used in singer ID \cite{DBLP:conf/dafx/Kruspe14, DBLP:conf/ijcnn/ZhangWCX22c}. With the advancement of neural networks, it has become possible to extract latent features from general audio representations, such as spectrograms \cite{DBLP:conf/ismir/LeeN19, DBLP:conf/icassp/ZhangQYSL21}, using CNNs and CRNNs. 
Recent methods propose using singing source separation to suppress background music \cite{DBLP:conf/interspeech/SharmaD019a} or data augmentation \cite{DBLP:conf/icassp/HsiehCFYY20}. These approaches have improved accuracy and F1 scores on the Artist20 \cite{DBLP:conf/ismir/Ellis07}. However, these methods have been tested on a small, fixed number of singers, and their ability to scale up and generalise to unseen singers remains unknown.

There have been several attempts to approach singer ID through representation learning. \cite{DBLP:conf/ismir/LeeN19} proposed a triplet network that learns a joint embedding for monophonic and mixture singing. \cite{DBLP:conf/icassp/ZhangQYSL21} transformed the classification problem into a query searching task by replacing the softmax layer with a k-nearest neighbours (KNN) layer. \cite{DBLP:conf/ismir/TorresLR23} trained SSL features on source-separated singing voices and compared them with speech SSL features in terms of classification accuracy.

\paragraph{Automatic Lyrics Transcription (ALT) \& alignment:} These are two closely related tasks aimed at recognising and locating lyrics within singing. Lyrics transcription focuses on identifying the linguistic content of the singing voice, while lyrics alignment involves retrieving the timestamps of each text unit of lyrics, typically words. Both depend on an acoustic model that captures the relationship between audio signals and phonetic information. Previous work has utilised the same training pipelines for both tasks to develop an effective acoustic model.

Early approaches to automatic lyrics transcription and alignment utilised a Gaussian Mixture Model-Hidden Markov Model (GMM-HMM) design or a factorised time-delay neural network model, often combined with a language model \cite{DBLP:conf/interspeech/DabikeB19}. Building on the traditional modular system, some researchers proposed enhancements in architecture  \cite{DBLP:conf/ijcnn/DemirelAD20} and multi-domain training \cite{DBLP:conf/ismir/DemirelAD21}. \cite{DBLP:conf/icassp/StollerDE19} introduced end-to-end training using the connectionist temporal classification (CTC) loss function. Subsequent works explored various aspects of the task using end-to-end methods, including multilingual application \cite{DBLP:conf/ismir/VaglioHMRd20, huang2024towards}, training domain \cite{gao2022music}, joint training \cite{DBLP:journals/taslp/GaoGL23, DBLP:conf/icassp/HuangBE22}, etc.

Given the great success of pre-trained self-supervised learning (SSL) features in speech recognition, researchers have begun applying these features to automatic lyrics transcription (ALT). \cite{DBLP:conf/ismir/OuGW22} were the first to finetune wav2vec 2.0 features on source-separated singing, achieving significant improvements. Following this, \cite{DBLP:conf/icassp/GaoYL23} applied self-training with noisy student augmentation. Whisper, a robust speech recognition model released by OpenAI \cite{radford2023robust}, has also proven effective for singing voices. Efforts have been made to create datasets using Whisper \cite{DBLP:conf/ismir/ZhuoYPMLZLDFLBC23} and to adapt it for low-resource language ALT \cite{DBLP:conf/asru/WangLLSJ23}.

\cite{DBLP:conf/icassp/DurandSE23} trained a similarity model between audio and text using contrastive loss and applied it to lyrics alignment. There have also been attempts to apply SSL features trained on music signals to ALT, such as in the MARBLE benchmark \cite{DBLP:conf/nips/YuanMLZCYZLHTDW23}, although these efforts have had limited success.

\paragraph{Singing Transcription: }
Singing transcription aims at estimating the musical notes of a sung melody. Vocal F0 estimation, or melody extraction, is a closely related task that focuses on extracting the fundamental frequency of the singing voice and often plays a role in typical cascading systems for singing transcription. Early approaches for singing transcription consist of an F0 estimation module and a quantisation module with segmentation to obtain note-level transcription. In \cite{DBLP:journals/taslp/MolinaTBB15, DBLP:journals/taslp/KroherG16}, the singing voice is first processed through pitch extraction algorithms, followed by note segmentation and quantisation modules that generate symbolic notations.
To avoid error propagation by modules, end-to-end approaches using encoder-decoder models have been proposed  \cite{DBLP:conf/waspaa/NishikimiNGY19, DBLP:conf/icassp/NishikimiNFGY19}. \cite{DBLP:journals/taslp/WangJ23} proposed using both Connectionist Temporal Classification (CTC) loss and cross-entropy loss to facilitate learning from weakly labelled data.
To leverage information such as beat and phoneme, these elements are incorporated either through multi-task learning \cite{DBLP:conf/waspaa/NishikimiNGY19, DBLP:journals/tomccap/GuOZZWW24} or conditioning mechanisms \cite{DBLP:conf/icassp/YongSN23}. 
Inspired by object detection and sound event detection work, MusicYOLO \cite{DBLP:journals/taslp/WangTYXC23} retrieves note objects from a macro perspective through an object detection model. 

Different training methods for singing transcription have also been investigated.  \cite{DBLP:conf/icassp/NishikimiNFGY19} proposed a loss function on the attention weights using semi-supervised alignment labels. \cite{DBLP:conf/icassp/KumLKKN22} introduced self-training with the noisy student framework \cite{DBLP:conf/cvpr/XieLHL20}, while \cite{DBLP:conf/ismir/Hsu021} utilised virtual adversarial training \cite{DBLP:journals/pami/MiyatoMKI19}, both leveraging unlabelled singing data.

\paragraph{Vocal Source Separation:} The singing voice is one of the four primary stems targeted in source separation, with applications such as karaoke and as a crucial component in various vocal understanding tasks. Therefore, researchers have focused on improving singing voice separation.
To enhance the performance of vocal source separation (VSS) and related singing analysis tasks, multi-task learning approaches have been proposed. These include joint vocal activity detection \cite{DBLP:conf/ica/StollerED18}, pitch estimation \cite{DBLP:conf/eusipco/JanssonBEW19}, and lyrics transcription \cite{DBLP:journals/taslp/GaoGL23}. Additionally, some methods incorporate lyrics information into the model to improve intelligibility \cite{DBLP:conf/ismir/Meseguer-Brocal20, DBLP:conf/ismir/JeonCL20}.
The literature has also explored generative models for VSS, including approaches using GANs \cite{DBLP:conf/icassp/FanLJ18} and normalising flows \cite{DBLP:journals/spl/ZhuDJSD22}. Furthermore, semi-supervised training frameworks have been proposed to enhance SVS performance \cite{DBLP:conf/icassp/StollerED18} and \cite{DBLP:conf/icassp/WangGIVK21}.

Several previous works have focused on training self-supervised or unsupervised learning features for VSS. \cite{DBLP:conf/eusipco/MimilakisDS20} and \cite{ioannis2020revisiting} conducted representation learning via a denoising objective using autoencoders. \cite{DBLP:journals/taslp/YuanWWUW23} introduced deep unfolding to learn a latent representation of vocal signals for VSS. Additionally, MERT \cite{li2023mert}, a general-purpose representation for acoustic music understanding, has been successfully applied to source separation.

Inspired by denoising diffusion models, \cite{DBLP:conf/ismir/Plaja-RoglansMS22} have proposed unconditional signal modelling to gradually convert a mixture into a singing voice or accompaniment. \cite{DBLP:conf/ismir/Plaja-RoglansMS23} introduced the use of a cold diffusion process as a feature followed by unsupervised clustering.
More challenging tasks, such as the separation of multiple singing voices, have also been investigated \cite{DBLP:conf/ismir/PetermannCCBG20, yu2023zero}.

Recently, methods based on Differentiable Digital Signal Processing (DDSP) are also being proposed. Unlike the above methods, DDSP-based models can easily be trained on unlabelled music data and adapted to separate homogeneous sources like lead and backing voices. \cite{DBLP:journals/taslp/SchulzeForsterRKDB23} proposed a unified framework that can be used for both music stems and homogeneous mixtures via cascaded pitch extractors and deep neural networks. \cite{richard2024fully} extends this method on leading and backing voice separation to a fully differentiable one via HCQT and voice assignment,

\paragraph{Lyrics Interpretation:}
Lyrics interpretation involves the text summarisation of song lyrics, aiding the quick understanding of the singing voice and the management of music based on its linguistic content. This task can be approached using text summarisation methods applied solely to the text \cite{son2018music, DBLP:conf/ranlp/FellCGG19}. Incorporating audio provides an additional dimension, enhancing the understanding of the song. \cite{DBLP:conf/ismir/ZhangJXD22} proposed a multimodal generative model with representation fusion for lyrics interpretation, leveraging both text and audio information.

\subsection{Music Generation}\label{sec:generation}
\subsubsection{Symbolic Music Generation}

In the area of symbolic music generation, whether it involves generating score-level symbolic music (e.g., ABC notation) or performance-level music (e.g., MIDI), the approach generally falls into two categories: generating from scratch or based on given conditions like chords, tracks, textual descriptions, or other musical properties.

\paragraph{From scratch.} Some early works explored \textbf{monophonic/melody music generation} from scratch as well as polyphonic music.
In early research stage, using RNNs to generate melodies was the most common approach. Folk-RNN~\cite{sturm2016music} trained LSTM networks on 23k music pieces represented in ABC notation in two approaches: one predicts characters based on the previous 50 characters, and the other predicts tokens based on all previous tokens of a transcription. Anticipation-RNN~\cite{hadjeres2017interactive} generates melodies through a RNN-based generative model while allowing user-defined positional constraints. Besides RNNs, some works utilise GAN-based models, such as SeqGAN~\cite{yu2017seqgan}, and VAE-based models, such as the hierarchical model proposed by Roberts et al.~\cite{roberts2018hierarchical}, to generate music melodies.

For \textbf{polyphonic music generation}, MuseGAN~\cite{dong2018musegan} was the first to propose models to generate multi-track symbolic music using the jamming model, the composer model, and the hybrid model. This method generates music with five tracks—bass, drums, guitar, piano, and strings in MIDI format. To generate more natural multi-track symbolic music, DMB-GAN~\cite{guan2019gan} employs a self-attention mechanism to extract both spatial and temporal features, building dual GANs for each branch to create a more harmonious structure.

With the advent of recent generative models, more advanced methods have been researched for generating symbolic music. 
Multitrack Music
Transformer (MMT)~\cite{dong2023multitrack} introduced a Transformer-based multitrack music representation that accommodates a diverse array of instruments while keeping the sequence length short, which aims to effectively address the limitations on the number of instruments supported by previous models.
Mittal et al.\cite{mittal2021symbolic} proposed a method to train diffusion models using symbolic music data, leveraging a pre-trained VAE to map the discrete domain to the continuous latent domain.
Building on the Transformer model, Museformer~\cite{yu2022museformer} introduced a fine-coarse-grained attention mechanism to handle the challenge of long music sequences. In this approach, fine-grained attention captures structural information, while coarse-grained attention gathers contextual information, improving the model's ability to generate coherent and complex musical pieces.

\paragraph{Based on condition.} Generating music from scratch can often result in music that follows a certain pattern, limiting the creativity of music generation. By incorporating various types of input conditions, such as chords, melody tracks, lyrics, and text descriptions, not only can users interact with the music generation process more dynamically, but they also gain higher and more fine-grained control over the output. Consequently, there is a growing group of research focused on generating symbolic music under different input conditions.

\textbf{Conditioned on chord sequences}, MIDINet~\cite{yang2017midinet} utilises a generative adversarial network to create music with multiple MIDI channels. This network is capable of generating music by adhering to a specified chord sequence or by conditioning on the melody established in preceding bars.
Choi et al.~\cite{choi2021chord} introduced a Transformer model conditioned on chords for generating K-POP melodies. This model generates rhythm and pitch components while adhering to a specified chord progression.
MelodyDiffusion~\cite{li2023melodydiffusion} introduces a transformer-based diffusion model specifically designed for chord-conditioned melody generation with discrete musical data. Unlike traditional U-nets, this model leverages transformers to capture long-range dependencies via attention mechanisms and parallel processing.
Besides, DeepChoir~\cite{wu2023chord} can generate a four-part chorale based on given melody and chord progression.

\textbf{Melody harmonisation} refers to crafting a chord progression that complements a given melody. This progression must harmonise with the melody while also aligning with its rhythmic pattern.
Lim et al.~\cite{lim2017chord} propose a technique for generating chord sequences from symbolic melodies by leveraging bidirectional long short-term memory (BLSTM) networks. These networks are trained on a database of lead sheets to achieve the desired chord generation.
AutoHarmoniser~\cite{wu2024generating} is a system designed for melody harmonisation with controllable harmonic density and rhythm. It features an extensive vocabulary of 1,462 chord types, enabling it to generate chord progressions with varying harmonic density for a given melody.
Not limited to melody, Multi-Track Music Machine (MMM)~\cite{ens2020mmm}, which is based on the Transformer model, treats each musical note as a time-ordered sequence. In this method, notes from different tracks are interleaved into a single sequence.
GETMusic~\cite{lv2023getmusic} employs a different strategy by representing musical notes as tokens and arranging them in a 2D grid, with tracks stacked vertically and time progressing horizontally. This diffusion-based model randomly designates each track of a music piece as either the target or the source during training.

\textbf{Generating symbolic music from text descriptions} with the advancement of pre-trained models and LLMs in recent years has been more and more popular. Wu et al.\cite{wu2022exploring} initiated the first investigation into generating complete and semantically coherent symbolic music scores from textual descriptions. They explore the efficacy of leveraging publicly available NLP checkpoints for the task of text-to-music generation. Music Composition Copilot (Musecoco)\cite{lu2023musecoco} is a two-stage framework for music generation. In the first stage, ChatGPT synthesises and refines the input text into musical attributes such as instrument, time signature, tempo and pitch, etc. In the second stage, these attributes are used to generate the symbolic music.

ChatMusician~\cite{yuan2024chatmusician} proposes a novel approach that incorporates music as a secondary language in Large Language Models (LLMs). By utilising ABC notation, this method effectively merges music and text, allowing for internal music composition and analysis without the dependency on external multimodal frameworks.
In contrast to ChatMusician, SongComposer~\cite{ding2024songcomposer} employs MIDI for representing symbolic music and introduces a unique tuple structure. This structure formats lyrics alongside three specific note attributes: pitch, duration, and rest duration, ensuring accurate musical symbol interpretation and precise alignment of lyrics with the melody.
MuPT~\cite{qu2024mupt}, a generative pre-trained transformer for symbolic music, design the Synchronised Multi-Track ABC Notation (SMT-ABC Notation) to maintain coherence across multiple tracks. This allows the model to handle up sequences containing up to 8192 tokens, enhancing its capability to generate complex musical pieces.
Apart from directly generating music from text descriptions, there are also efforts focused on using LLM agents to create symbolic music. Examples include Musicagent~\cite{yu2023musicagent}, ComposerX~\cite{deng2024composerx}, and ByteComposer~\cite{liang2024bytecomposer}.

\textbf{Music generation conditioned on video.}
Music, integral to video production, has sparked researchers' interest in generating music that is conditioned on video content.

CMT~\cite{DBLP:conf/mm/DiJ0WZHLY21}, Video2Music~\cite{DBLP:journals/corr/abs-2311-00968}, and Video background music generation~\cite{DBLP:conf/iccv/ZhuoWWLBPHZFL23} all focus on creating background music with general video content. CMT initially establishes rhythmic connections between video and background music, offering local manipulation of these rhythmic elements and comprehensive control over music genre and instrument choices. Video2Music~\cite{DBLP:journals/corr/abs-2311-00968} and Video background music generation~\cite{DBLP:conf/iccv/ZhuoWWLBPHZFL23} extract both low-level and high-level semantic features for generating music. Diff-BGM~\cite{li2024diff} generate background music using a diffusion-based method. 

 Different from the broader scope of music generation for general video content, some works specifically focus on generating music tailored for dance videos.
Research explores translating human motions from dance videos into musical notation \cite{su2021does}, enhancing rhythmic sound generation. 
Multi-Instrumentalist Net innovates by generating instrumental music directly from videos of musicians without relying on supervised learning \cite{DBLP:journals/corr/abs-2012-03478}. 
Similarly, Dance2Music~\cite{DBLP:journals/corr/abs-2107-06252} utilises dance movements as a foundation for generating music 
\cite{DBLP:journals/corr/abs-2107-06252}, illustrating the potential of dance-centric approaches in music creation.
Efforts extend to synthesising music from silent videos of musical performances, converting body movements into MIDI sequences for realistic music synthesis \cite{DBLP:conf/eccv/GanHCTT20} and generating music from silent piano performances \cite{DBLP:conf/nips/SuLS20}. Advances include deep learning frameworks for transcribing piano music from visual data \cite{DBLP:conf/icassp/KoepkeWMZ20}.
To generate background music corresponding with the body movements of musicians in video clips, Foley Music ~\cite{gan2020foley} utilises a Graph-Transformer architecture, which includes a Graph Convolutional Network (GCN) encoder and a Transformer decoder. It learns to map the relationship between human body key points detected in videos and MIDI events.

\subsubsection{Language Model for Acoustic Music Generation}

The state-of-the-art (SOTA) generative models various fields are predominantly transformer-based language models (LMs) \cite{transformer}.
These LMs have revolutionised natural language processing \cite{gpt2, brown2020language,lamda,llama}, demonstrating extraordinary capability in capturing contextual dependencies and relationships within sequences efficiently.
In light of this, the application of LMs has extended beyond natural language processing to various tasks, including image generation \cite{esser2021taming,yu2022scaling,yu2022vectorquantized} ,video generation \cite{ge2022long,yu2023language}.
Furthermore, this breakthrough in capturing sequence relationships has been extensively studied in speech and audio generation and achieves competitive performance.

\textbf{Application in speech synthesis.} GSLM \cite{2021-generative} and pGSLM \cite{kharitonov-etal-2022-text} proposed to quantise speech representations derived from the self-supervised learning models for speech (such as CPC \cite{oord2018representation}, wav2vec 2.0 \cite{baevski2020wav2vec} and HuBERT \cite{hubert}) and employed a generative language model for both conditional and unconditional speech generation.
Similarly, approaches that combine LMs with discrete representations have also been widely applied in speech resynthesis \cite{polyak21_interspeech}, speech emotion conversion \cite{kreuk-etal-2022-textless}, spoken dialog system \cite{nguyen2023generative} and speech-to-speech translation \cite{lee-etal-2022-direct,popuri22_interspeech}.
Recently, Encodec \cite{encodec}, SoundStream \cite{zeghidourSoundStreamEndtoEndNeural2021a}, and their derivative works \cite{kumar2023high, yang2023hifi} proposed representing audio signals as multiple streams of discrete tokens by utilising residual vector quantizer (RVQ).
This approach allows for the generation of high-quality audio from quantised tokens.
Subsequently, several studies combined such discrete representation of audio with LMs for text-to-speech synthesis \cite{wang2023neural,zhang2023speak,kharitonov2023speak,soundstorm}.

\textbf{Unconditional music generation.} To achieve music generation, ADAS \cite{dieleman2018challenge} has already attempted to use hierarchical VQ-VAEs \cite{vqvae, razavi2019generating} to learn discrete representations of music samples at various temporal resolutions before RVQ was proposed.
They combined these with LMs to generate music with high temporal coherence, but the audio quality remains limited.
With the advent of RVQ, Perceiver AR \cite{preceiverar} sought to model the flattened multi-stream discrete token sequences of SoundStream \cite{zeghidourSoundStreamEndtoEndNeural2021a} directly with LMs, achieving high-quality piano music generation. 
However, due to the excessive length of the flattened sequences, these methods face challenges in maintaining long-term temporal coherence.
AudioLM \cite{audiolm} addressed this issue by using semantic, coarse acoustic and fine acoustic tokens to represent audio and employing multiple cascaded LMs to generate different token sequences.
This allows AudioLM to generate coherent, high-quality piano music unconditionally. Besides the GPT style models, DrumGAN \cite{DBLP:conf/ismir/NistalLR20} is a progressive generative adversarial network (GAN) model designed for the synthesis of drum sounds. It leverages conditional inputs based on perceptual features, enabling intuitive and musically relevant control over the generation process, thereby enhancing both the quality and the expressiveness of the synthesized drum sounds.

\textbf{Text to music.} Compared to unconditional music generation, recent works have introduced descriptive text as a condition to achieve controllable music generation, also known as text-to-music generation.
AudioGen \cite{audiogen} introduced textual information through T5 \cite{t5} encoder, enabling text-to-audio generation.
Inspired by CLAP \cite{clap}, MusicLM \cite{mulan} further introduces descriptive text as a condition through MuLan \cite{mulan}, which is trained with contrastive loss, and is capable of modelling a large variety of long music sequences beyond piano music
SingSong \cite{singsong} followed a similar modelling approach but for producing instrumental music that harmoniously aligns with the provided vocals.
Additionally, some works have explored visual-conditioned music generation.
In generating background music for videos, models introduce controllable features \cite{DBLP:conf/mm/DiJ0WZHLY21} and explore mapping visual arts to music \cite{DBLP:journals/corr/abs-2211-05543}. 
A Transformer-based model aligns music with video content \cite{DBLP:journals/corr/abs-2311-00968}, showcasing diverse applications.
Innovative methods focus on generating music audio with style control from silent videos \cite{DBLP:journals/corr/abs-2305-06594} and complex music samples from dance videos \cite{DBLP:conf/eccv/ZhuOWACYT22}, highlighting the integration of visual content into music generation.

\textbf{Long high-quality music generation.} However, the cascaded language models also present computational challenges. 
In the context of modelling the token sequences of neural audio codecs \cite{encodec, soundstorm, kumar2023high, yang2023hifi}, efficiently generating long, high-quality music segments remains an unresolved issue. 
SoundStorm \cite{soundstorm} utilises a non-autoregressive decoding scheme \cite{maskgit} to significantly accelerate the AudioLM \cite{audiolm}, enabling the acoustic LM to complete decoding within 27 forward passes.
Similarly, VAMPNET \cite{garcia2023vampnet} uses the Descript Audio Codec (DAC) \cite{kumar2023high} as the audio tokeniser and incorporates parallel iterative decoding to achieve music generation.
Parallel to this work, MusicGen \cite{copet2024simple} introduces efficient token interleaving patterns that eliminate the need for cascading multiple LMs, achieving the generation of high-quality music samples with a single-stage LM.
Similarly focused on producing high-quality music samples with a single-stage model, but in contrast to MusicGen which conditions on text descriptions, VidMuse~\cite{tian2024vidmuse} generates music based on video input.

\textbf{Lyrics to singing}. All aforementioned methods only take the instrumental music generation into consideration, generating music with singing remains a significant challenge.
To this end, Jukebox \cite{jukebox} can be seen as the first and only attempt from published literature so far to simultaneously generate music with singing from lyrics using a single LM.
Recently, while the industry has seen the emergence of song generation tools like Suno\footnote{\url{https://suno.com/}} and Udio\footnote{\url{https://www.udio.com/}}, neither has disclosed their methodologies. For more information, please refer to section\ref{subsec:vocal_generation}

\textbf{Music \& general audio generation.} Furthermore, some research attempts to design universal audio generation methods rather than solely focusing on music generation. 
WavJourney \cite{wavjourney} proposed to utilise Large Language Models (LLMs) to connect various audio generation models across different tasks, enabling the generation of comprehensive audio content (covering speech, music, and sound effects) from textual story narratives. 
Meanwhile, SpeechGPT \cite{speechgpt} and UniAudio \cite{yang2023uniaudio} strive to implement universal models based on textual instructions, utilising a single LM to execute various tasks with correct output on different instructions.

\subsubsection{Audio Diffusion Model}

In addition to the language model-based approaches, diffusion models \cite{sohl2015deep, ddpm, vdm}, as a competitive class of generative models, have recently delivered impressive results in various domains of generative modelling.
These models, through a series of diffusion and reverse diffusion processes, effectively formulate the mapping between data and latent distributions, enabling the generation of highly realistic and quality outputs.
Diffusion models have demonstrated exceptional capabilities in tasks such as text-to-image generation \cite{dhariwal2021diffusion, ldm, ho2022cascaded}, image super-resolution \cite{saharia2022image, li2022srdiff} and image inpainting \cite{song2021scorebased}, showcasing their ability to generate high-quality and diverse samples for images.
Furthermore, advancements such as DDIM \cite{ddim} and progressive distillation \cite{salimans2022progressive} have introduced sampling acceleration techniques that significantly reduce the long sampling time, a known drawback of traditional diffusion models, while maintaining the quality of generation.
This improvement in efficiency makes diffusion models more practical for a wider range of applications.

\textbf{Application in speech \& general audio.} The adaptability and versatility of diffusion models have impressively extended beyond image generation to a variety of modalities, including speech and audio.
In speech, diffusion models are widely applied in vocoders \cite{wavegrad, diffwave, priorgrad, bddm, fastdiff}, to transform the mel-spectrograms into high-quality speech waveforms.
WaveGrad \cite{wavegrad} and DiffWave \cite{diffwave} based on the diffusion process to incrementally generate speech waveforms by gradually adding noise and reversing this noise.
FastDiff \cite{fastdiff} enhances the speed of diffusion by decreasing the number of sampling steps while maintaining high generation quality.
Additionally, some works \cite{difftts, gradtts, wavegrad2, guidedtts, guidedtts2, naturalspeech2} also apply diffusion models to acoustic models.
The most significant contribution of diffusion models in this field is the ability to generate high-fidelity speech, which is a critical advancement for realistic and naturalistic speech generation.

\textbf{Text-to-music generation. }This capability is particularly salient in audio generation and music generation, as these models excel in modeling complex temporal dynamics and generating high-quality audio.
Diffsound \cite{diffsound} proposed to quantise the mel spectrogram by VQ-VAE \cite{vqvae} and apply a discrete-diffusion-model-based non-autoregressive decoder \cite{austin2021structured, gu2022vector} instead of the traditional autoregressive (AR) decoder for generating quantised tokens from textual CLIP \cite{clip} embeddings, effectively enhancing generative performance.
However, the introduction of VQ-VAE, while beneficial for quantisation, has compromised the generation quality.
AudioLDM \cite{audioldm} and Make-An-Audio 1 \& 2 \cite{makeanaudio,makeanaudio2} have advanced text-to-audio generation by employing latent diffusion models (LDMs)\cite{ldm}.
By utilising a latent space derived from contrastive language-audio pre-training (CLAP) \cite{clap} to learn continuous audio representations, these models achieve superior generation quality and computational efficiency compared to Diffsound.
Notably, both AudioLDM and Make-An-Audio 2 have extended their capabilities to generate text-conditional music by incorporating music data into training processes.
TANGO \cite{tango} employs an instruction-tuned LLM Flan-T5 \cite{flant5} in place of CLAP, providing the model with advantages in the domain of audio and music generation from text input.
Furthermore, AudioLDM 2 \cite{audioldm2} introduces a novel approach by using the self-supervised learning model AudioMAE \cite{huang2022masked} to extract a general representation of audio, called "language of audio" (LOA), achieving unified generation of speech, music, and sound effects.

Compared to these works that generate audio, including some music, some works specifically focus on music generation, exploring generating long-term music with high-quality and high-fidelity.
Riffusion \cite{riffusion} represents an attempt to apply diffusion models in music generation, directly fine-tuning Stable Diffusion \cite{ldm} on mel-spectrograms of music pieces from a paired text-music dataset to generate 5-second music clips.
Building on this, MusicLDM \cite{musicldm} further incorporates the framework of AudioLDM \cite{audioldm} and refined the CLAP model \cite{clap} on music dataset, enabling the model to generate a more diverse and richer variety of music based on the provided text input.
However, MusicLDM can only generate music with a 16 kHz sampling rate, while most standard music productions are 44.1 kHz, which falls short of the requirements for producing high-quality, high-fidelity music.

\textbf{Long high-quality music generation.} To address this limitation, Mo{\^u}sai \cite{mousai} not only utilises text-conditioned LDMs to learn and generate the reduced latent representations, but also employs a novel diffusion autoencoder to directly compress and generate audio.
By cascading two LDMs, this model can handle the long-term structure of music and generate high-quality stereo music with a 48kHz sampling rate condition on a given textual description.
Simultaneously, Noise2Music \cite{noise2music} proposed an alternative approach to use cascaded diffusion models, employing multiple diffusion models to gradually increase the sampling rate the generated music, which show SOTA generative performance with high-fidelity. 
However, due to the employment of multiple cascaded diffusion models, the success of these two approaches incurs substantial computational costs, which would be a serious impediment to their practicalities.

MeLoDy \cite{melody} introduces a novel solution by combining the advantages of language models and diffusion models. It employs a semantic LM based on MuLan \cite{mulan} to capture the semantic structures of music, and proposes a dual-path diffusion (DPD) model to simultaneously model both coarse and fine acoustic information conditions on the semantic LM. 
This strategy enables the efficient generation of competitively high-quality music.
JEN-1 \cite{jen1} introduces an omnidirectional diffusion model that integrates both autoregressive and non-autoregressive modes, enhancing sequence generation while improving sequence dependency. 
Furthermore, the model has been extended to music continuation and music inpainting.

\textbf{Visual-music generation.}
Some studies have explored acoustic music generation conditioned on visual information. V2Meow~\cite{DBLP:journals/corr/abs-2305-06594} generates music through video input and style control via text input. M$^2$UGen~\cite{hussain2023m} employs LLMs as a bridge between visual and audio features, integrating a video module and a music generation module. VidMuse~\cite{tian2024vidmuse} uses a long-short-term modeling strategy to generate music conditioned on input videos autoregressively. MELFUSION~\cite{chowdhury2024melfusion} introduces a model that leverages cues from textual descriptions and corresponding images to synthesise music.
Another line of research in multimodal musical understanding leverages denoising diffusion models to learn the recovery of latent features from noise. DiffMAViL~\cite{nunez2023diffusion} integrates denoising diffusion processes with the MAViL~\cite{huang2024mavil} in a masked encoder-decoder framework to learn multimodal representations. It processes audio and video by masking pixels, adding noise, and using diffusion. This helps it pick up detailed features and rebuild the original audio-visual content.
Through coupled architectures and bidirectional diffusion processes, models such as EasyGen \cite{zhao2024easygen} and MM-Diffusion \cite{DBLP:conf/cvpr/RuanMYH0FYJG23} combine audio and visual modalities, improving the quality of joint generation. With an emphasis on audio-video synchronisation, MM-Diffusion's unified architecture includes multimodal attention techniques and a connected U-Net \cite{DBLP:conf/miccai/RonnebergerFB15}. EasyGen uses BiDiffuser, which combines LLMs and diffusion models for efficiency in a variety of generative applications.
Differently, Seeing and Hearing~\cite{DBLP:journals/corr/abs-2402-17723} build aligners across text, audio, and video modalities to jointly generate audio and video.
Conditional Discrete Contrastive Diffusion (CDCD) \cite{zhu2023discrete} utilises contrastive learning within diffusion frameworks to align music with specific images, enhancing cross-modal content generation and creative processes.

\textbf{Beyond text condition.} Beyond text-to-music generation, many works focus on other tasks of music generation based on diffusion models.
AUDIT \cite{audit} and InstructME \cite{instructme} acheieved zero-shot music editing by feeding text instructions into LDMs.
Music ControlNet \cite{musiccontrolnet} proposes a method similar to the ControlNet \cite{zhang2023adding} approach in the image domain, to offers multiple precise, time-varying controls over the generated music.

\subsubsection{Vocal Music Generation}\label{subsec:vocal_generation}

The human voice is the most natural musical ``instrument'' possessed by humans. Singing voices, rich with emotive lyrics and diverse melodies, vividly convey explicit human feelings and are an essential element of music. Vocal music generation focuses on producing singing voices under various conditions, broadly classified into the following key tasks: \textbf{Singing Voice Synthesis (SVS)}, which generates voices from input lyrics and music scores; and \textbf{Singing Voice Conversion (SVC)}, which adapts the singing voice from one person to match the timbre of another using a timbre prompt. %

\paragraph{Singing Voice Synthesis (SVS):} SVS can be viewed as an extension of Text-to-Speech (TTS), incorporating not only text input (lyrics) but also musical notation to mediate the speech generation process.

Like TTS, the popular framework in SVS usually involves a two-stage process.  Initially, an ``acoustic model'' converts lyrics and musical notation into an intermediate acoustic-level representation, usually in the form of a mel spectrogram.  The ``vocoder'' then renders these acoustic representations into audio waveforms in a second stage.
Given that singing voices exhibit a wider tonal range and greater dynamics than speaking voices, SVS presents a more complex challenge than TTS. Considerable research efforts have focused on developing effective acoustic models capable of producing expressive acoustic characteristics.  In \cite{gu2021bytesing}, autoregressive (AR)-based Bytesing is developed, which adopts a Tacotron\cite{wang2017tacotron}-like structure and can generate high-fidelity speech, but at a slow speed.  Later research showed that non-autoregressive (NAR) models such as FastSpeech 1\&2 \cite{ren2019fastspeech,ren2020fastspeech} can also generate high-quality speech for TTS but at a much faster speed.  Therefore, FastSpeech-based SVS methods have also been developed, such as XiaoiceSing 1\&2 \cite{lu2020xiaoicesing,wang2022xiaoicesing}.  To further improve the quality of synthesised singing voice, methods based on generative adversarial networks (GAN) are proposed, including HifiSinger \cite{chen2020hifisinger}, WGANSing \cite{chandna2019wgansing}, XiaoiceSing2 \cite{wang2022xiaoicesing}  and SingGAN \cite{huang2022singgan}.  

For the vocoder of SVS, being data-driven, a significant challenge with conventional TTS vocoders like MelGAN \cite{kumar2019melgan} and HiFi-GAN \cite{kong2020hifi} is their tendency to generate fragmented tones at unseen pitches. However, the ability to accurately render wide-pitch singing is a crucial requirement for SVS systems. To address this issue, neural source filter models \cite{wang2019neural,yoneyama2023source,wu2022ddsp} have been developed, which treat the singing voice as a filtered version of the stimuli source signal. By incorporating the parametric digital signal processing, such vocoders could generalise well to unseen pitch ranges.
It has been argued that the two stages of SVS, i.e., acoustic modeling and vocoding, are separately optimised, the cascade combination may lead to a sub-optimal overall system.  End-to-end (E2E) SVS systems, such as VIsinger \cite{zhang2022visinger} and UniSinger\cite{hong2023unisinger}  are proposed to produce more realistic and expressive audios by leveraging the learned hidden representations. 

Large speech language models have been employed to enhance TTS systems \cite{DBLP:conf/interspeech/DuGC022, DBLP:conf/interspeech/SiuzdakDRJ22, DBLP:conf/nips/LeeKLSHL22, fish-speech-v1}. These approaches either train their own SSL feature or use pre-trained features as guidance for their acoustic model. Similar techniques have been applied to SVS. Similar techniques have been applied to SVS. TokSing \cite{wu2024toksing} proposes token blending across models and layers to capture better singing semantics and acoustics. StyleSinger \cite{DBLP:conf/aaai/ZhangHLHXCDHZ24} captures the timbre and emotion with the pre-trained wav2vec 2.0 feature from the reference recording. However, due to the lack of singing-dedicated pre-trained models, these approaches rely on pre-trained speech models, which may lead to sub-optimal performance.

Beyond the above single-purpose methods, NANSY++ \cite{DBLP:conf/iclr/ChoiYLK23} is a unified voice analysis and synthesis framework built with normalizing flow, that is able to perform 4 tasks including SVS, SVC and TTS.
It is widely recognised that diffusion models can produce excellent performance in terms of generation quality of images \cite{dhariwal2021diffusion,rombach2022high}, videos \cite{harvey2022flexible,ho2022imagen}, and audio \cite{audioldm,audioldm2,diffsound}, therefore, Diffsinger \cite{liu2022diffsinger} and HiddenSinger \cite{hwang2023hiddensinger} are proposed to apply the diffusion to singing voice generation. However, since diffusion sampling typically involves tens to hundreds of steps, acceleration has become a critical issue. In \cite{ye2023comospeech}, CoMoSpeech is proposed to apply the consistency model \cite{song2023consistency} to TTS and SVS with only one step of sampling to achieve fast speed. 

Besides generating high-fidelity singing voices, many other researchers have been focusing on modelling the ``singing technique'', which is crucial to vocal performances. Typical works include \cite{lee2022expressive} for intensity and breath modelling, \cite{kim2023muse,nose2015hmm,song2022singing} for vibrato control, \cite{lee2020disentangling,nose2015hmm} for timbre modelling and \cite{song2022singing} that also enables controlling of the singing emotions. 

\paragraph{Singing Voice Conversion (SVC):}

SVC, analogous to voice conversion for singing, involves modifying various attributes of voices, most commonly changing the identity of the singer/speaker \cite{DBLP:journals/taslp/SismanYKL21}. 
AutoVC \cite{DBLP:conf/icml/QianZCYH19} is an autoencoder designed for speech conversion conditioned on speaker embeddings, where tuning the bottleneck dimension allows the encoder to focus on speaker-independent features. This model was adapted for SVC in \cite{DBLP:conf/ismir/Nercessian20}, where AutoVC applies only on the harmonic spectral envelope produced by WORLD \cite{DBLP:journals/ieicet/MoriseYO16}. More recent work further enhanced the AutoVC model with a latent regressor loss for improved identity embedding \cite{o2023comparative}.
Other approaches based on disentanglement use Variational Autoencoders (VAE), phonetic posteriorgrams (PPG), and adversarial training. For example,  \cite{DBLP:conf/icassp/LuoHAH20} achieves many-to-many SVC by training a Gaussian mixture variational autoencoder (GMVAE) on non-parallel data. Another method \cite{DBLP:conf/icassp/LiTYWXSM21} condenses phonetic information into PPGs while encoding other content information with a separate encoder. In \cite{DBLP:conf/apsipa/LuZS020}, the authors adapted VAW-GAN \cite{DBLP:conf/interspeech/HsuHWTW17} to the SVC task. The adversarial losses in \cite{DBLP:conf/icassp/LiTYWXSM21} and \cite{DBLP:conf/apsipa/LuZS020} enhance the robustness. 

Similar to SVS, self-supervised features have been applied to SVC for extracting rich and meaningful features. \cite{DBLP:conf/icassp/JayashankarWSKMH23} explores the use of wav2vec 2.0 \cite{baevskiWav2vecFrameworkSelfSupervised2020a} and HuBERT \cite{hsu2021hubert} together with an $f_0$ harmonic generation module.  Additionally, \cite{DBLP:conf/interspeech/WangLT0WYM22} demonstrates that HuBERT features outperform Mel-spectrogram and PPG-like features when using a contrastive predictive coding module. \cite{DBLP:journals/ejasmp/HuangWYHH23} replaces the traditional acoustic features input with WavLM \cite{DBLP:journals/jstsp/ChenWCWLCLKYXWZ22} to extract content features and for reconstruction. \cite{DBLP:conf/asru/NingJWZX23} utilises Whisper's \cite{DBLP:conf/icml/RadfordKXBMS23} encoder to extract bottleneck features as content representation.

The introduction of diffusion decoders has significantly enhanced conversion quality in SVC. DiffSVC \cite{DBLP:conf/asru/LiuCSM21} is the first SVC method to use a diffusion decoder, though it is limited to any-to-one conversion. CoMoSVC \cite{lu2024comosvc} extends this capability to many-to-many conversion using a consistency model, achieving fast and high-quality inference.
The so-vits-svc \cite{so_vits_svc} is an impactful open-source project providing state-of-the-art singing voice conversion tools based on VITS \cite{DBLP:conf/icml/KimKS21}. It supports training with SSL features including HuBERT, Whisper PPG, and WavLM. Shallow diffusion \cite{diffsinger} is provided as a post-processing method.

\paragraph{Lyrics Generation}
The LOAF-M2L model \cite{ou2023loaf} presents a melody-to-lyrics generation method. Through a hybrid training strategy that combines unsupervised learning with supervised training objectives based on musicological insights, LOAF-M2L improves the structure and syllable alignment of lyrics with respect to the melody, and it maintains a high level of text fluency. Significant improvements in both objective metrics and subjective assessments were validated.

\subsection{Music Therapy \& Medical Applications}\label{sec:music_therapy}

In the last three decades, music therapy research has seen a growing number of publications investigating the role of music in non-pharmacological care. Music has shown positive effects in supporting a number of different healthcare conditions and challenges such as depression \cite{tangEffectsMusicTherapy2020}, stress and anxiety \cite{dewitteMusicTherapyStress2022}, pain analgesia \cite{leeEffectsMusicPain2016}, dementia \cite{moreno-moralesMusicTherapyTreatment2020}, and Alzheimer disease \cite{matziorinisPromiseMusicTherapy2022}.

Although the use of music foundation models for healthcare is currently under-explored, we argue here that it could potentially boost the domain due to the generalisation capability acquired by unlabelled data and the enormous integrated domain knowledge learned by foundation models in other modalities. In other words, the use of foundation models can help both improve our understanding of the therapeutic effects of music while helping us generate better musical content to support music therapy interventions. By either implicitly learning interdisciplinary tasks by updating parameters during pre-training and fine-tuning, or explicitly specifying mapping relationships in a hand-crafted pipeline, foundation models can be utilised as a powerful plug-and-play component in music healthcare applications for a variety of tasks like developing bespoke technologies, improving accessibility, or building musical databases for specific interventions \cite{agres2021music}.

The existence of universal traits in the effects of music on human cognition has been challenged in recent years \cite{swaminathanCurrentEmotionResearch2015}, and foundation models could help instead foster multimodal approaches that take into account a plurality of aspects in the care process \cite{rucsandaExploringRelationshipMusic2024}. The rich semantics embedded in representations during pre-training \cite{li2022map,li2023mert,liu2024music} can help identify relationships between different modalities of treatment and foster personalised treatment \cite{pingleHarmonicHealingNeural2024}, thus directing music therapy interventions towards new paradigms in the music generation process. Foundation models can also support the creation of interventions rooted in the biopsychosocial model of illness \cite{engelNeedNewMedical1977}, that proposes a multifaceted view of health conditions including biological, social and psychological factors such as personality traits, stress or socioeconomical status, that can improve the outcomes of music therapy interventions, in particular for chronic illnesses \cite{wadeBiopsychosocialModelIllness2017}. 

The relationship between multimodal biosignals, emotions and music has  been explored in recent works \cite{rahmanEffectiveMusicTherapy2021} with the objective to further guide generative models and recommendation systems. Studies share a similar purpose when identifying audio features or metrics to select music tracks in playlists to improve for example sleep quality \cite{10.1145/1873951.1874154} or alleviate dementia \cite{nunes2024automatic}, looking to find optimal machine learning methods and suitable music representations. Multimodal datasets involving for example kinematic measures or biosignals such as electro-encephalogram (EEG) can further help explore the embodied relationship music has with our body and psychology, particularly when related to perceived and induced emotions \cite{santanaAffectiveComputingContext2021}, and affective computing applications involving recommendation systems have already been proposed in recent literature \cite{guoEMOMusicEmotionRecognition2024, joelEmotionBasedMusic2023}.

A natural step forward from recommendation systems would be to potentially leverage automatic music generation to meet the diversified needs of music creation in healthcare applications. For instance, generative foundation models could be introduced to enhance the production of a diversity of existing automatic improvisation techniques \cite{xia2017improvised} or during therapy sessions involving improvisations \cite{wigram2004improvisation}. Music generation can also be used to evaluate and analyse interventions in different scenarios \cite{liLongShortTermMemoryBased2022} or complement specific treatments \cite{chenAutomaticMethodDevelop2020}. Fine grain controls, such as phrase or bar level generation, can be leveraged to guide the music generation process when designing a music therapy intervention aimed at eliciting specific functions or emotions dynamically \cite{wangSongDriver2RealtimeEmotionbased2023}. Music generation can be a powerful aid also in broader healthcare and social contexts, such as supporting people on the autism spectrum \cite{jacobSystematicReviewScientific2023} or movement rehabilitation \cite{friedman2011musicglove, beveridge2018rhythmic}, and already showing comparable results to traditional methods for binaural beats music therapy \cite{yangResearchImprovementChildren2024}, an auditory illusion technique created by playing slightly different frequencies in each ear that the user will perceive as a single, pulsating tone.

MIR techniques have recently been applied to sonification (i.e., the translation of information from a given domain to the audio domain \cite{kramerSonificationReportStatus2010}), and can potentially improve a range of tasks including cancer diagnosis \cite{walkerDermoscopyDiagnosisCancerous2019}, sleep stage classification \cite{moradiNovelMethodSleepStage2020}, recognising significant status changes during therapies \cite{fachner2019telling}, and other biosignal-related tasks that could benefit not only from analysis techniques but, potentially, from data augmentation or the production of ex-novo synthetic data, similar to analogue applications in protein sequencing \cite{yuSelfConsistentSonificationMethod2019}. Analysis techniques can also be applied a-posterior to allow for a reacher and more fine-grained analysis of the outcomes of a Music Therapy intervention \cite{zalkow2017exploring}. Sonification has been seen as well to support movement and rehabilitation \cite{newboldMusicallyInformedSonification2016}, and generative foundation models can further improve the ability of wearable and integrated systems to interact with the body in both its physical and psychological facets.

The introduction of AI tools in Music Therapy should prompt also to accelerate the critical discussion around how music is currently conceptualised and applied in healthcare interventions \cite{mainsMusicTherapyVery2024}, using AI also to challenge the Western perspective that currently monopolises music therapy interventions \cite{kwonModernizationOrientalMusic2024}. Participation and collaborative design are a key asset to ensure both therapists and patients are directly involved into the development of AI-driven Music Therapy interventions \cite{sunUnderstandingHumanAICollaboration2024, delsetteSoundCareCoOperative2023}, mitigating some of the ethical and social issues that are discussed later in this work.

\section{Technical Details of Foundation Models}
\label{sec:technical-details-foundation-models}

\begin{table*}[!htp]\centering
\caption{Music Foundation Model Trained with Contrastive Learning \& Instance Contrastive Learning.}\label{tab:music_fm_cl}
\scriptsize
\begin{tabular}{c|ccccc}
\toprule
\toprule
\textbf{Model} &\textbf{Modality} &\textbf{Application} &\textbf{Training Paradigm} &\textbf{Music Tokeniser} &\textbf{Architecture} \\\midrule\midrule
COLA &Audio (Speech, Sound \& Music) &Understanding &Contrastive Learning &spectrum &CNN Encoder \\
MULE &Audio (Music) &Understanding & Contrastive Learning &spectrum &CNN Encoder \\
CLAP &Audio (Sound), Text &Understanding &Contrastive Learning & spectrum%
&Transformer Encoder \\
MusCALL &Audio (Music), Text &Understanding &Contrastive Learning &spectrum &CNN Encoder \\
MuLan& Audio (Music), Text& Understanding &Contrastive Learning& Spectrum  & CNN Encoder \& Transformer Encoder\\
CLAMP &Symbolic (MIDI), Text &Understanding &Contrastive Learning &MIDI &Transformer Encoder \\
Wav2CLIP &Audio (Sound), Text, Image &Understanding &Contrastive Learning &spectrum &CNN Encoder \\
AudioCLIP &Audio (Sound), Text, Image &Understanding &Contrastive Learning &spectrum &CNN Encoder \\
\midrule
vq-wav2vec &Audio (Speech) &Understanding &MLM (Clustering via CL.) &1-D CNN &CNN Encoder \\
wav2vec 2.0 &Audio (Speech) &Understanding &MLM (Clustering via CL.) &1-D CNN &Transformer Encoder \\
HuBERT &Audio (Speech) &Understanding &MLM (Clustering via CL.) &1-D CNN &Transformer Encoder \\
BEST-RQ &Audio (Speech) &Understanding &MLM (Clustering via CL.) &Spectrum &Transformer Encoder \\
musicHuBERT &Audio (Music) &Understanding &MLM (Clustering via CL.) &1-D CNN &Transformer Encoder \\
MERT &Audio (Music) &Understanding &MLM (Clustering via CL.) &1-D CNN &Transformer Encoder \\
MusicFM &Audio (Music) &Understanding &MLM (Clustering via CL.) &Spectrum, BEST-RQ &Conformer Encoder \\

\bottomrule
\bottomrule
\end{tabular}
\end{table*}

In this section, we introduce key technical aspects of developing foundation models, drawing from research in relevant fields. 
We start by detailing the main pre-training paradigms for music PLMs and multimodal foundation models (FMs), including LLMs and LDMs in Section \ref{subsec:pretraining}. 
We then cover domain adaptation techniques such as finetuning on downstream tasks, instruction tuning and in-context learning in Section \ref{sec:adaptation}, and outline the design of audio tokenisers and model architectures in Sections \ref{sec:audio_tokenizers} and \ref{subsec:architecture} respectively. We conclude with a discussion of interpretability and controllability (Section \ref{sec:interpretability}), LLMs for music agents (Section \ref{sec:agents}) and scaling laws (Section \ref{sec:scaling_laws}), before looking at future work in Section \ref{sec:future_improvements}. 
We note that many of the techniques developed for foundation modelling in other fields, such as instruction tuning and in-context learning, have not yet been fully explored in music. Similarly, research on scaling laws and emergent abilities of music foundation models is still in its infancy. In addition to this, there remain several open problems specific to music, such as long-sequence modelling (Section \ref{subsec:long_sequence_modelling}), that require foundation model innovations in pre-training strategies or model architecture methodologies etc. We touch upon these open issues throughout this section.

\subsection{Model Design \& Pre-training Strategies}\label{subsec:pretraining}

Foundation models are pre-trained in a self-supervised fashion on large-scale datasets, avoiding or minimising the need for labelled data. As different pre-training strategies and model designs can lead to different capabilities and performance on different downstream tasks \cite{yuan2024marble, ravanelli2020multi, wu2021multi}, essential design considerations regarding tokenisation, architectures, training protocols, and other methodologies vary across foundation models, studies, and implementations. This subsection covers pre-training strategies for music foundation models, which we categorise into \textbf{Contrastive Learning and Clustering}, \textbf{Generative Pre-training}, and \textbf{Masked Modelling}. While many architectures and paradigms have been adapted to deal with music data, we observe that limited work has investigated the integration of music domain knowledge in the pre-training paradigm and the technique of instruction tuning remains largely unexplored.

\begin{figure*}[ht]
    \centering
    \begin{subfigure}[b]{0.31\textwidth}
        \centering
        \includegraphics[width=\textwidth]{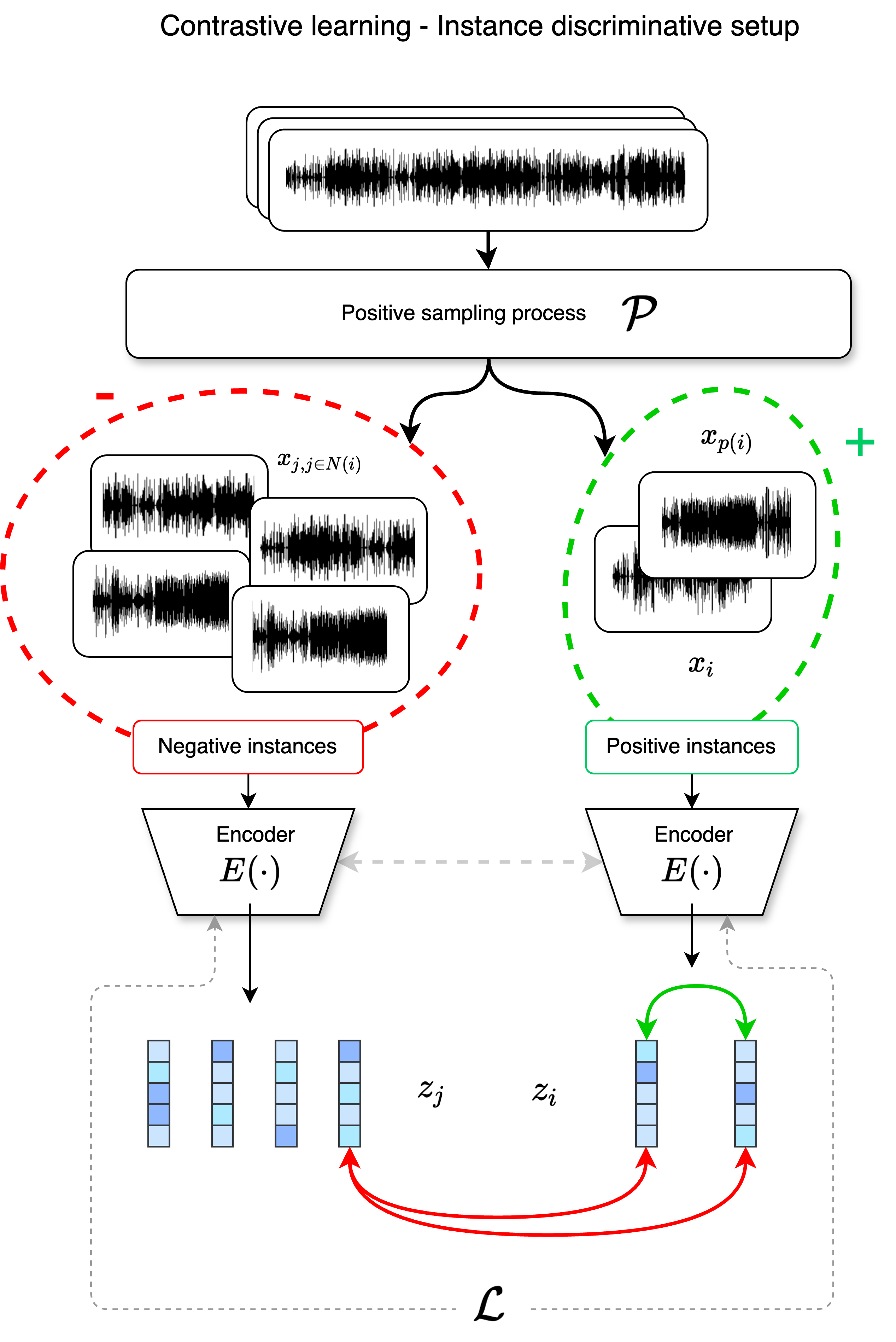}
        \caption{Contrastive learning, which relies on an instance-discriminative framework to learn general instance-level or context-level representations.}
        \label{fig:subfig1}
    \end{subfigure}
    \hfill
    \begin{subfigure}[b]{0.40\textwidth}
        \centering
        \includegraphics[width=\textwidth]{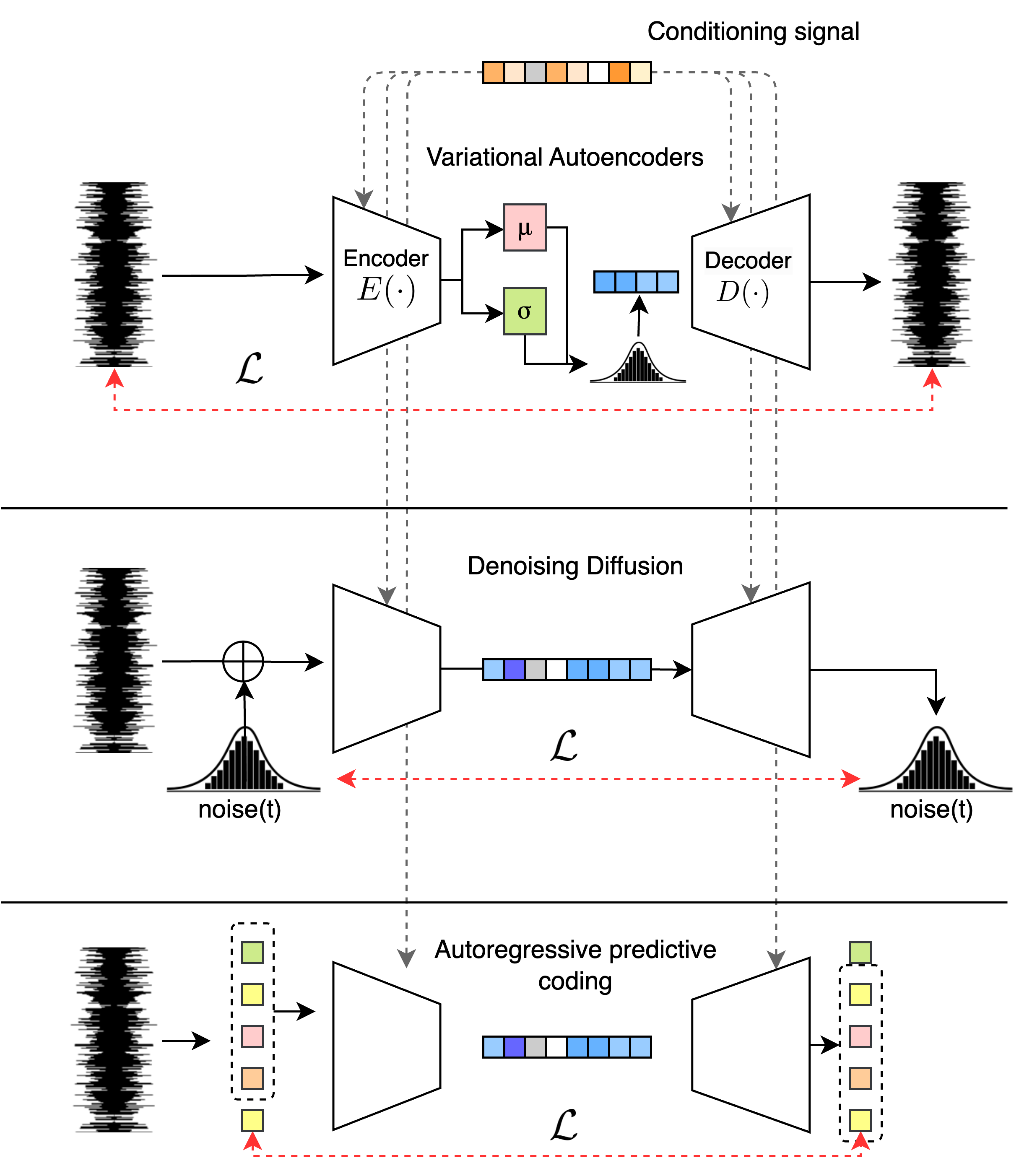}
        \caption{Generative Pre-training, which we categorise into three main families for music foundation models : VAEs, Diffusion models, and Autoregressive Predictive Coding (APC).}
        \label{fig:subfig2}
    \end{subfigure}
    \hfill
    \begin{subfigure}[b]{0.25\textwidth}
        \centering
        \includegraphics[width=\textwidth]{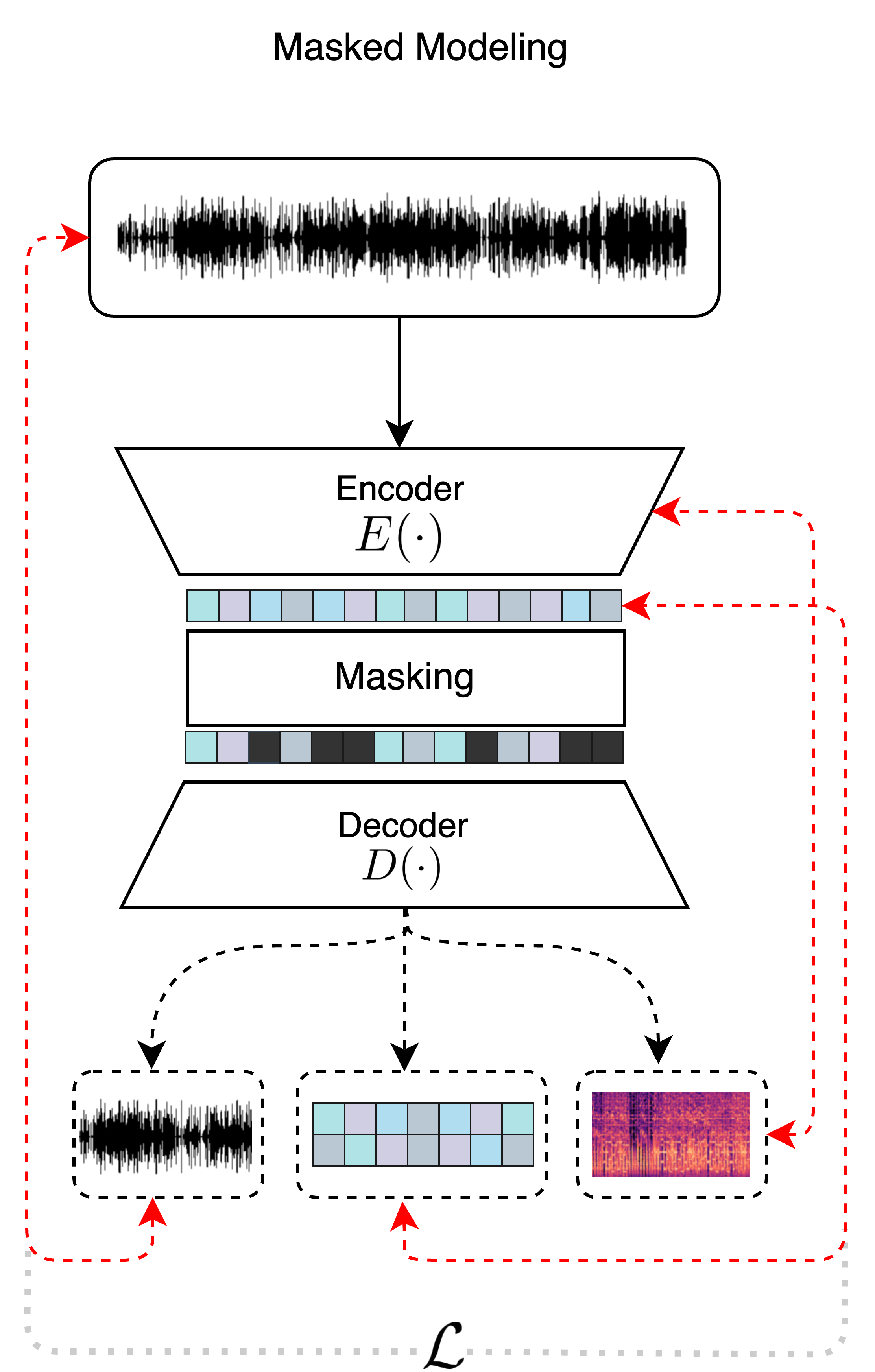}
        \caption{Masked Modeling, where representations of music are learned through reconstructing masked inputs. Reconstruction targets vary by method.}
        \label{fig:subfig3}
    \end{subfigure}
    \caption{A broad taxonomy of pre-training strategies for Music Foundation Models. We categorise these strategies into Contrastive Learning (\ref{fig:subfig1}), Generative Pre-training (\ref{fig:subfig2}), and Masked Modelling (\ref{fig:subfig3}).}
    \label{fig:three_subfigures}
\end{figure*}

\subsubsection{Contrastive Learning \& Clustering}\label{sec:contrastive_learning}
\paragraph{Definition of contrastive learning}
Contrastive learning is a machine learning paradigm that typically provides impressive results in self-supervised learning. Models learn meaningful semantic representations by maximising the similarity between similar samples and minimising the similarity between dissimilar samples. In doing so, it learns high-level discriminative semantic information. Maximising similarity agreement was implemented prior to contrastive learning through Siamese networks \cite{hadsell2006dimensionality}, which suffer from representation collapse without intervention. Prominent solutions that emerged to combat this trivial solution include Bootstrap Your Own Latent (BYOL \cite{grill2020bootstrap}), Barlow twins \cite{zbontar2021barlow} (which both have applications in audio representation learning \cite{niizumi2021byol,anton2023audio}), and triplet networks \cite{hoffer2015deep}. Contrastive learning is a generalisation of the triplet framework to N pairs, which has seen great success in general representation learning in the fields of Computer Vision and NLP, but also in environmental audio, speech, and music.

\textbf{Formalization of contrastive learning}
Concretely, contrastive learning teaches general representations of data by contrasting representations of instances and maximising the agreement between positive representations. At the same time, within a set of samples, representations of negative instances are pushed away from positive pairs. Formally, we consider a general form of the contrastive learning framework.

Let $\{z_i\}_{i \in [1...N]}$ be a set of embedding representations in the latent space of points $\{x_i\}$ in the data space, encoded by a model $E : x_i \mapsto z_i$ that maps the data space to the latent space. For each index $i$, let $p(i)$ be the index of the positive sample corresponding to index $i$. This positive sample is determined by a positive sampling process $\mathcal{P}$ that is task and design-dependent. The Info-NCE loss $\mathcal{L}_{infoNCE}$, introduced in \cite{sohn2016improved} is used as the objective function for a given pair

\begin{equation}
    \mathcal{L}^{infoNCE}_i = - \log \frac{\exp(sim(z_i,z_{p(i)})/\tau)}{ \sum_{j \in N(i)}\exp(sim(z_j,z_i)/\tau)}
\end{equation}

Here $sim$ is a similarity function between representations and $\tau$ is a temperature hyperparameter. $N(i)$ is the set of negative indices within the batch for index $i$. In this formulation, the network learns to maximise the similarity between positive samples while minimising that of mutually negative samples. In doing so, the encoder captures high-level semantic information about samples necessary to perform the discriminative task.

\textbf{Positive instance sampling in contrastive learning}
The notion of ``positive sampling process'' has been intentionally left vague previously, as it has varied significantly from study to study from the first implementation in SimCLR \cite{chen2020simple}. Contrastive learning, rather than a specific method, is a set of instance-discriminative approaches, which can be generalised across modalities, learning setups, architectures, and granularities. The canonical approach which was introduced in SimCLR is to implement a stochastic augmentation chain that generates two augmented versions of the same sample as positives and treats other augmented samples as negatives. However, the following work has explored alternatives such as masking inputs \cite{huang2023contrastive}, generating a different version of the anchor through generative models \cite{jahanian2021generative}, or using classifier-guided sampling \cite{guo2023ultimate,yang2022class,huynh2022boosting}. Some approaches consider a student-teacher setup in which positive pairs are determined by encoded representations from two encoders of the same sample \cite{atito2022asit,chung2021w2v}.

In multimodal contrastive learning, a key development of the field which has enabled numerous foundation models, positive samples are determined from multimodal pairs \cite{clip,clap,mulan,manco2022contrastive}. Further, we make the distinction between \textbf{instance contrastive learning} and \textbf{context contrastive learning}: In addition to parameterising the global similarity of data samples (context contrastive learning), contrastive learning can also parameterise the similarity between granular parts (instances) of a data sample, i.e. individual tokens of a sequential representation. \cite{baevski2020wav2vec,li2023mert,hubert,oord2018representation,gong2022ssast,atito2022asit}

Apart from data augmentation, the more general implementation of a positive sampling strategy rejoins a key consideration of contrastive learning: When teaching a model a similarity metric, how does one determine the best similarity metric to learn? Different positive sampling strategies lead to different similarity metrics, for which the model might need to capture different semantic information about the underlying data. Studies have explored the effect of different augmentation chains and sampling strategies on the learned representations, and there exists a general understanding that contrastive learning is largely dependent on the choice of positive sampling methods, which is a key design consideration for contrastive training pipelines. 

\textbf{Batch size considerations for contrastive training}
Another key consideration is the batch size. Larger batches generally provide better downstream performance for contrastive foundation models in optimal settings \cite{le2020contrastive,balestriero2023cookbook} because they provide more negative samples per batch. However, the batch formulation of the InfoNCE loss prevents gradient accumulation, which inherently bottlenecks contrastive learning approaches to computational requirements. As models scale up, training pure contrastive foundation models becomes more prohibitive. One framework that alleviates these computational costs is MoCo \cite{he2020momentum}, which reframes contrastive learning as a key-query problem and introduces a gradient-free momentum encoder and key queue. Later implementations of the framework have since discarded such devices \cite{chen2020improved,chen2021empirical}.
If the similarity evaluation procedure is done during preprocessing before pre-training, such as the clustering of audio features in HuBERT\cite{hubert}, then the restriction of batch size can be mitigated by gradient accumulation.

\paragraph{context contrastive learning}
\textbf{Examples in audio \& music}
Contrastive learning has been used for large-scale training approaches in environmental audio and speech representation learning. Notable examples of context contrastive learning for learning global audio representations include COLA \cite{saeed2021contrastive}, CLAR \cite{al2021clar} and CL-SER \cite{fonseca2021unsupervised}. The most recent framework leverages a Normaliser-Free (NF) SlowFast (SF) Convolutional network and is currently SOTA across multiple tasks for context contrastive learning for speech and environmental audio \cite{wang2022towards}.

In music, contrastive learning was first implemented with a small sampleCNN \cite{lee2017sample} network following the SimCLR approach on raw waveforms with Contrastive Learning of Musical Representations \cite{spijkervet2021contrastive}, which was later adapted with a Tailed U-Net architecture \cite{vasquez2022tailed}. S3T implements a MoCoV2 setup with Swin transformer encoders for music representation learning \cite{zhao2022s3t}. Arguably, the only unimodal contrastive foundation model for music is MULE, which was trained on 1.7M tracks from a private distribution catalog \cite{mccallum2022supervised}, both in a self-supervised and a supervised fashion. MULE uses the same SF-NFNet as \cite{wang2022towards} and approaches current SOTA on multiple tasks \cite{yuan2024marble}. As in other domains, key considerations for contrastive learning for music is the positive sampling strategy, including augmentation strategies \cite{spijkervet2021contrastive,mccallum2024effect}, relative position of the samples within the track \cite{choi2022towards}, and use of generative or multi-source strategies \cite{ciranni2024cocola,garoufisMultiSourceContrastiveLearning2023}.

\textbf{Multimodal contrastive learning}
Multimodal contrastive foundation models are arguably the most widely used large-scale context contrastive learning approaches. The core principle of maximising similarity between positive samples remains the same, but in multimodal contrastive learning (MMCL), positive pairs originate from different modalities (e.g. text and images \cite{clip}, text and audio \cite{clap}, text and music \cite{mulan,manco2022contrastive}, image and audio \cite{wu2022wav2clip,guzhov2022audioclip}). This requires projecting them into a shared latent space, which requires modality-specific encoders. 
Furthermore, as batches are better represented as from the same modality, we reformulate the contrastive loss for its general usage in MMCL, with modalities $m_1$ and $m_2$ corresponding to encoders $E_1, E_2$ which map samples from modality 1 $x^{m_1}$ and 2 $x^{m_2}$ to a shared latent space: $E_1 \mapsto z_{m_1}, E_2 : x \mapsto z_{m_2} | z_{m_1},z_{m_2} \in \mathbb{R}^d$, where $d$ is the dimensionality of the shared latent space.

\begin{equation*}
    \mathcal{L}_{i}^{m_1 \rightarrow m_2} = -\log \frac{\exp (sim (z_{m_1,i},z_{m_2,i})/ \tau)}{\sum\limits_{z \ \in \{z_{m_2}\}}  \exp (sim (z_{m_1,i},z)/ \tau)  }
\end{equation*}

and

\begin{equation*}
    \mathcal{L}_{i}^{m_1 \leftrightarrow m_2} = \mathcal{L}_{i}^{m_1 \rightarrow m_2} + \mathcal{L}_{i}^{m_2 \rightarrow m_1}
\end{equation*}

The first implementation of MMCL is CLIP \cite{clip}, which has been instrumental in the development of text-image generation models as well as captioning and retrieval approaches. Analogous applications have resulted from CLAP \cite{clap} and derivative work, which applies MMCL between audio and text. The original CLAP \cite{clap} uses a simple PANN \cite{kong2020panns} as the audio encoder and BERT as the text encoder \cite{bert}. The following studies scale up both the encoders to an HTSAT architecture \cite{elizalde2023clap,wu2023large,chen2022hts} for audio and Respectively RoBERTA \cite{wu2023large} and GPT-2 \cite{elizalde2023clap,gpt2} for text, as well as the pre-training data using large-scale datasets \cite{wu2023large}. In addition to keyword-to-caption augmentation, \cite{wu2023large} introduces feature fusion using attention pooling in the audio branch of the model.

In music specifically, key models include MusCALL \cite{manco2022contrastive} and MuLan \cite{mulan}, which have been leveraged in generation and have been shown to hold meaningful representations for general music understanding tasks. MusCALL implements a ResNet50 encoder and a simple BERT text encoder and includes intramodality supervision and semantic weighing of negatives by textual similarity as training design devices \cite{manco2022contrastive}. MuLan, in contrast, uses a straightforward multimodal contrastive training setup, with an Audio Spectrogram Transformer \cite{gong2021ast} audio encoder and a BERT text encoder. However, MuLan drastically scales up training data to close to 40 million text-music pairs. While MusCALL \cite{manco2022contrastive} uses average + attention pooling of audio as the contrastive embedding, MuLan uses the \texttt{[CLS]} token from the AST and the BERT encoder for the contrastive task. In the field of symbolic music, recent work on contrastive learning between textual descriptions and textual representations of music has bridged the gap between symbolic music and text \cite{wu2023clamp}. The encoders used for this work are a fine-tuned DistilRoBERTa and a MAE-style BERT encoder pre-trained on a large-scale ABC notation dataset (see sections \ref{subsec:symbolic-acoustic}, \ref{subsec:musicdatasets}) \cite{wu2023clamp}). Other modalities have been leveraged for MMCL for music such as metadata \cite{favory2020coala} and playlists \cite{alonso2023pre}, but never to the scale of text-music contrastive approaches.

Finally, efforts have been made to bridge the representations learned by multimodal contrastive models by either distilling CLIP knowledge into audio or creating any-to-any joint embedding spaces. Key approaches include Wav2CLIP \cite{wu2022wav2clip} and AudioCLIP \cite{guzhov2022audioclip}

\paragraph{Instance Contrastive Learning}
Instance contrastive learning has had high success rates in the field of speech representation learning, where numerous foundation models apply contrastive learning loss granularly between individual tokens from extracted sequential representations. 
The similarity of such tokens can lead to pseudo-labels for pre-training, making the contrastive learning loss as classification of mask modelling discussed in \ref{subsubsec:masked_modeling}.

The first implementation of such approaches was Contrastive Predictive Coding, which uses a contrastive loss between an aggregated context and future samples as positives and past samples as negatives \cite{oord2018representation}, which was applied for speech and image representation learning. The following approaches for speech and general audio include Wav2Vec \cite{schneider2019wav2vec}, which uses larger front-end and context encoders as well as a convolutional context model rather than a GRU. VQ-Wav2Vec follows by introducing a quantisation module after the convolutional frontend (Gumbel-softmax or K-means clustering) for downstream BERT training. Wav2Vec2.0 \cite{baevski2020wav2vec} leverages a transformer architecture as a context model. Embeddings from the frontend encoder are masked before the transformer module and a contrastive loss is applied to identify the GS-quantised frontend embedding corresponding to each masked transformer output SSAST \cite{gong2022ssast} also applied an auxiliary contrastive loss to masked modelling between masked transformer tokens and reconstructions of other tokens using an AST decoder-only setup. W2V-BERT \cite{chung2021w2v} extracts quantised convolutional embeddings in the same fashion as \cite{baevski2020wav2vec}. the unquantified embeddings are masked and passed to a conformer stack, from which an embedding sequence is extracted after $N$ out of $N+M$ blocks. A contrastive loss is applied between tokens of this intermediary sequence and corresponding target embeddings.

Finally, instrumental to MERT, a key recent music foundation model, is the contrastive learning approach in HuBERT \cite{hubert}. HuBERT and musicHuBERT\cite{ma2023effectiveness} apply masked modelling on discrete tokens to learn representations (see section \ref{subsubsec:masked_modeling}). To obtain these discrete tokens, the authors use an offline convolutional encoder to obtain continuous embedding sequences. the distribution of embeddings from these sequences is parameterised with a contrastive loss and token indices are obtained through K-Means clustering. MERT \cite{li2023mert} applies the same contrastive approach to discover its token vocabulary for one of its acoustic teachers (see Section \ref{subsubsec:masked_modeling}).

\begin{table*}[!htp]\centering
\caption{Music Foundation Model with Generative Pre-training including VAE, GPT, multimodal GPT, and Audio Diffusion.}\label{tab:music_fm_diffusion_gpt_pt}
\scriptsize

\begin{tabular}{c|ccc}
\toprule \toprule
\textbf{Model} &\textbf{Modality} &\textbf{Application} &\textbf{Training Paradigm} \\
\midrule
\midrule
Jukebox, JukeMIR &Audio (Music) &Both &VAE, GPT \\
MusER &Symbolic (MIDI) &Generation &VAE \\
\midrule
Singsong &Audio (Music) &Generation &GPT \\
AudioLM &Audio (Sound), Text &Generation &GPT \\
MusicGen &Audio (Music), Text &Generation &GPT \\
MusicLM &Audio (Music), Text &Generation &GPT \\
Music Transformer &Symbolic (MIDI) &Generation &GPT \\
pop music Transformer &Symbolic (MIDI) &Generation &GPT \\
Jazz Transformer &Symbolic (MIDI) &Generation &GPT \\
MelodyGLM &Symbolic (MIDI) &Generation &GPT \\
MUPT &Symbolic (ABC) &Generation &GPT \\
\midrule
SpeechGPT &Audio (Sound), Text &Both &GPT \\
LauraGPT &Audio (Sound), Text &Both &GPT \\
Audio-PaLM &Audio (Sound), Text &Both &GPT \\
MuseCoCo &Symbolic (MIDI), Text &Generation &GPT \\
ChatMusician &Symbolic (ABC), Text &Both &GPT \\
\midrule
AudioLDM &Audio (Sound), Text &Generation &Diffusion \\
AudioLDM2 &Audio (Sound), Text &Generation &Diffusion \\
Make-An-Audio 1 &Audio (Sound), Text &Generation &Diffusion \\
Make-An-Audio 2 &Audio (Sound), Text &Generation &Diffusion \\
Stable Audio Open &Audio (Sound), Text &Generation &Diffusion \\
CRASH &Audio (Music), Score &Generation &Diffusion \\
Noise2Music &Audio (Music), Text &Generation &Diffusion \\
Mousai &Audio (Music), Text &Generation &Diffusion \\
MusicLDM & Audio (Music), Text &Generation &Diffusion \\
TANGO &Audio (Music), Text &Generation &Diffusion \\
JEN-1 &Audio (Music), Text &Generation &Diffusion \\
Diff-A-Riff &Audio (Music), Text &Generation &Diffusion \\
GETMusic &Symbolic (MIDI) &Generation &Diffusion \\
whole-song-gen &Symbolic (MIDI) &Generation &Diffusion \\
\bottomrule
\bottomrule
\end{tabular}
\end{table*}

\subsubsection{Generative Pre-training}\label{susbubsec:generativepretraining}

\paragraph{Autoencoders}
Autoencoders (AEs) \cite{rumelhartLearningInternalRepresentations1986} are designed to learn lower-dimensional feature representations of data in an unsupervised manner. The basic architecture consists of an encoder compressing the input data into a lower-dimensional latent space and a decoder attempting to reconstruct the original input from it. Formally, given an input $x$, an encoder function $E$, and a decoder function $D$, an autoencoder aims to minimise the reconstruction loss
\[\mathcal{L}(x, D(E(x))),\]
where $\mathcal{L}$ is typically a mean squared error loss for continuous data or cross-entropy loss for discrete data. Given the sole focus on reconstruction, conventional AEs don't inherently model the data distribution, nor do they guarantee that their latent space is continuous or well-structured. As such, sampling arbitrary points from their latent space and decoding them is not likely to generate sensible outputs in many cases.

Variational Autoencoders (VAEs) \cite{kingmaAutoEncodingVariationalBayes2014} address these limitations by introducing a probabilistic framework scalable to large datasets. Instead of encoding inputs to fixed points in latent space, VAEs encode them as probability distributions, typically Gaussian. The encoder outputs parameters of this distribution (mean $\mu$ and variance $\sigma^2$), and the decoder reconstructs from samples drawn from this distribution. VAEs are trained to minimise both reconstruction loss and the Kullback-Leibler divergence between the encoded distribution and a prior (usually a standard normal distribution):
\[\mathcal{L} = \mathbb{E}[\log p(x|z)] - \mathbb{D}_{\text{KL}}[q(z|x) || p(z)],\]

where $x$ is the input, $z$ is a latent variable sampled from the encoded distribution, $\mathbb{E}[\cdot]$ denotes the expected value, and $D_{\text{KL}}(\cdot||\cdot)$ is the Kullback-Leibler divergence term acting as a regulariser, encouraging the encoded distributions to be close to the prior distribution. This formulation allows VAEs to generate novel samples by sampling from the prior distribution and decoding, making them more promising than conventional AEs for generation. Still, because they optimise for the average reconstruction error over the latent distribution, they often tend to produce blurry or averaged-out reconstructions. They also suffer from posterior collapse, because the KL divergence term can be minimised by making the encoder output match the prior regardless of the the input. VAEs are typically used in combination with diffusion models, providing better results \cite{rombach2022high, riffusion}.

Vector Quantised VAEs (VQ-VAEs) \cite{oordNeuralDiscreteRepresentation2017} instead use a discrete latent space, in which the encoder output is mapped to the nearest vector in a learned codebook, preventing collapse by forcing the encoder to output semantically meaningful, discrete codes. The discrete latent space can better capture structured representations (which is particularly useful for modalities with inherently discrete elements like symbolic music) and often produces more detailed reconstructions. The VQ-VAE training process involves optimising a loss function with three main components:
\begin{itemize}
 \item Reconstruction loss: $\mathcal{L}_{recon} = \log p(x|z_q(x))$, where $x$ is the input and $z_q$ is the quantised latent vector.
 \item Codebook loss: $\mathcal{L}_{codebook} = ||sg[z_e(x)] - e||_2^2$, where $z_e$ is the encoder output, $e$ is the selected codebook vector, and $sg[\cdot]$ is the stop-gradient operator.
 \item Commitment loss: $\mathcal{L}_{commit} = \beta||z_e(x) - sg[e]||_2^2$, where $\beta$ is a hyperparameter.
\end{itemize}
The total loss is thus $\mathcal{L} = \mathcal{L}_{recon} +\mathcal{L}_{codebook} + \mathcal{L}_{commit}$.
During training, the encoder learns to map inputs to continuous latent vectors that are close to codebook entries. The codebook itself is trained by moving its vectors towards the encoder outputs that select them. This is achieved through an exponential moving average update or by directly optimising the codebook loss. The commitment loss encourages the encoder to ``commit'' to codebook vectors, preventing its output from fluctuating excessively.

Residual Vector Quantised VAEs (RVQ-VAEs) \cite{juangMultipleStageVector1982} address limitations of VQ-VAEs, namely the fixed codebook size and potentially limited codebook utilisation. RVQ-VAEs extend the VQ-VAE concept by using multiple VQ layers in a hierarchical fashion. Each layer quantises the residual error from the previous layer, allowing for more fine-grained representations. This approach aims to increase representational capacity by using multiple codebooks, encourage better use of all codebook entries across different levels of abstraction, and allow for better preservation of fine details through the residual nature of the quantisation.

Some works in the music domain have explored the use of conventional AEs and particularly VAEs for primarily the tasks of music generation \cite{robertsHierarchicalLatentVector2018, yangDeepMusicAnalogy2019, wangPIANOTREEVAEStructured2020, jiangTransformerVAEHierarchical2020, caillonRAVEVariationalAutoencoder2021, tanMusicFaderNetsControllable2020} and style transfer \cite{brunnerMidiVaeModelingDynamics2018, wuMuseMorphoseFullSongFineGrained2023}, although they don't fit most definitions of being a `foundation model' due to their relatively small size and limited amount of training data used. These approaches are primarily in the symbolic music domain \cite{robertsHierarchicalLatentVector2018, brunnerMidiVaeModelingDynamics2018, yangDeepMusicAnalogy2019, wangPIANOTREEVAEStructured2020, jiangTransformerVAEHierarchical2020, caillonRAVEVariationalAutoencoder2021, wuMuseMorphoseFullSongFineGrained2023}, leveraging architectures such as RNNs, GRUs, LSTMs, and even transformers. Some work exists on audio as well \cite{luoLearningDisentangledRepresentations2019a, tanMusicFaderNetsControllable2020} using a Gaussian Mixture distribution rather than a Gaussian one in order to further aid the disentanglement of musical attributes in the VAE's latent space.

More relevant for music foundation models has been the work on neural audio codecs (NACs), which are a machine learning-based alternative to audio compression. NACs aim to learn a codebook of discrete audio tokens that can efficiently represent audio signals with the highest fidelity possible for a given codebook size. Representations from NACs are often used for quantisation and tokenisation, as described further in Sections \ref{sec:contrastive_learning} and \ref{subsubsec:masked_modeling}. One of them, EnCodec \cite{encodec}, uses an RVQ-VAE with convolutional layers and a time- and frequency-domain reconstruction loss along with multiple codebooks to support variable bandwidth training. It further uses a multi-scale spectrogram discriminator as a perceptual loss, and, optionally, entropy coding with a small transformer to further compress the RVQ representation. SoundStream \cite{zeghidourSoundStreamEndtoEndNeural2021a} uses a similar architecture but proposes an end-to-end training procedure that incorporates the reconstruction and adversarial losses. The Descript Audio Codec \cite{kumar2023high} employs a range of further optimisations to increase compression, such as a periodic activation function, a modified quantiser dropout, a multi-scale STFT discriminator, and a multi-scale mel reconstruction loss with varying mel bin sizes.

\paragraph{Diffusion models}

\begin{figure*}
    
    \centering
    \includegraphics[width=0.7\linewidth]{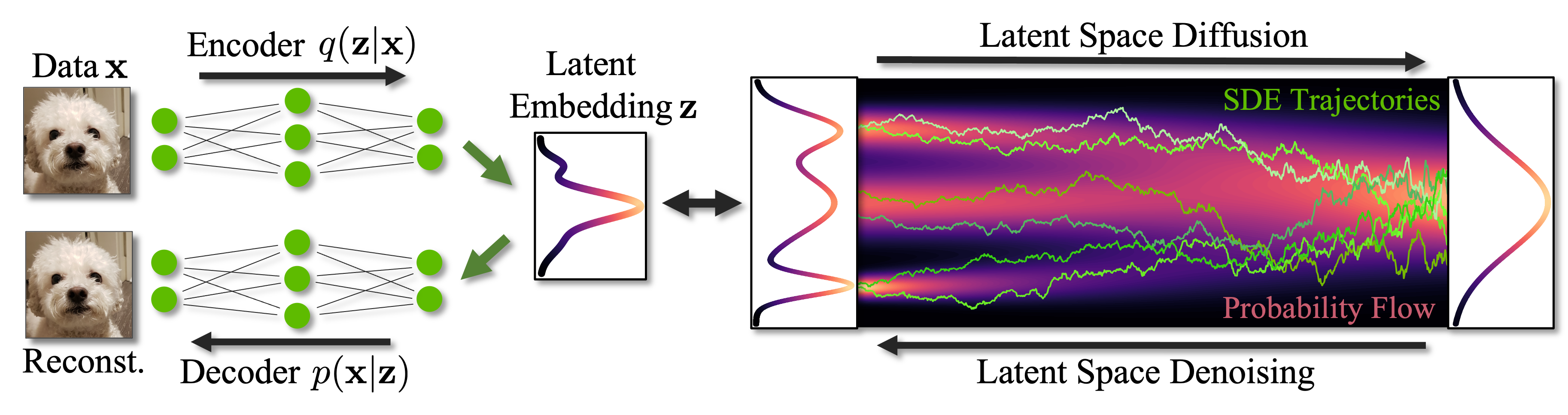}
    \caption{Paradigm of latent diffusion models. \cite{tutorial}}
    \label{fig:enter-label}
\end{figure*}

Diffusion models are a class of probabilistic generative models that have recently become popular due to the generation quality, particularly in the domain of image generation, for their capacity to generate high-quality synthetic data \cite{esser2021taming,yu2022scaling,yu2022vectorquantized,ddpm}. The core idea behind diffusion models is to model the data distribution by simulating a diffusion process that gradually transforms data into noise, and then learn to reverse this process. This approach contrasts with other generative models like Generative Adversarial Networks (GANs) \cite{goodfellow2020generative} and Variational Autoencoders (VAEs), which directly map random noise to data samples. Diffusion models have made notable progress in the fields of audio and music generation. Section \ref{sec:generation} explores these applications thoroughly. In this paragraph, we focus on the fundamentals of diffusion models and delve into design choices for prominent foundation models utilising diffusion for audio and music. 

\textbf{Fundamentals of diffusion models}: diffusion models generate samples by learning to iteratively reverse a diffusion process: The diffusion process can be understood as a sequence of $T$ steps where a data sample $\mathbf{x}_0$ is progressively transformed into a noisy sample $\mathbf{x}_T$ through a series of Gaussian noise additions. The forward process is defined by a Markov chain:

\begin{equation*}
    q(\mathbf{x}_t | \mathbf{x}_{t-1}) = \mathcal{N}(\mathbf{x}_t, \sqrt{\alpha_t}\mathbf{x}_{t-1}, (1 - \alpha_t)\mathbf{I})
\end{equation*}
which can be reparameterised as: 

\begin{equation*}
    \mathbf{x}_t = \sqrt{\alpha_t}\mathbf{x}_0 + \sqrt{1 - \alpha_t}\epsilon_t, \epsilon \sim \mathcal{N}(0,\mathbf{I})
\end{equation*}

where $\alpha_t$ are determined by a noise schedule controlling the amount of noise added at each step. The reverse process, which is the core of the diffusion model, aims to denoise $\mathbf{x}_t$ to $\mathbf{x}_0$. It is parameterised as:

\begin{equation*}
    p_\theta(\mathbf{x}_{t-1} \mid \mathbf{x}_t) = \mathcal{N}(\mathbf{x}_{t-1}; \mu_\theta(\mathbf{x}_t, t), \Sigma_\theta(\mathbf{x}_t, t))
\end{equation*}

The parameters \( \theta \) are learned to approximate the reverse diffusion process using variational inference. The objective is to minimise the variational bound on the negative log-likelihood of the data, which translates to:

\begin{equation*}
    \mathcal{L} = \mathbb{E}_q \left[ \sum_{t=1}^T D_{KL}(q(\mathbf{x}_{t-1} \mid \mathbf{x}_t, \mathbf{x}_0) \parallel p_\theta(\mathbf{x}_{t-1} \mid \mathbf{x}_t)) \right]
\end{equation*}

Denoising diffusion probabilistic models \cite{ddpm} introduce reparameterisation devices that largely simplify the formulation of the objective and the sampling strategy by reframing the model as a noise-predictive model $\epsilon_\theta$ :

\begin{equation*}
    \mathcal{L}_{\text{simple}} = \mathbb{E}_{t, \mathbf{x}_0, \boldsymbol{\epsilon}} \left[ \left\| \boldsymbol{\epsilon} - \boldsymbol{\epsilon}_\theta(\mathbf{x}_t, t) \right\|_2^2 \right]
\end{equation*}

Among key considerations for diffusion models, the noise schedule (i.e.  $s : t \in [1  \cdots T] \mapsto \alpha_t \in \mathbb{R}$) , the network architecture, the conditioning mechanism (covered in a following paragraph), and the guidance mechanism.

\textbf{Accelerated sampling strategies}: As diffusion models typically require numerous diffusion steps to reach generated samples of a desired quality, a prevalent focus of research on diffusion models has been requiring fewer sampling steps to reach a given generation quality. Approaches such as truncation \cite{zheng2022truncated,lyu2022accelerating} and knowledge distillation \cite{salimans2022progressive,huang2024knowledge} have been generally successful. However, perhaps the most widely used is the Denoising Diffusion Implicit Model (DDIM) sampling procedure \cite{ddim}, which is also largely used in audio diffusion models \cite{audioldm,audioldm2,musicldm,mousai, tango}. Briefly, DDIM approximates the stochastic sampling process of DDPM by a simple probability flow ODE:

\begin{equation*}
    d \Bar{x}(t) - \epsilon_\theta^{(t)} \Bigg( \frac{\Bar{x}(t)}{\sqrt{\sigma^2 +1}} \Bigg ) d \sigma
\end{equation*}

Where $\sigma = \sqrt{(1-\alpha)/ \alpha} $, $\Bar{x} = x / \sqrt{\alpha}$. The deterministic nature of solving this equivalent ODE greatly speeds up the sampling process with minimal quality loss.

\textbf{Latent diffusion models}, which have been leveraged to great effect in image generation \cite{rombach2022high} and further in audio and music generation \cite{audioldm,audioldm2,mousai,ghosal2023text, tango, makeanaudio}, apply the same principles as diffusion models with a key difference that has been foundational in recent studies for music and audio diffusion. While traditional DDPMs model diffusion directly in the data space (i.e. pixel or sample space), latent diffusion models do so in the latent space of a pre-trained autoencoder. This provides the distinct computational advantage of not modelling expensive diffusion steps in a high-dimensional space and has sped up audio and music generation systems considerably, making diffusion systems tractable from an application standpoint.

\textbf{Conditional diffusion models}: Diffusion models, like many generative models, can be conditioned on various signals to guide the diffusion process. Most notable, perhaps, is the use of text conditioning for image, audio, and music. Generally, we consider a conditioning signal $C$, which is used to influence the generation process. Commonly, conditioning is applied through cross-attention in the diffusion architecture (U-Net \cite{ddpm} or Transformer \cite{peebles2023scalable}).

Finally, diffusion models have largely adopted classifier-free guidance as a conditioning strategy for training and generation, after some works used ``vanilla'' guidance such as addition or self-attention \cite{avrahami2023blended,hong2023improving} or classifier guidance \cite{dhariwal2021diffusion}. which leverages a pre-trained classifier's gradient to change the denoising direction. Classifier-free guidance simplifies the guidance process by alternating between unconditional and conditional denoising during training with a ``conditioning dropout probability'' \cite{ho2022classifier}:

\begin{equation*}
    \nabla_x \log p(x|c) = w \nabla_x \log p(x|c) + (1-w) \nabla_x \log p(x)
\end{equation*}

This, at inference time, is simple to implement by denoising a weighted average of conditional and unconditional weights. This has been shown to produce superior results to other guidance mechanisms. In audio and music specifically, it is the predominant mechanism of guidance for diffusion models.

\textbf{Diffusion design choices in audio and music generation}: We now refer back to Section \ref{sec:generation} for a thorough overview of key diffusion models for audio and music. This paragraph covers specific design choices of interest for these models. 

Controllable Raw Audio Synthesis with High-Resolution (CRASH) \cite{DBLP:conf/ismir/RouardH21} is a score-based generative model tailored for the unconditional synthesis of raw audio. Leveraging the early approach of diffusion processes modelled by stochastic differential equations, CRASH is engineered to produce high-fidelity drum sounds at 44.1 kHz.

 Noise2Music \cite{noise2music} utilises an efficient 1D UNet conditioned on text through cross-attention and classifier-free guidance for diffusion on raw audio. It is the first diffusion model to generate 3.2kHz audio conditioned on text. It includes an upsampler diffusion model conditioned on the LoFi audio and text that upsamples the audio to 16kHz, and a super-resolution model produces the final 24kHz audio output. 

Mousai \cite{mousai} leverages a similar setup with latent diffusion: it uses cascaded diffusion U-Nets conditioned with cross-attention and classifier-free guidance on text encodings. Authors pre-train a diffusion autoencoder on magnitude spectrograms and then condition An efficient 1D U-Net on the obtained latent. At inference, the latent and the audio are generated through diffusion with a DDIM Sampler. This cascaded architecture allows for long-form coherence and fast generation. 

In the field of audio generation, AudioLDM and \cite{audioldm} subsequently AudioLDM2 \cite{audioldm2} have had influence on music generation through derivative work MusicLDM \cite{musicldm}.  AudioLDM uses a mel-spectrogram VAE and a HiFiGan \cite{hifigan} vocoder as frozen components and trains a classifier-free guided diffusion UNet to generate audio. One specificity of AudioLDM which is often echoed in other work is the use of CLAP global audio and text conditioning during training instead of token sequence conditioning through cross-attention. This is due to the notable scale difference between text-image and text-audio datasets, which allows text-guided image generation to exploit text for training, while such augmentation strategies as proposed in AudioLDM are beneficial to alleviating the lack of captioned data. AudioLDM2 uses another approach to circumvent the lack of captioned data. By finetuning a GPT2 model to predict the sequential output of an AudioMAE encoder conditioned on text, audio or phoneme data, AudioLDM2 creates a shared sequential conditioning space dubbed "Language of Audio". They then employ cross-attention CFG conditioning to train the LDM. The authors of MusicLDM largely take inspiration from AudioLDM, but retrain CLAP on music data and propose a beat-synchronous mixup strategy to enhance diversity and novelty in output generations. 

TANGO \cite{tango} trains a VAE on mel-spectrograms and employs cross-attention to condition the same latent diffusion UNet as AudioLDM1 \cite{audioldm} on T5 text embeddings, with classifier-free-guidance.  In contrast to AudioLDM, TANGO does not leverage CLAP encodings for text conditioning, removing the limitations induced by singe-element embeddings and trusting T5 to hold necessary information for the latent diffusion model to learn the appropriate intermodal mapping.

Diff-A-Riff \cite{DBLP:journals/corr/abs-2406-08384} is a latent diffusion model specifically designed to generate high-quality instrumental accompaniments suitable for different musical contexts. With integrated controls for audio referencing and text cueing, Diff-A-Riff enhances compositional flexibility, generates 48kHz pseudo-stereo audio, and significantly reduces inference time and memory usage.

Other key work with diffusion models includes JEN-1 \cite{jen1}, which employs a standard diffusion UNet conditioned on T5 embeddings with the added specificity that authors include an autoregressive training objective by including causal padding in self-attention UNet blocks, which allows for multiple generation use-cases. Make-an-audio 1  \cite{makeanaudio} and Make-an-audio  2 \cite{makeanaudio2} focus on captioned data scarcity and temporal alignment respectively. Make-an-audio 1 leverages a standard cross-attention Diffusion UNet with classifier-free guidance and trains on CLAP-score selected captions generated from pre-trained automatic captioning models. Make-an-audio leverages temporal encoding of audio events through LLM augmentation to enforce temporal coherence of occurrences between prompts and generated audio.  Most recently, Stable audio \cite{evans2024fast} uses a proprietary-trained CLAP model and timing embedding information to condition a latent diffusion UNet for timing-conditioned audio generation. specifically, authors make generation faster by using a memory-efficient attention implementation. Specific details include the use of the next-to-last layer of the CLAP text encoder for conditioning and FiLM conditioning of the diffusion timestep. Finally, Stable audio 2 \cite{evans2024long} upscales stable audio 1 by using a Diffusion Transformer Architecture with 1.1 Billion parameters. Further, authors employ a DAC-like autoencoder \cite{kumar2023high}. This architecture upscaling allows Stable Audio 2 to generate long-form (upwards of 3 minutes) music with better objective metrics on music generation compared to previous models. 

Finally, Diffusion has also emerged as a promising paradigm in symbolic music generation. Authors of \cite{Generating_symbolic_music} use a standard diffusion UNet pipeline with DDPM sampling to generate binary piano rolls representing MIDI data. \cite{wang2024whole} leverages a similar approach with hierarchical generation to first generate form, lead sheets structure, lead sheets, and accompaniment with conditioning from previous stages. All four stages are diffusion UNets with classifier-free guidance from previous stages.  Finally, latent diffusion is used to generate MusicVAE \cite{roberts2018hierarchical} latent embeddings in \cite{mittal2021symbolic}. Authors leverage a transformer as diffusion backbone for this approach with musical scale conditioning through a FiLM layer.

\paragraph{Autoregressive Predictive Coding}

Autoregressive Predictive Coding (APC) has had immense success in the field of natural language processing, largely facilitated by the widespread usage of the transformer architecture \cite{vaswani2017attention}. Key applications include the GPT language model family \cite{gpt2, brown2020language}, which uses APC among other techniques as a pre-training paradigm. Conceptually, APC trains a model to predict the next element in a sequence of discrete tokens using an autoregressive architecture. In doing so, the model learns contextual representations over the whole sequence that contain high-level semantic information. Overwhelmingly, the architecture used for APC pre-training is the transformer, although recent advances in state space models (SSMs) have begun encroaching on the territory of the transformer model \cite{gu2023mamba}. For more information, please refer to subsection \ref{subsec:architecture}

\textbf{Fundamentals of APC:} Formally, consider a sequence of discrete tokens (codes) \( X = (x_1, x_2, \ldots, x_n) \). The goal of Autoregressive Predictive Coding (APC) is to model the joint probability distribution of the next token given the previous tokens in the sequence using an autoregressive model \( A_\theta \):

\[ P(X) = \prod_{t=1}^n P(x_t \mid x_{<t}; A_\theta) \]

where \( x_{<t} = (x_1, x_2, \ldots, x_{t-1}) \) represents the tokens preceding \( x_t \). The model \( A_\theta \) is applied to the sequence \( x_{<t} \) to predict \( x_t \) \( (x_{\leq t} = A_\theta(x_{<t})) \). Optionally, the model can be conditioned on an additional sequence of context tokens \( C = (c_1, c_2, \ldots, c_m) \) pertaining to any conditioning signal or modality:

\[ P(X \mid C) = \prod_{t=1}^n P(x_t \mid x_{<t}, C; A_\theta) \]

During training, the model is optimised to maximise the conditional log-likelihood, formulated as the cross-entropy loss:

\[ \mathcal{L} = - \mathbb{E}_{X, C} \left[ \sum_{t=1}^n \log P(x_t \mid x_{<t}, C; A_\theta) \right] \]

\textbf{Model design in APC for audio, symbolic music, and acoustic music:} For audio and music specifically, a key design consideration beyond the usual context length, attention mechanism, model size, choice of positional encodings, and conditioning mechanism, is the choice of the discrete audio tokeniser. APC requires discrete tokens for prediction, so to perform APC pre-training on audio sequences, these sequences must be converted into discrete tokens (codes). Varieties of tokens and tokenisers are covered in Subsection \ref{sec:audio_tokenizers}, but in this section, we briefly delve into the design choices of foundation APC models in audio, speech, and music for which tokeniser to leverage for APC pre-training as well as specific training details.
The range of applications for APC-based models in audio and music generation and understanding as well as multimodal approaches are well covered in Section \ref{sec:applications}. Here, we cover specific architectural details as well as training devices used for these models. 

In generic audio, AudioLM \cite{audiolm} utilises a Soundstream encoder \cite{zeghidourSoundStreamEndtoEndNeural2021a} to extract acoustic tokens and k-means-clustered w2v-BERT \cite{chung2021w2v} representations to extract semantic tokens. AudioLM then cascades three autoregressive transformer models to model semantic tokens, then coarse acoustic tokens conditioned on semantic tokens, and finally fine acoustic tokens conditioned on coarse acoustic tokens.  The sample rate of acoustic vs semantic tokens for AudioLM is 50Hz for acoustic tokens and 25Hz for semantic tokens, both with a vocabulary size of 1024.

In acoustic music, a key foundation model that employs APC for pre-training is Jukebox \cite{jukebox}. The authors train three VQ-VAE (vocabulary size 2048) models to reconstruct three temporal resolutions of music with losses accounting for reconstruction accuracy, codebook usage, and embedding stability. Three autoregressive transformers are used to model these sequences, each conditioned on upsampled tokens from the one-level-coarser transformer model. Jukebox is notable as the largest foundation model for music, sitting at 5B billion parameters and trained on the largest scale amongst other approaches. It has notably served as a foundation model for many downstream applications and has been shown to hold competitively informative representations for downstream probing, for which it is notably state of the art for many tasks to this day \cite{castellon2021codified}. Other APC-trained acoustic music models include MusicGen \cite{musicgen}, which is trained to model sequences of Encodec \cite{encodec} tokens, with novel token interleaving patterns to alleviate the computational costs of generating multiple codebook streams. MusicLM \cite{agostinelli2023musiclm} implements a similar setup to audioLM with soundstream acoustic tokens, w2vBERT semantic tokens, and possible text conditioning using MuLan \cite{mulan} embeddings and a similar hierarchical transformer cascade to AudioLM \cite{audiolm} to predict semantic tokens then fine-grained acoustic tokens. 

In symbolic music, APC has also been used for understanding and generation, where tokenisers are key to generation performance as they encode information from MIDI signals into tokens. Key approaches include Music Transformer \cite{huang2018musictransformer}, pop music transformer \cite{huang2020pop}, Jazz transformer \cite{wu2020jazz}, And MuseCoCo \cite{lu2023musecoco}, which pre-trains a BERT extractor for text-attribute pretext tokens and then trains an autoregressive transformer on text-conditioned symbolic music generation to great effect.

\textbf{Model design in APC for multimodal audio and music understanding:}
APC-trained audio and music language models have also been leveraged for LLM-based audio and music understanding. leveraging representation sequences from these models has proven to be effective for multimodal music understanding and captioning models, also trained or fine-tuned using APC. Most notable examples in the audio domain include listen, think, understand \cite{gong2023listen}, which makes use of LoRa adapters applied to a pre-trained LLaMa \cite{llama} LLM with acoustic tokens obtained from a pre-trained AST model to generate responses to user text queries including descriptive questions and chain-of-thought reasoning. Audio-PaLM \cite{rubenstein2023audiopalm} use a pre-trained PaLM \cite{anil2023palm} model adapted to generate text as well as audio tokens, which is done by adding audio-token rows to the token embedding matrix and fine-tuning the whole architecture. In doing so, AudioPaLM is able to generate both audio and/or textual responses to audio and/or textual user inputs, augmenting its versatility. Qwen-audio \cite{chu2023qwen} integrates generative pre-training with an audio encoder and frozen LLM as a natural language decoder. This architecture strategically leverages multi-task training to handle a variety of audio types and tasks, such as speech transcription and translation. By adjusting the decoder of hierarchical label sequences, Qwen-Audio can effectively cope with the challenges caused by label changes in different datasets. Besides, it can also handle basic music classification and description tasks on singer identification, emotion, genre, instrument, etc. Qwn2-audio \cite{chu2024qwen2} has a similar pre-training process with expanded data volume and better natural language prefix for different data and tasks, providing better results in audio.

APC has also been used for LLM-based music understanding with great success. Music-Understanding LLaMA \cite{liu2023music} adopts a pre-trained MERT encoder and a LLaMA LLM. The system is trained with APC with only the music understanding adaptor from the MERT embedding space to the LLaMA embedding space being unfrozen, and all other elements of the system frozen. LLaRK \cite{gardner2023llark} uses Jukebox \cite{jukebox} as an audio encoder and a fine-tuned LLaMA2 model to generate responses to textual queries. One specificity of LLaRK is that it augments training data with extracted musical features such as chord sequences, Key, tempo, and tags. GPT-4 is then used to create question-answer pairs using the extracted features to bolster the size of the training data for LLaRK. M$^2$UGEN \cite{hussain2023m} generalises music multimodal understanding with LLMs beyond audio and text to video and images. M$^2$UGEN also employs a frozen LLaMA model to generate textual responses. For this approach, live modules include adaptors from frozen MERT, ViT, and ViViT encoders for various modalities, as well as a text-audio token adaptor trained from scratch to generate audio tokens for either MusicGen or AudioLDM2. Specifics of M$^2$UGEN include special understanding/generation tokens appended as prefixes to generate responses to distinguish between audio-generative and text-generative tasks. 
Finally, recently, in the field of Symbolic Music, ChatMusician \cite{yuan2024chatmusician} makes use of LoRA adaptors on a pre-trained LLaMA model to generate answers including ABC notation as a second language for the model in the training set. It is clear through these studies that a key consideration for multimodal LLMs that use pre-trained textual LLMs for music understanding is the adaptation of encoded audio tokens to the LLM vocabulary, which might include LoRa adaptors, from-scratch external adaptors, or fine-tuning the LLM itself.

\subsubsection{Masked Modeling}\label{subsubsec:masked_modeling}
The main idea behind masked (language) modelling (MM/MLM) pre-training is to mask a portion of the input data and train the model to predict the original content of the masked parts using the remaining context. This is usually done with Transformer models, which naturally deal with sequential data and can effectively capture long-range dependencies. Practically, input sequences are represented as a series of tokens, with different tokenisation strategies existing, ranging from learnable embeddings to a token codebook learned from a different model (see Subsection \ref{sec:audio_tokenizers}). Depending on whether the tokens are continuous or discretised, the masked modeling objective is typically formulated as a regression loss between the predicted and true tokens or a classification loss between the predicted probabilities and the true tokens, respectively. As will be described in the following paragraphs, this loss is often used in conjunction with others, such as a contrastive loss.

\begin{table*}[!htp]\centering
\caption{Music Foundation Model Trained with Masked Language Modeling.}\label{tab:music_fm_mlm}
\scriptsize
\begin{tabular}{c|ccccc}
\toprule
\toprule
\textbf{Model} &\textbf{Modality} &\textbf{Application} &\textbf{Training Paradigm} &\textbf{Tokenizer} &\textbf{Architecture} \\\midrule \midrule
MAE-AST &Audio (Speech\&Sound) &Understanding &MLM &Spectrum &Transformer Encoder Decoder \\
Audio-MAE &Audio (Speech\&Sound) &Understanding &MLM &Spectrum &Transformer Encoder \\
SSAST &Audio (Speech\&Sound) &Understanding &MLM &Spectrum &Transformer Encoder \\
Beats &Audio (Sound) &Understanding &MLM &Spectrum &Transformer Encoder \\
DiscreteBERT &Audio (Speech) &Understanding &MLM &vqwav2vec &Transformer Encoder \\
WavLM &Audio (Speech) &Understanding &MLM &1-D CNN &Transformer Encoder \\
w2v-BERT &Audio (Speech, Audio, Music) &Understanding &MLM, Contrastive Learning &Spectrum &Transformer Encoder \\
ampNet &Audio (Music) &Generation &MLM & Discrete Tokens (DAC) &Transformer Encoder Decoder \\
MidiBERT-Piano&Symbolic (REMI) &Understanding &MLM &REMI, compound word &Transformer Encoder \\
MusicBERT &Symbolic (MIDI) &Generation &MLM &MIDI (OctupleMIDI) &Transformer Encoder Decoder \\
MRBERT & Symbolic (MusicXML) &Generation & MLM & \makecell{MusicXML Note Event, \\Compound Word} &Transformer Encoder Decoder \\
\midrule
EAT &Audio (Sound) &Understanding &MLM (Online Distillation) &Spectrum &Transformer Encoder \\
A-JEPA &Audio (Speech\&Sound) &Understanding &MLM (Online Distillation) &Spectrum &Transformer Encoder \\
data2vec &Audio (Speech) &Understanding &MLM (Online Distillation) &1-D CNN &Transformer Encoder \\
MT4SSL &Audio (Speech) &Understanding & MLM, MLM(Online Distillation) &1-D CNN &Transformer Encoder \\
data2vec 2.0 &Audio (Speech) &Understanding & MLM (Online Distillation) &1-D CNN &Transformer Encoder \\
M2-Duo &Audio (Speech, Audio, Music) &Understanding &MLM (Online Distillation) &Spectrum &Transformer Encoder \\
music2vec &Audio (Music) &Understanding &MLM (Online Distillation) &1-D CNN &Transformer Encoder \\
\midrule
MuLaP &Audio (Music), Text &Understanding &MLM &1-D CNN &Transformer Encoder \\
JMLA &Audio (Sound), Text &Understanding &MLM (Online Distillation) &Spectrum &Transformer Encoder Decoder \\
MusIAC &Symbolic (REMI), Text &Generation &MLM &REMI &Transformer Encoder Decoder \\
AV-HuBERT &Audio (Speech), Image &Understanding &MLM &1-D CNN &Transformer Encoder \\
\bottomrule
\bottomrule
\end{tabular}
\end{table*}

\paragraph{Continuous Masked Modelling}
Continuous Masked Modelling approaches generally operate on frequency representations of audio, learning continuous (i.e. not discretised) ``tokens'' of the input. At their core, they are inspired by the Vision Transformer (ViT) \cite{dosovitskiy2020image}, which divides an input image in fixed-size 16x16 patches, flattens them and linearly projects them to 768 dimensions to obtain patch embeddings, adds positional embeddings to maintain spatial information, and processes them through a standard Transformer encoder used for supervised pre-training. ViT was later adapted to the Audio Spectrogram Transformer (AST) \cite{gong2021ast} for audio classification, which demonstrated the necessity of image pre-training to achieve good performance. For that reason, SSAST \cite{gong2022ssast} proposed a self-supervised adaptation of AST by introducing two two-layer MLPs at the Transformer output. The first, a reconstruction head, tries to predict the original masked patch and, as is now typical in continuous masked modelling, uses the Mean Squared Error (MSE) loss, defined as $\mathcal{L} = (r - x)^2$, where $x$ is the true patch and $r$ is the reconstructed patch. The other, a classification head, tries to match the correct spectrogram patch for each masked position, which is a token-wise contrastive loss (see Subsection \ref{sec:contrastive_learning}). The losses are summed, with the reconstruction loss being weighted ten times the classification loss. SSAST also proposes a clustering method for choosing which patches to mask with controllable granularity, rather than randomly masking them, which aims to guide modelling at both a local and a global level.

Many audio-based approaches followed replacing the reconstruction head with a transformer decoder and using high masking ratios (between 70\% and 85\%). Among other design choices, these approaches defer in the exact patch masking strategy, which patches are fed to the encoder, and what loss is used. MaskSpec \cite{chongMaskedSpectrogramPrediction2023} and MSM-MAE \cite{niizumiMaskedSpectrogramModeling2022} use a simple encoder-decoder architecture, where both masked and unmasked embedded patches are given to the encoder. MAE-AST \cite{baadeMAEASTMaskedAutoencoding2022} uses a similar regression and contrastive loss to SSAST, but only unmasked tokens are given to the encoder. The encoder outputs are then given to the decoder together with the masked embedded patches. This approach results in a significant training speedup and memory usage reduction. Audio-MAE \cite{huang2022masked} only uses unmasked patches in the encoder and the MSE loss, but also shows improvements by using a local attention mechanism that groups patches into local windows for decoding. More recently, Dasheng \cite{dinkel_scaling_2024} scales up the MAE paradigm to 1.2 billion parameters and 272,356 hours of audio, using densely extracted mel spectrogram frames as input and learnable position embeddings, achieving several state-of-the-art audio classification results.

Some other approaches further tweak the patch masking strategy and architecture used. Audio-GMML \cite{atitoGroupMaskedModel2023a} uses a Group Masked Model Learning (GMML) strategy \cite{atitoGMMLAllYou2023} to mask patches that are deemed significant in the input, and combines intermediate representations from the encoder with its outputs as an input to a shallow decoder. MW-MAE \cite{yadavMaskedAutoencodersMultiWindow2024}  uses a multi-window, multi-head attention module in the decoder aiming to capture local and global context levels simultaneously. Finally, ASiT \cite{atitoASiTLocalGlobalAudio2024a} employs GMML and self-distillation with a teacher-student architecture, using a reconstruction loss alongside a local and global similarity contrastive loss.

More recently, interest has been shown in predicting latent representations rather than reconstructing the original input, particularly to encourage modelling of more abstract, higher-level features. data2vec \cite{baevskiData2vecGeneralFramework2022} proposed this concept with a self-distillation setup where a teacher transformer encoder model processes the full, unmasked input, while a student model with an identical architecture learns to predict the teacher's latent representations from a masked view using MSE. After each student model update, the teacher's weights are updated using an Exponential Moving Average (EMA) of the student's weights. data2vec demonstrated the potential of this approach for self-supervised learning in any data modality. MAP-Music2Vec \cite{li2022map} adapted this framework specifically to music, using a waveform input with a 1D convolutional embedder for both the teacher and student transformer models. They pre-train various versions of the model with different input length, mask span, and masking probability with 130k hours of music audio, and probe various transformer layers on clip-level music tasks, achieving comparable performance to other approaches.

A similar latent prediction architecture was proposed in JEPA \cite{assranSelfSupervisedLearningImages2023} for images, which was later adapted to audio \cite{feiAJEPAJointEmbeddingPredictive2024}. JEPA contains a so-called context encoder and a target encoder, similar to a student and teacher setup, respectively. A large part of the image or spectrogram is given to the context encoder, while the target encoder sees the whole input. A predictor network, which is a lighter-weight ViT, then attempts to reconstruct the target encoder output via the context encoder output and positional conditioning. Again, the target encoder weights are updated through an EMA of the context encoder weights. The audio adaptation of JEPA also introduces a curriculum masking strategy, where masking in pre-training is initially done randomly, but gradually time-frequency aware masking takes over, which should be a harder task for the model. Finally, M2-Duo \cite{niizumiMaskedModelingDuo2024a} uses a similar setup to JEPA in the audio domain, but masking is also applied to the target encoder input, opposite to that applied to the context encoder input, which they claim encourages the teacher to also model the input well and produce better encodings.

Masked modelling has also been adapted to a multimodal setting. AV-MAE \cite{georgescuAudiovisualMaskedAutoencoders2023} proposes a simple audio-video adaptation using just the reconstruction loss. They experiment with different encoder and decoder architectures, including how cross-modal fusion happens, and investigate finetuning applications, particularly unimodal ones. Specifically, they identify and experiment with \textit{early fusion}, where audio and vision embeddings are concatenated before being passed to the transformer, \textit{separate} processing, where the two streams are encoded separately and are fuse before being decoded (\textit{late fusion}), \textit{shared fusion}, in which weights are shared between the two encoders, and \textit{mid-fusion}, in which the output of two separate encoders is fused and fed to a joint encoder. pre-training with random initialisation, they find minor differences between fusion strategies for audio, although larger ones are identified for video and audiovisual applications. Those were favored by parameter sharing in the decoder and mid- or late-fusion in the encoder, possibly because they allowed stronger coupling between the modalities.

CAV-MAE \cite{gongContrastiveAudioVisualMasked2023} proposes an audiovisual approach using mid-fusion, with separate audio and video encoders proceeded by a joint encoder. Three partially masked streams are passed through the joint encoder: an audio encoding, a visual encoding, and a concatenated audiovisual encoding, all stemming from patch representations of the inputs. The inter-modal contrastive loss is then computed between mean average pooled embeddings of each single-modality stream. The encoder output of the joint encoding is decoded with a joint decoder, with the reconstruction loss computed between its outputs and the original input patches. MaViL \cite{huang2024mavil} follows a similar mid-fusion encoding setup. In addition to using the inter-model contrastive loss, it also uses an intra-model contrastive loss between different views of the input. pre-training is first done by training MaViL with the input reconstruction objective for each modality, and then the model is used as a teacher to a student that learns to predict the aligned, contextualised audiovisual representations produced by the teacher.

Finally, MuLaP \cite{DBLP:conf/icassp/MancoBQF22} proposes a music audio multimodal approach using weakly-aligned captions. Specifically, it follows a similar setup to ViLBERT with one branch for audio and one for language, which are encoded separately using a CNN encoder and a BERT-style tokeniser and encoder respectively. The outputs are then fed to two co-attentional layers, in which the key and value vectors are exchanged between modalities, allowing each modality to attend to relevant information in the other modality. In addition to a reconstruction and a classification loss for the audio and text masked modelling respectively, MuLaP also employs an audio-text matching objective by processing both true and fake audio-text pairs.

\paragraph{Discrete Masked Modelling}
Many music modelling approaches work with discretised representations. This is the case with symbolic music but also with several notable music foundation models that operate on audio by transforming it into a sequence of discrete tokens. Several approaches to discretised masked modelling have been proposed, including for audio, differing in their discretisation method, as well as whether the input or only the prediction targets are discretised. Unlike continuous masked modelling, which typically uses a regression loss, discrete masked modelling often employs a classification loss like the categorical cross-entropy:
\[\mathcal{L} = -\sum_{c=1}^{C}y_c \log(p_c),\]
where $C$ is the number of classes (discrete tokens), $y_c$ is the true label (1 if the sample belongs to class $c$, 0 otherwise), and $p_c$ is the predicted probability of class $c$ for the sample.

The first approach to BERT-like discretised masked modelling for audio is HuBERT (Hidden-Unit BERT) \cite{hsuHuBERTSelfSupervisedSpeech2021}, applied to speech processing. The input to HuBERT’s transformer encoder is still continuous, being a waveform encoded through CNN layers. The encoder outputs are then used to predict a discretised target. During the first iteration, k-means is run on MFCC features, and the assigned cluster IDs are used as the targets. The loss is then computed between predictions and assigned cluster IDs only for masked regions. During subsequent iterations, k-means is instead run on features from an intermediate layer of the transformer. The authors also demonstrate performance improvements with a multitask learning setup, in which clustering ensembles with multiple k-means models are used to create multiple codebooks corresponding to different granularities. WavLM \cite{chenWavLMLargeScaleSelfSupervised2022}, with the same architecture, effectively improves performance by increasing the data scale and mixing noise in the input during pre-training. Fast-HuBERT \cite{yang2023fast} accelerates the training of HuBERT using Fbank features with a lower sampling rate and simplifies the HuBERT loss to cross-entropy loss. HuBERT-AP \cite{ren2022speech} and CTCBERT \cite{fan2023ctcbert} change how the Transformer output is aligned with the unsupervised pre-training targets, making the pre-training process more like the practice of ASR task. PBERT \cite{wang2022supervision}, MonoBERT \cite{ma2023pushing}, and Poly-BERT \cite{ma2023pushing} introduce phoneme-like information to the SSL pre-training to improve performance. 

w2v-BERT \cite{chungW2vBERTCombiningContrastive2021a} uses a similar setup to HuBERT, but proposes and end-to-end training process. It utilises a learnable wav2vec 2.0 quantisation module using the features encoded by the CNN as input and trained on the contrastive loss to the context vectors (outputs of the context encoder, which in this case is a conformer). The context vectors are passed through another conformer that predicts the token ID, using the target IDs learned by the quantisation module in the same iteration. DiscreteBERT \cite{baevskiEffectivenessSelfsupervisedPretraining2020}, on the other hand, shows that discretisation of the input, not just the targets, can be more effective. Specifically, they use a pre-trained vq-wav2vec model to quantise the input, which is subsequently masked, as well as to create the targets, and perform standard BERT pre-training. They also try using k-means cluster assignments from MFCCs and filterbank features, but vq-wav2vec significantly outperforms them. 
  
More recently, EncodecMAE \cite{pepinoEnCodecMAELeveragingNeural2024} combines the HuBERT approach with the EnCodec quantiser and a masked encoder-decoder setup. Specifically, it uses EnCodec's encoder and RVQ layer with 8 out of its 32 codebooks to produce discrete targets. The MAE, operating on spectrogram frames rather than patches to increase its time resolution, is then trained to predict those targets. EnCodecMAE is shown to perform well across audio modalities, including music. BEATs \cite{chen2022beats} proposes a different setup, where a random codebook is initialised, and on the first iteration projected patches are matched to the nearest vector in the codebook. Their assigned ID is then used as a target for masked modelling. In subsequent iterations, the model acts as a teacher, producing prediction targets, with the tokeniser trained to predict them using cosine similarity.

A few approaches have been proposed specifically in the music domain. VampNet \cite{garcia2023vampnet} operates on discretised inputs and targets using the Descript Audio Codec (DAC). DAC compresses the 44.1kHz music audio input to a bitrate of 8kpbs using 14 codebooks, 4 of which are for coarse information and 10 for detailed. VampNet then uses two transformers, one for each token type, to learn to predict the masked set of tokens conditioned on the other token type. MERT \cite{li2023mert} instead proposes a multitask approach, trying to predict discrete RVQ EnCodec tokens from 8 codebooks and trying to reconstruct continuous CQT frames with masked modelling. It uses a continuous input of 24kHz waveforms encoded through a CNN. As an alternative to EnCodec tokens, MERT experiments with k-means clustering of log mel spectrograms and chroma features, however those generally perform slightly worse. It also shows that mixing short audio excerpts from the same batch to an input as an augmentation can improve the robustness of the model to common audio perturbations. Finally, another approach \cite{wonFoundationModelMusic2024} investigates various discrete token prediction setups using BEST-RQ \cite{chiuSelfsupervisedLearningRandomprojection2022a}, a method where the projection matrix and codebook are randomly initialised and not updated during training.  BEST-RQ provides a simpler alternative to discretisation that doesn't require a separate training phase, while performing comparably when enough training data is used.

While symbolic music can express continuous values (e.g. pitch bending between notes, or gradual dynamics changes such as crescendos), it's effectively usually treated as a discretised representation with a given level of rhythm and pitch quantisation. As such, discrete masked modelling has prevailed for symbolic music. MusicBERT \cite{zengMusicBERTSymbolicMusic2021}, aimed at symbolic music generation, uses the BERT setup with cross-entropy loss with a custom symbolic tokenisation called OctupleMIDI. The tokenisation creates separate tokens for time signature, tempo, bar number, position within the bar, instrument, pitch, duration, and velocity, which they show performs better than other tokenisations at both phrase- and song-level tasks. Instead of random masking, they mask tokens of the same type that belong to a bar, which they claim encourages the model to learn better representations than it would with adjacent token prediction. Another approach \cite{chouBERTlikePretrainingSymbolic2024} uses the same setup as MusicBERT, but uses the more established REMI and CP symbolic representations rather than OctupleMIDI, and tries to establish a symbolic piano music classification benchmark on public data, rather than the private dataset used to train MusicBERT. 

MRBERT \cite{liMRBERTPreTrainingMelody2023} follows a similar setup, but treats the melody and rhythm separately by using a ``semi-cross attention'' mechanism to exchange information between the two streams. This is done by allowing the query from one stream to interact with the other through a softmax operation, and then proceed to attend to the key-value pairs of the original stream. MRBERT is then finetuned for autoregressive, conditional, and seq2seq generation. Finally, MusIAC \cite{guoMusIACExtensibleGenerative2022}, focused on music infilling, uses an adapted version of the REMI tokenisation with multiple levels of control tokens such as track-level and bar-level controls. It uses BERT-like pre-training, but with an encoder-decoder transformer setup.

Finally, a few approaches have applied discrete masked audio modelling to a multimodal setting. AV-HuBERT \cite{shi_learning_2022} proposes an audio-visual adaptation with speech data using cluster assignment targets. The two modalities are initially encoded separately, with the encoded features then concatenated and passed through a joint transformer encoder. During the first iteration, MFCC features from the audio are clustered, and the assignment IDs are used as prediction targets. During subsequent iterations, features from an intermediate layer of the preceding model are instead used. The authors note the model's over-reliance on speech audio during pre-training, and counter it with two main methods: firstly, while visual data is encoded with a ResNet, audio data is projected with a fully connected network, forcing the audio encoder to learn simple features, and, secondly, during pre-training, an entire modality is randomly dropped out, with a likelihood of 0.5. u-HuBERT \cite{hsu_u-hubert_2022}, TESSP \cite{yao2022tessp} and VATLM \cite{zhu_vatlm_2023} further extend this approach by introducing a text stream as well as fusion techniques to allow pre-training with examples that don't have all modalities available.

\subsection{Music Domain Adaptation for Foundation Models}\label{sec:adaptation}

At the core of foundation models' versatility lie their few-shot, zero-shot, and in-context learning capabilities. Supervised finetuning paradigms enable foundation models to perform tasks without needing extensive task-specific training or instruction data, enabling dealing with use cases in the music domain that traditionally have limited data available. In this subsection, we will discuss finetuning methods for music FMs such as prefix tuning and adaptor tuning, followed by discussion on the zero-shot and in-context learning capabilities of FMs.
We note that limited work in the music domain investigates the impact of projection layers like MLP, cross-attention, and Q-formers in prefix tuning, as well as prompting for zero-shot learning, such as chain-of-thought.

\begin{table*}[!htp]\centering
\caption{Music Foundation Model Trained with Multimodal Adapters.}\label{tab:adaptor}
\scriptsize
\begin{tabular}{c|ccccc}
\toprule
\toprule
\textbf{Model} &\textbf{Modality} &\textbf{Application} &\textbf{Training Paradigm} &\textbf{Tokenizer} &\textbf{Architecture} \\\midrule \midrule
Qwen-Audio &Audio (Speech, Sound \& Music), Text &Understanding &prefix tuning, GPT & 1-D CNN &Transformer Encoder Decoder \\
LLaRK &Audio (Music), Text &Understanding &prefix tuning, GPT &Pre-trained model (CLAP, Jukebox) &Transformer Decoder \\
Musilingo &Audio (Music), Text &Understanding &prefix tuning, GPT &Pre-trained model (MERT) &Transformer Decoder \\
\midrule
MU-LLaMA & Audio (Music), Text &Understanding &adapter tuning, GPT &Pre-trained model (MERT) &Transformer Decoder \\
M2UGen &Audio (Music), Image, Text &Both &adapter tuning, GPT & Pre-trained model (MERT) &Transformer Decoder \\
SALMONN & Audio (Sound \& Speech), Text & Understanding & adapter tuning, GPT & Pre-trained model (Whisper, BERT) & Transformer Decoder \\
LTU &Audio (Sound), Text &Understanding &adapter tuning, GPT &Pre-trained model (Whisper) & \\
\bottomrule
\bottomrule
\end{tabular}
\end{table*}
\subsubsection{Prefix Tuning \& Prompt Tuning}
\begin{figure}[ht]
    \centering
    \includegraphics[width=0.75\linewidth]{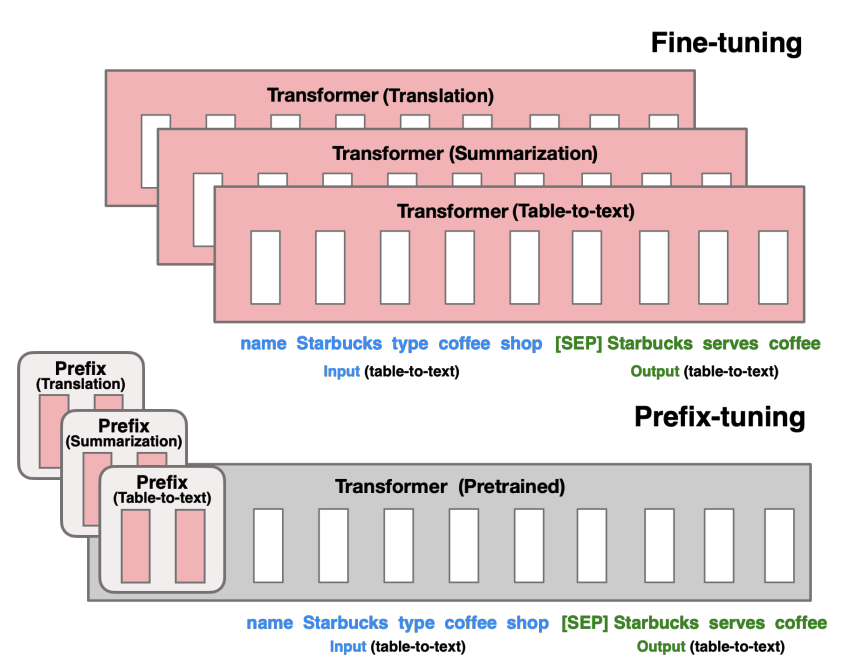}
    \caption{The top subfigure demonstrates the finetuning that updates all parameters shown in red. The button one shows prefix-tuning, which only optimises the parameters of input prompts or prefix blocks\cite{li2021prefix}.
}
    \label{fig:prefix}
\end{figure}
Various approaches, including prompt tuning~\cite{lester2021power} and prefix tuning~\cite{li2021prefix}, have been developed for large FMs to enable tuning on a relatively small set of downstream tasks or for modality alignment. 
Prefix tuning involves adding a series of trainable vectors (prefixes) to the input, thus effectively preparing the model's context for task-specific processing without changing its underlying parameters. This approach allows for a focused tuning of the model's behaviour while retaining its extensive pre-training knowledge. It can also be used to connect an image or music encoder to the LLM to enable it to understand information in other modalities. On the other hand, prompt tuning refines model performance by crafting and tuning task-specific prompts that guide model response generation. This approach exploits the inherent ability of LLM to generate context-appropriate responses based on the cues provided in the prompts and, therefore, requires minimal data for effective learning.

MusiLingo \cite{deng2023musilingo} and Llark\cite{gardner2023llark} seamlessly integrate pre-trained music encoders into LLMs using prefix tuning. MusiLingo uses a single projection layer to connect a pre-trained music model, MERT, to a Llama, while Llark uses CLAP and Jukebox to Llama2, enabling efficient task-specific adaptation through prefix adjustment. These allow such models to excel at generating music subtitles and processing music instruction, demonstrating a scalable approach to augmenting LLM in music multimodal applications. While these approaches use an MLP projection that aligns different modalities, some other visual-language models such as BLIP-2\cite{li2023blip} and Flamingo\cite{alayrac2022flamingo} also use Q-former and cross-attention for the modality alignment, respectively.
\subsubsection{Adaptors and Full Tuning}
Model adaptation~\cite{farahani2021brief,csurka2017domain} involves adapting a pre-trained model to perform better in specific, often narrower, domains and tasks. Adaptors\cite{houlsby2019parameter, zhang2023llama} have been widely adopted in language modelling and other modalities like computer vision. For example, MINERVA~\cite{lewkowycz2022solving} and Galactica~\cite{taylor2022galactica} use continued pre-training to adapt existing LLMs to the domains of science and math. Unlike prefix tuning or prompt tuning that freezes or only finetunes the initial transformer layers of LLMs, adaptors generally modify the deeper transformer layers.

In the realm of music foundation models, an example of domain adaptation is when a model trained on a wide variety of music is adapted to specialise in particular types of music with, for example, unique rhythms or instrumentation. On the other hand, task adaptation usually takes place when a model is adapted to solve new tasks, often narrower than the original. One such example demonstrates the use of a music tagging model, which hopefully captures rich semantic information about various musical aspects,   to other music classification and regression tasks ~\cite{choi2017transfer}.
In terms of finetuning PLMs in the music domain, MertTech~\cite{Dichucheng2024Measurement} uses multi-task finetuning to improve instrument playing technique detection in world music, of which limited examples existed in the pre-training dataset. 
For music-related multimodality,
Mu-llama \cite{liu2024music} utilises llama-adapter \cite{zhang2023llama} to align music recordings and their text captions.
SALMONN \cite{tangsalmonn} utilises a window-level Q-Former to convert variable-length sequences from speech and audio encoders output into variable numbers of augmented audio tokens for input to the Vicuna LLM and to finetune the LLM's parameters using Low-Rank Adaptation (LoRA)\cite{hu2021lora}, allowing for a range of audio-speech and music-understanding tasks to be performed.  Its music comprehension capability even exceeds that of a model designed for music \cite{weck2024muchomusic}.

\subsubsection{ Zero-Shot Learning \& Instruction Tuning}

\begin{figure}[hb]
\centering
    \includegraphics[width=0.75\linewidth]{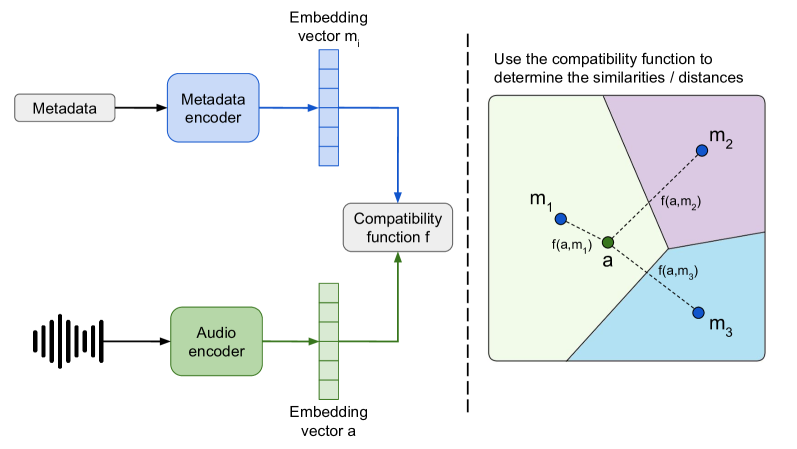}
    \caption{Zero-shot learning with PLM\cite{triantafyllopoulos2024computer}.}
    \label{fig:zero-shot}
\end{figure}

Zero-shot learning~\cite{romera2015embarrassingly,xian2017zero} takes the concept of minimal examples even further, allowing models to perform tasks without having been explicitly trained on them. 
In music and audio, supervised learning has been well studied to tackle different classification tasks such as scene classification, environmental sound classification, etc. However, the current supervised learning techniques require large amounts of annotated data from target sound classes. It can become expensive for humans to collect sufficient labelled data. Therefore, zero-shot learning has been favoured in many realistic scenarios without human annotation. As shown in Fig \ref{fig:zero-shot}, traditional zero-shot classification in audio leverages semantic embeddings to enable sound recognition without direct training examples. This method maps pre-trained audio features and textual descriptions into a unified semantic space, utilising acoustic embeddings from VGGish and textual embeddings from pre-trained language models like Word2Vec, GloVe, and BERT \cite{choi2019zero, xie2021zero2}. By evaluating compatibility between these embeddings, the approach classifies sounds effectively, even for untrained classes, significantly improving classification performance.

Instruction tuning is typically used to fine-tune language models on a wide variety of tasks, enabling the language model to generalise to new tasks with only instructions and no examples. 
FLAN (Finetuned Language Net) ~\cite{wei2021finetuned} utilises the concept of instruction tuning to enhance the zero-sample learning capabilities of language models. By pre-training LLMs with finetuned 137B parameters on more than 60 NLP tasks, FLAN demonstrates significant improvements in handling unseen tasks. The approach organises NLP tasks into clusters and selectively trains the model on some tasks while retaining others for evaluation. This approach allows FLAN to outperform larger models such as 175B GPT-3 on a variety of datasets, demonstrating its efficacy in generalising instruction tasks without prior direct contact. 
Supernatural Instruction \cite{wang2022super} introduces a comprehensive benchmark of 1,616 NLP tasks of 76 different types, providing a powerful platform for assessing the generalisation capabilities of NLP models in an instructional setting. The tasks are described through expert-written natural language instructions across multiple languages and task formats, enabling detailed evaluation of models such as Tk-Instruct. This smaller-scale Transformer model outperforms traditional large-scale models by being trained specifically for these different instruction sets, revealing the potential of targeted instruction-based training to achieve superior cross-task generalisation.
Instruction Tuning in music such as ChatMusician \cite{DBLP:journals/corr/abs-2402-16153} represents a significant advancement in the application of FMs to music. ChatMusician, through continuous pre-training and supervised finetuning of text and ABC notation, has gained the ability to perform complex music-generation tasks such as generating music based on given chord progressions, keys, motifs, and music structures, along with understanding music theory at the zero-sample level, effectively surpassing the musical capabilities of LLMs such as GPT-4. Prompt tuning with multi-tasking has also been shown to help with zero-shot learning\cite{sanh2021multitask}, while prompt writing techniques such as chain-of-thought \cite{wei2022chain} have remained underexplored.

\subsubsection{Few-shot Learning \& In-context Learning}

\begin{figure}[h]
\centering
    \includegraphics[width=0.75\linewidth]{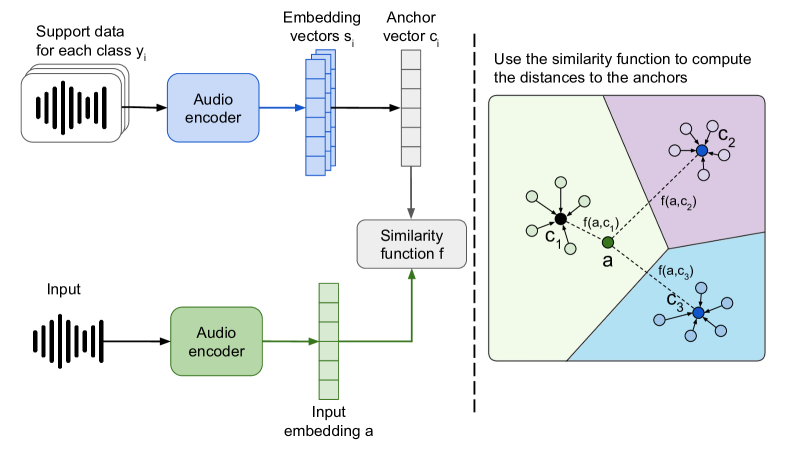}
    \caption{Few-shot learning with PLM \cite{triantafyllopoulos2024computer}.}
    \label{fig:few-shot}
\end{figure}

Few-shot learning refers to the ability of a model to learn and adapt to new tasks by observing only a small number of examples. As shown in Fig \ref{fig:few-shot}, traditional few-shot algorithms evaluate the pre-trained features of all examples and merge them into an anchor embedding for classification.
Few-shot learning has been widely adopted in fields such as natural language processing and computer vision. 
In the context of audio and music, few-shot learning enables foundation models to understand different properties of audio and music to perform certain tasks on sound event detection, transcription and source separation ~\cite{wang2020few, Wang2020FewShotDT,wang2022few}. 

Recently, in-context learning (ICL) was proposed, in which a model uses exemplars in the input to make predictions or generate outputs for unseen tasks ~\cite{brown2020language, dong2022survey}. 
This approach has not yet been fully explored in music. ICL typically requires a generative model capable of handling interleaved textual and musical data, which is a way of combining instructions and musical examples in a format in the input sequence. However, the development of generative pre-trained models that effectively handle this interleaved musical text format remains scarce, with both AnyGPT and ChatMusician lacking the nuanced understanding required to interpret and generate complex musical content based on contextual cues.

\subsection{Audio Tokenisers}\label{sec:audio_tokenizers}

The length of a music waveform sequence is typically long. A music recording may have a sampling rate of up to 48 kHz and several minutes or hours long. Such long sequences pose challenges for training machine learning systems. To address this issue, researchers use tokenisers to compress audio signals into latent representations with shorter sequence lengths than raw audio signals. Audio tokenisers can be categorised from different perspectives as follows.

\subsubsection{Hand-crafted versus Learning-based Tokenisers}

Before the advent of learning-based tokenisers, researchers developed hand-crafted tokenisers for audio processing. The most representative tokeniser is the Short-Time Fourier Transform (STFT) \cite{griffin1984signal}. An audio signal is denoted as $ x \in \mathbb{R}^{L} $, where $L$ is the number of samples. The audio signal $x$ is split into short-time frames. Then, a Fourier transform is applied on each frame to obtain the \textit{spectrogram} with a shape of $ \mathbb{R}^{T\times F} $, where $T$ is the frames number and $F$ is the frequency bins number. Later on, the \textit{mel spectrogram} was proposed to compress the STFT by applying the mel scale \cite{stevens1937scale}. The mel spectrogram has a shape of $T \times M$ where $ M $ is the mel frequency bins and there is $ M \ll F $. Mel spectrogram has been widely used in audio pattern recognition tasks such as automatic speech recognition. Based on mel spectrograms, researchers proposed MFCCs features (c.f. subsection \ref{subsec:symbolic-acoustic}). The advantage of hand-crafted features is that they have explicit explanations - for example, MFCCs decouple speech contents and the timbre of speakers. However, hand-crafted features may lead to information loss. Recently, learning-based representations have been proposed to address audio understanding and generation problems. 

\subsubsection{Continuous Audio Tokens } 
{\color{blue} to include more info}
Audio tokens can be continuous or discrete. Continuous tokens are usually used for audio understanding and diffusion model-based audio generation tasks. For example, continuous tokens can be wav2vec \cite{schneider2019wav2vec}, wav2vec 2.0 \cite{baevski2020wav2vec}, HuBERT \cite{hsu2021hubert}, and HIFI-codec \cite{yang2023hifi}. The continuous tokens usually have a shape of $ \mathbb{R}^{T \times C} $, where $ C $ is the channels of the representation. Later on, modified discrete cosine transform (MDCT) methods inspired from audio compression algorithms were proposed \cite{davidson2023high}. FunCodec is a frequency-domain codec that can achieve higher compression ratio than time-domain codecs. Recently, semantic audio codecs such as \cite{liu2024semanticodec} have been proposed to introduce both semantic and acoustic information into audio representations by using AudioMAE and k-means clustering algorithms. 

Many codecs are trained in an unsupervised way. We denote the input to a mask-based regression system as $ X $ and the mask as $ M $, where $M$ has the same shape as $X$. The mask-based regression system takes $ (1 - M) \odot X $ as input to predict $ (1 - M) \odot X $ $ M \odot X $. The represented works include SSAST \cite{gong2022ssast}, Audio-MAE \cite{xu2022masked}, MAE-AST \cite{baadeMAEASTMaskedAutoencoding2022}, MSM-MAE \cite{niizumi2022masked}, Audio-MAE \cite{xu2022masked}, MAE-AST \cite{baadeMAEASTMaskedAutoencoding2022}, BEATs \cite{chen2022beats}, and EAT \cite{chen2024eat}. The advantage of regression-based training is that they can reconstruct fine-grained audio signals. On the other hand, the contrastive training strategy uses semantically rich discrete label prediction to encourage the tokens to have high-level semantic information and discard redundant details of audio signals. For example, Wav2Vec \cite{schneider2019wav2vec} and Wav2Vec 2.0 \cite{baevski2020wav2vec}, HuBERT \cite{hsu2021hubert}, w2v-BERT \cite{chung2021w2v}, MERT \cite{li2023mert} apply contrastive pre-training for automatic speech recognition and music recognition. Recently, SemantiCodec \cite{liu2024semanticodec} has been proposed by using a k-means clustering algorithm to extract audio semantic information and use a diffusion model to reconstruct audio. BEST-RQ \cite{chiu2022self} applies a random projection quantiser to train audio tokenisers in an unsupervised way. MT4SSL \cite{ma2022mt4ssl} combines both k-means clustering offline targets from HuBERT and the teacher model's online targets from data2vec as the pre-training pretext task. 

\subsubsection{Discrete Audio Tokens} Discrete tokens are commonly used for language model-based audio generation tasks. For instance, in VQ-VAE \cite{van2017neural}, the encoder outputs discrete tokens. These discrete tokens can be input to language models such as Transformers to predict the next discrete token for audio generation. In SoundStream, a residual vector quantizer (RVQ) \cite{zeghidourSoundStreamEndtoEndNeural2021a} is introduced to represent audio signals in a hierarchical structure. Discrete tokens usually have a shape of $ \{0, 1, ..., D-1\}^{T} $, where $D$ is the vocabulary size of the tokeniser. Discrete audio tokens have the advantage of representing audio signals with a finite vocabulary. Therefore, discrete audio tokens can be seamlessly integrated with large language models in the natural language processing domain. DiscreteBERT \cite{baevski2019effectiveness} applies a VQ-wav2vec vocabulary to extract audio tokens. RVQGAN has been proposed for music generation in \cite{kumar2024high}. VampNet \cite{garcia2023vampnet} applied Descript audio codec (DAC) to extract tokens for music generation. MusicGen \cite{copet2024simple} proposes a simple and controllable music generation system by using Encodec.

\subsubsection{Symbolic Music Tokeniser}
Symbolic music tokenisers encode music information into a structured and symbolic format, such as MIDI \cite{yang2017midinet, chou2021midibert}, MusicXML \cite{good2001musicxml}, ABC notation \cite{geerlings2020interacting}, Humdrum \cite{huron2002music}, LilyPond \cite{nienhuys2003lilypond}, and Octopus \cite{liu2020octopus}. Symbolic music datasets have the advantage of containing compact information of music for editing than audio representations. Symbolic music representations high-level musical information that is useful for tasks such as music theory analysis and algorithmic music generation. Music Byte Pair Encoding (BPE) \cite{fradet2023byte} compresses symbolic tokens into a latent space to further reduce the latent dimension. Those representations have been used in symbolic music generation systems such as ChatMusician \cite{yuan2024chatmusician}.

\subsection{Model Architectures}\label{subsec:architecture}
After tokenising the audio signals, the representations or tokens are input into audio understanding or generation models. Many of these systems have an encoder-only, decoder-only, or encoder-decoder architecture. The audio encoders can be trained separately on audio tasks or jointly trained with the decoders. We describe the encoders and decoders as follows.

\subsubsection{Encoder}
An audio encoder is used to extract representations or tokens of audio signals. We describe the architecture details in this section. For the tokeniser of decoder-only models such as Encodec for MusicGen, please refer to the previous subsection.

\paragraph{Convolutional Neural Networks (CNNs)-based Audio Encoder}
Convolutional neural networks have been widely used in audio pattern recognition \cite{choi2016automatic, hershey2017cnn, kong2020panns}. We denote the audio signal as $ x \in \mathbb{R}^{L} $ and its time-frequency representation such as log mel spectrogram as $X \in \mathbb{R}^{T \times F}$. CNNs consist of convolutional layers, downsampling layers, and pooling layers to extract high-level features fromthe mel spectrogram. The CNNs have the advantage of extracting both time-domain patterns and time-frequency domain patterns. CNNs have been widely used in automatic speech recognition, music tagging, audio tagging, and source separation \cite{chandna2017monoaural}. A variety of CNNs, including ResNet and EfficientNets \cite{verbitskiy2022eranns} have been used to improve the vanilla CNN architectures. CNN-based encoders also include Musicset Unsupervised Large Embedding (MULE) \cite{mccallum2022supervised} and universal audio representations \cite{wang2022towards}. The advantage of CNNs trained on time-frequency domain representations is that they are relatively easier to train compared to time-domain systems and Transform architectures. 

Researchers have also investigated time-domain-only audio encoders. The time-domain audio encoders are directly applied on the waveform $ x \in \mathbb{R}^{L} $ without transforming it into a time-frequency representation such as a spectrogram. The time-domain encoders have the advantage of extracting non-harmonic time-domain features. For one-dimensional CNNs, time-domain convolutional layers are applied to the waveform to extract high-level features. To reduce sequence length, time-domain CNNs use dilated convolution layers to capture a larger receptive field, such as WaveNet \cite{van2016wavenet}. SampleRNN uses hierarchical recurrent neural networks to split long sequences into short sequences \cite{mehri2016samplernn}. The advantage of time-domain encoders is that they can be used to reconstruct audio signals without estimating the phases of spectrograms. Time-domain encoders have been widely used in source separation such as ConvTasNet \cite{luo2019conv} and 1D convolutional LSTMs \cite{defossez2019music}. Audio codec algorithms such as SoundStream \cite{zeghidourSoundStreamEndtoEndNeural2021a} also apply convolutional layers to extract latent representations from waveforms.

\paragraph{Transformer-based Audio Encoder}
The Transformer \cite{vaswani2017attention} is an architecture that utilises self-attention mechanisms to capture contextual relationships in input sequences. The core part of a transformer is self-attention: $ Attention(Q, K, V) = Softmax(\frac{QK^T}{\sqrt{d_{k}}}V) $, where $Q$, $K$, and $V$ are query, key, and value with shapes of $ N \times d_{k} $. The variable $N$ is the sequence length and $d_{k}$ is the dimension of the tensor. Since the successful applications of Transformer in natural language processing and computer vision, Transformers have also been widely used in audio processing. In audio tagging, Transformer-based systems usually take the time-frequency representation of an audio file such as a spectrogram $ X \in \mathbb{R}^{T \times F} $ as input. Audio spectrogram transformer (AST) \cite{gong2021ast} proposes to split a spectrogram into $16 \times 16$ patches. The patches are linearly projected to a sequence of 1-D embeddings. The embeddings are inputted to a transformer for classification. HTS-AT \cite{chen2022hts} is a hierarchical audio transformer with semantic tokens for sound event classification and detection. To reduce the sequence length of embeddings, efficient training of Transformers with patch-out \cite{koutini2021efficient} was proposed. SpecTNT \cite{lu2021spectnt} is a frequency-dependent architecture. In each SpecTNT block, a Transformer is used to extract frequency-dependent features into the tokens. In source separation, band-split Transformer \cite{lu2024music} was proposed to extract frequency-dependent features as input to Transformers. %

\subsubsection{Decoder}
\paragraph{Linear Projection} is typically applied to the encoder output to predict pre-training targets like tags or tokens in encoder-only models. Early works of decoders apply linear projection on the latent embeddings, such as PANNs \cite{kong2020panns}, and HTS-AT \cite{chen2022hts}. Multiple linear projection (MLP) has also been used as a decoder in audio tagging tasks. We denote the embedding extracted from the audio encoder as $h$ and the output of the decoder as $y = f_{\text{dec}}(h)$. The advantage of applying linear projections is that they are lightweight and require little data to train or finetune. The projection has been widely used in audio tasks with simple output formats, such as audio tagging.

\paragraph{Transformer.}
A linear projection is not enough to map from audio representations to a desired task. To address this problem, Transformer-based decoders were proposed as decoders to map from the audio representations to desired tasks and support flexible output formats. Inspired by the T5 architecture \cite{raffel2020exploring} in natural language processing, SpeechT5 \cite{ao2021speecht5} applies a Transformer-based symmetric encoder and decoder to autoregressively decode audio representations to the outputs of a variety of tasks. Sparse attention \cite{zhang2021sparse} methods such as Deepspeed \cite{aminabadi2022deepspeed} have been proposed to reduce the computational cost in Transformers. Long attention models such as block-wise self-attention \cite{qiu2019blockwise} and local attention \cite{dubois2019location} have been proposed to model long sequences.

\paragraph{State-Space Models} have also been used to model audio signals. In \cite{goel2022s}, a SaShiMi system was proposed to introduce state into audio representations. SSamba \cite{shams2024ssamba} is an unsupervised and attention-free system that uses states. Audio Mamba \cite{erol2024audio} investigated using state space models to replace self-attention for audio classification. Those state-based systems have the advantage of removing the quadratic computation complexity in Transformer models. 

Recently, decoder-only architectures such as LLaMA \cite{touvron2023llama} have been proposed to address the natural language processing and multimodality problems. The decoder-only systems interleave the audio representations with natural language. 
The representative works include Listen, Think, and Understand (LTU) \cite{gong2023listen}, Pengi \cite{agre1987pengi}, Salmonn \cite{tang2023salmonn}, LauraGPT \cite{chen2023lauragpt} and Qwen-Audio \cite{chu2023qwen}, where automatic recognition systems and audio tagging systems are used as an encoder, and a language model is used to predict the outputs according to the input questions. Decoder-only architectures have the advantage of allowing using arbitrary interleaved modalities as input and output.

\subsection{Interpretability \& Controllability on Music Generation}\label{sec:interpretability}

From a musical perspective, the \emph{goal} of controllable music generation is \emph{to make the music in \textbf{certain locations} follow \textbf{certain features} during the generation}. The location of music is multidimensional – it usually contains time-axis, stem-axis, and pitch-axis. For example, a specific location can be "notes within the first four bars in the piano track between C2 and C5", or "the last ten seconds of the drumset". The features of music include both information \textit{of} music and information \textit{about} music. Information about music is usually described by intrinsic music language, such as rhythm, pitch, chord etc., while information about music is usually described by natural language, such as “a pop style”, “a happy emotion”, “syncopated with the melody track”, or even the synesthesia of a particular visual scene.

To achieve the goal, most machine-learning methods follow a conditional generation approach, modelling $P(target|control)$ where $target$ is the music within the specified location and $control$ is the information based on which based on the \emph{methods} of control includes to \emph{add}, \emph{subtract}, and \emph{edit} notes or features of certain locations.

Interpretability could lead to better controllability of foundation models. For an end-to-end model, the controllability can be added when the music control coincides with the mathematical properties. These mathematical constraints can be used to guide the gradient descent path\cite{lattner2018imposing}, to evaluate the sampling process \cite{dai2021controllable}, or constrain the learned distribution \cite{Hadjeres2017DeepBachGeneration}. These can be used to control the factors that we can mathematically define, but the performance differs in cases. For end-to-end models, another approach is to use prefix control which defines tokens, such as MuseCoCo\cite{luMuseCocoGeneratingSymbolic2023}. All these show that control can only be applied to those concepts that can be explicitly defined.  

In the interpretability-oriented approach, one of the common approaches is to use representation learning models to learn a latent space of implicit music concepts, such as pitch contour, accompaniment texture \cite{wangPIANOTREEVAEStructured2020} and timbre \cite{lin2021unified}. These concepts are usually hard to define by rules, but we encode them in the latent space. It has been shown that these learned representations can be recombined for style transfer, and we can produce variations or new music via sampling or interpolating the latent space. We have also shown these latent codes can be used in longer-term prediction, and infilling \cite{zhao2021accomontage, wei2022learning}. As another approach, interpretability can be achieved by defining an interpretable workflow. For example, whole-song \cite{wang2024whole} defines a general hierarchical music language, so that the generation process is interpretable with respect to the defined workflow. Accomontage3 \cite{zhao2023accomontage} uses an interpretable architecture to generate two layers of latent codes in order to achieve a multi-track arrangement. Under current-generation fashion, the control of chords and rhythm is still external, and current methods cannot be learned well. The interpretability of such a concept is for future research. 

On the other hand, we believe that aligning these external controls with implicit music knowledge in a non-interpretable foundation model helps enhance its interpretability. In recent years, we have seen significant progress in text-to-music generation models. These models generate symbolic or acoustic music based on given text descriptions. Many of these models leverage language models or diffusion models. For instance, MusicGen~\cite{copet2023musicgen} stands out as an exemplary audio text-to-music model combining a T5 encoder~\cite{t5} for text descriptions, a pre-trained Encodec~\cite{encodec} as a compressor for music audio signals, and an acoustic transformer decoder for generating Encodec tokens. Riffusion~\cite{Forsgren_Martiros_2022} represents a notable diffusion-based audio text-to-music model, employing a UNet-based stable diffusion~\cite{ldm} framework. Additionally, recent prominent models like MuseCoCo~\cite{musecoco}, MusicLM~\cite{agostinelli2023musiclm}, and MusicLDM~\cite{musicldm} demonstrate promising results in symbolic and acoustic text-to-music generation.
 
All these models are trained in a supervised manner on extensive datasets comprising pairs of textual descriptions and corresponding music. This training enables them to generate symbolic or acoustic music based on textual inputs during inference. Despite their impressive capabilities, these advanced models remain black boxes, making it challenging to extract embedded musical knowledge from them or determine their understanding of musical concepts.
 
The opacity of these models presents significant challenges for interpretability. Understanding how these models implicitly translate textual descriptions into musical concepts and subsequently produce music remains an unsolved task. However, recent developments in content-based controllable music generation models offer a promising solution to enhance interpretability. This approach involves explicitly defining musical contents and fine-tuning the models to generate music aligned with these predefined content-based controls.
 
This methodology assumes that large models, when generating music, employ musical elements as internal intermediaries; in other words, interpretable prior musical knowledge exists hidden within the large models. By finetuning these models to generate music in a flexible and controlled manner based on diverse musical elements, researchers aim to leverage the interpretability of these models in music generation tasks.
 
A notable related work is the emergence of Parameter-Efficient Fine-Tuning (PEFT) methods, including Low-Rank Adaptor (LoRA)~\cite{hu2021lora} and LLaMA adapter~\cite{llama-adaptor}, which provides efficient ways to adjust pre-trained large models via tuning just a few parameters. These methods offer cost-effective ways to manipulate large pre-trained generation models, paving the way to transforming a standard text-to-music model into a content-based controllable generation model.
 
Based on PEFT techniques, recent works such as CocoMulla~\cite{cocomulla} and AirGen~\cite{airgen} focus on generating music from chords, drums, piano rolls, and text inputs. Similarly, MusicControlNet~\cite{musiccontrolnet} considers rhythm, melody dynamics, and textual cues to generate music. These efforts demonstrate a promising way to improve the model's interpretability, facilitating more nuanced and controlled music generation based on explicit musical controls. Meanwhile, they also suggest the growing trend toward enhancing the controllability of music generation through interpretable model adjustments.

\subsection{Foundation Models as Music Agents}\label{sec:agents}

\paragraph{What is an agent?} Broadly speaking, an agent is a kind of entity that has desires, beliefs, intentions, and the ability to take actions~\cite{zalta1995stanford}. In the realm of AI, an agent is an artificial entity that can set objectives autonomously or semi-autonomously, perceive the environment, work out plans, reason and make decisions, and take actions to achieve these objectives, powered by the technologies of artificial intelligence~\cite{xi2023rise}. In terms of music agents, they refer to AI agents dedicated to music and related domains, which can understand, generate, and convert music in symbolic, acoustic, and other modalities.

The development of AI agents has undergone a long process. Early attempts at building AI agents focus on improving the specific capabilities of an agent, such as sensing the environment, symbolic reasoning, taking actions with reinforcement learning, or handling specific tasks. However, the keys to AI agents are the general capabilities of reasoning, planning, perception, and action. While focusing on specific aspects of AI agents can result in steady progress, it might not be able to make fundamental enough improvements to push AI agents to practical usage. 

Large language models (LLMs) that emerged in recent years have revolutionised the era of language and multimodality and brought strong capabilities in reasoning, planning, perception, and generation, making them a good fit as the core module of AI agents. Some pioneer investigations such as AutoGPT~\cite{autogpt}, HuggingGPT~\cite{shen2023hugginggpt}, and Visual ChatGPT~\cite{wu2023visual} leverage LLMs to understand user requests, reason and plan according to objectives, decompose complicated tasks into subtasks, revoke external tools for task execution, and response generation. On the one hand, LLM-based agents extend the capabilities of LLMs to handle complicated and/or multimodality tasks. On the other hand, LLMs empowered AI agents to a new level with strong general capabilities that have never been possible using traditional AI agent technologies. Due to the great potential of LLM-powered AI agents, a lot of research efforts have been made in this area, such as improving the reasoning and planning capabilities~\cite{wei2022chain,agentgpt}, benchmarking AI agents~\cite{shen2023taskbench}, enhancing tool use~\cite{schick2023toolformer,yuan2024easytool}, designing agents to handle challenging tasks~\cite{shen2023hugginggpt}, and applying AI agents to more domains, including audio \cite{wavjourney, liang2024wavcraft} and music~\cite{yu2023musicagent}. In the next paragraph, we will review the typical AI agent in music and audio domains.

\paragraph{Examples of music and audio agents}
\begin{itemize}
    \item MusicAgent~\cite{yu2023musicagent}, the first LLM-powered AI for music, integrates diverse models and an autonomous workflow to address various music tasks like generation, transcription, and conversion. It simplifies the complex process for professionals and amateurs by analyzing requests, decomposing them into subtasks, and invoking external tools to fulfil these tasks.
    \item AudioGPT~\cite{huang2023audiogpt} focuses on audio modality and leverages a large language model to process different audio modalities (speech, music, sound) and handle different audio understanding and generation tasks. For music tasks, it supports singing voice synthesis by calling external music models.
    \item SpeechAgents~\cite{zhang2024speechagents} is a multimodal multi-agent system based on LLMs to simulate human communication. In SpeechAgents, each agent leverages a multimodal LLM as the decision centre and uses multimodal signals to exchange with other agents. 
    \item Loop Copilot~\cite{zhang2023loop} introduces an innovative system combining large language models with specialised AI music models to streamline the collaborative creation of music loops. This system utilises a conversational interface for dynamic, iterative music editing, and a global attribute table to ensure consistency throughout the creation process. Besides, it is not limited to creating music based on vague text inputs but allows for fine-grained musical edits, including adding or removing tracks and making localised adjustments to modes and tempos. This capability enhances the system’s utility in detailed music production tasks.
    \item ComposerX \cite{deng2024composerx} introduces a novel multi-agent framework for polyphonic notated music composition, utilising the reasoning power of large-scale language models as well as extensive knowledge of music history and theory. This approach produces high-quality, coherent compositions better than traditional single-agent systems, and requires no specialised training or services, making it a cost-effective alternative.
    \item ByteComposer \cite{liang2024bytecomposer} pioneers a human-like melodic composition process using a four-step agentic framework: conceptual analysis, draft composition, self-assessment, and aesthetic selection. ByteComposer combines the interactive and knowledge-understanding capabilities of LLMs with symbolic music modelling to achieve performance comparable to that of a human composer and is extensively validated through professional feedback.

\end{itemize}

\paragraph{Future Work of Music Agents}
Although there are some preliminary investigations on music agents, there are still a lot of space to improve. We introduce some future work of music agents bellow.
\begin{itemize}
    \item Improve reasoning and planning ability. The key capability of the music agents is enabled by the foundation model behind it. When handling complex tasks, the reasoning and planning ability of the foundation model plays a critical role in understanding the user request, decomposing the whole task into subtasks, organizing complicated working flows, and calling suitable models and tools for task execution \cite{huang2023towards, hao2023reasoning}. Thus, improving the reasoning and planning capability has always been the top priority in building music agents as well as general artificial intelligence. 
    \item From semi-autonomous to full-autonomous agents. Nowadays, most music agents are semi-autonomous, requiring a human in the loop, such as initiating user requests, setting system configurations, providing suitable model and tool sets, and strong interactions in the agent workflows. While, Autonomous agents based on LLM are popular research directionss\cite{hao2023reasoning, wang2024survey, yang2024unified, yang2024react}. In order to play more important and critical roles in helping musicians and music consumers, music agents should evolve from semi-autonomous to full-autonomous, gradually reducing the degree of user interference when fulfilling complex user requests.    
    \item Versatile music agents supporting multiple modalities and tasks. Music does not appear alone but usually engages with other modalities, such as text in lyrics and comments, tags/taxonomy in genres and styles, symbols/MIDI in music scores, and image/video in album covers and music videos. Modelling music with other modalities and tasks together could greatly extend the scenarios that music agents can support. This either requires the foundation model to be a powerful multimodal model or the external tools and models to cover multiple modalities and tasks. 
\end{itemize}

\subsection{Scaling Laws for Music}\label{sec:scaling_laws}
Scaling laws~\cite{kaplan2020scaling,henighan2020scaling,hoffmann2022training,achiam2023gpt} characterise how the performance of large language models (LLMs) (usually measured with a cross-entropy loss on a held-out set) scales as a power law of model size, dataset size, and training compute. Scaling laws can predict the loss of LLMs with a given number of model parameters and data size  (the number of training tokens, i.e., batch size $\times$ training steps) and can determine the optimal configuration of model size and number of training tokens given a computational budget. There are different formulations for scaling laws, such as OpenAI's scaling laws~\cite{kaplan2020scaling}, Chinchilla scaling laws~\cite{hoffmann2022training}, and data-constraint scaling laws~\cite{muennighoff2024scaling}. OpenAI's scaling laws are more precise, while Chinchilla's laws are simplified based on some assumptions. For simplicity\footnote{Strictly speaking, Chinchilla scaling laws have obvious errors. For example, given two models with model size $N_1$ and $N_2$ where $N_1>N_2$, when increasing the same amount of training tokens $\Delta D$, the loss reductions of the two models are the same according to Chinchilla laws as shown in Equation~\ref{eq_scaling_law}. However, it is evidently not the case in practice since the larger model $N_1$ will have more loss reduction than $N_2$. But for ease of explanation, we just use Chinchilla scaling laws as an example here.}, we briefly describe Chinchilla scaling laws as follows: 
\begin{equation}
\mathcal{L}(N, D) = \frac{A}{N^{\alpha}} + \frac{B}{D^{\beta}} + E,
\label{eq_scaling_law}
\end{equation}
which describes the loss $\mathcal{L}$ with regard to the model size $N$ and the number of training tokens $D$. It consists of three terms: 
\begin{itemize}[leftmargin=2em]
\item Functional approximation error ($\frac{A}{N^\alpha}$), which describes a perfectly trained model with $N$ parameters underperforms the ideal generative process~\cite{hoffmann2022training}. As $N$ is increasingly large, this term will approach $0$. 
\item Convergence error ($\frac{B}{D^\beta}$), which reveals that the model is not trained to convergence, as we just use a certain number of training steps and training tokens. As $D$ is increasingly large to cover the true data distribution, this term will approach $0$.
\item Minimal achievable loss ($E$), which corresponds to the entropy of the natural text and characterises the loss for an ideal generative process on the true data distribution.
\end{itemize}

In Equation~\ref{eq_scaling_law}, $\alpha$ and $\beta$  describe the effectiveness of the model size and data size in reducing loss. A larger $\alpha$/$\beta$ means the model/data is more effective in loss reduction. $A$ and $B$ are constants that are dependent on many factors such as data corpus, tokenisation, and vocabulary size, and do not have a fundamental meaning. Estimating the parameters $(A, B, E, \alpha, \beta)$ of the above scaling law usually requires collecting some data points $(N, D, \mathcal{L})$ and fitting the predicted loss and observed loss using some criterion and optimisation algorithms (e.g., Huber loss~\cite{carver1930annals} and L-BFGS algorithms~\cite{nocedal1980updating}). 

There are several typical uses of scaling laws: 1) predict the loss of LLMs with a given model size and training tokens; 2) allocate the optimal model/data size given a compute budget; 3) determine the relative scaling of model and data with an increased compute budget; 4) determine the ratio of data to model given a compute budget.

Traditional scaling laws are mostly studied in language language models for text~\cite{kaplan2020scaling,henighan2020scaling,hoffmann2022training,achiam2023gpt,muennighoff2024scaling}, which are not applicable for music. MuPT~\cite{qu2024mupt} is the first work to study the scaling laws for music, with dedicated design on symbolic music data representations and scaling laws in the data-constraint and over-fitting scenarios. While the great benefits of MuPT in guiding the modelling scaling of symbolic music generation, there is still a lack of scaling laws for music understanding and generation in the audio domain. Later, some studies emerged investigating scaling laws for multimodality~\cite{henighan2020scaling,aghajanyan2023scaling} and especially for speech-language model~\cite{cuervo2024scaling}. Since multimodal data such as images, videos, and speech are high-dimensional and complex and not usually modelled by language models, it is not straightforward to study the scaling laws on multimodal data. Therefore, they usually tokenise multimodal data into tokens and model them as the next token prediction. In this way, they could study the scaling laws similarly to those in text-based large language models. 

In~\cite{cuervo2024scaling}, two different data representations for speech are studied: 1) semantic tokens extracted by HuBERT~\cite{hsu2021hubert}; 2) semantic BPE tokens which are obtained by applying byte-pair encoding~\cite{bpe} on the semantic tokens from HuBERT. The scaling laws show that given the same amount of tokens, semantic BPE tokens could consume more model capacity than semantic tokens. This is because semantic BPE tokens are semantically richer and have a larger coefficient $\beta$ as shown in Equation~\ref{eq_scaling_law}, which means by increasing the same amount of tokens, semantic BPE tokens could cause more loss reduction. In this way, in order to better coordinate with the data and reduce the loss, the model size should increase. An intuitive explanation is that as the data is semantically richer, it should consume more model capacity to process them. This also aligns with our intuition: in the audio domain, the vocoder or codec model usually generates the waveform data with either auto-reconstruction or from mel-spectrogram, which processes less semantic information, so the vocoder or codec model size is usually small. However, the language models that predict semantic tokens are usually very large.   

Since there is not much research on scaling laws for music in the audio domain (although the scaling laws from multimodal and especially speech domain could provide much insight into the audio music domain, they are not directly applicable), and the emergent ability of music FMs is under-explored, we call for actions to proactively study the scaling properties of different data representations and model paradigms for audio music understanding and generation in the future. Such research topics include, but are not limited to, which type of downstream capability of large models can be inferred by the small models, how the properties of audio tokenisers relate to the parameters of scaling law, and what the minimal number of  model parameters or training data is for a specific type of downstream tasks etc.

\subsection{Additional Future Improvements}\label{sec:future_improvements}

In this subsection, we will shortly discuss potential research topics related to foundation models for music. Other than the foundation model techniques discussed in previous subsections, the following paragraphs will include domain knowledge, long sequence modelling and causal modelling. For discussions on fairness, transparency, and potential issues from model bias, please refer to subsection\ref{subsec:bias}.
\subsubsection{Domain Knowledge}
The model architecture described in this section so far, apart from small modifications, can be construed as ‘standard’ in the sense that it has been explored for a plethora of other tasks both somewhat linked with music (e.g. speech) as well as entirely different from it (e.g. machine translation). Although such an approach is quite appealing \cite{humphrey2012deep}, the limited data circumstances  and inadequacies of architecture/optimisation in the music domain have led to a substantial interest in incorporating domain knowledge. This reversal towards an appreciation of domain expertise has brought about a revival in domain-specific engineering of machine learning approaches in many areas, including music \cite{serra2013domain}. There are quite a few disciplines from which domain knowledge for music foundation models can be sourced. Acoustics, for example, gave rise to spectrum-based representations, which continue to find their use in deep learning architectures \cite{dhariwal2020jukebox}. The same applies to psychoacoustic chroma tokens \cite{muller2015fundamentals} which can assist deep learning should it fail to automatically infer the importance of timbre information from the data \cite{li2023mert}. Music theory provides a further rich source of information which has inspired numerous approaches and will possibly continue to do so.

One key challenge here is how to integrate music theory and related concepts into existing standard architectures rather than designing custom architectures to fit one or more music theories, as was done before \cite{widmer1998applications} and still remains popular in the area \cite{olivan2023composition}, or uncover new ones by means of automatic theoretical analysis \cite{yu2017rover}. One popular direction for integrating music theory into the Transformer and related architectures is input modification. Although a simple approach, it provides opportunities for encoding a wide range of music-theoretical information. So far, the work in this direction has focused on encoding lower-level relative attributes, such as pitch interval, duration and onset \cite{guo2023relative}, and some higher-level attributes, such as bars \cite{waite2016magenta}. There remains to be an extension of this work to other higher-level relative attributes. The same applies to absolute attributes such as melody, harmony and rhythm. Methods that do not rely on input modifications to incorporate music theory are interesting but likely challenging due to changes required to the core Transformer implementation as was the case with the work of \cite{huang2018music} on relative positional encoding.

Another key challenge lies in bridging the gap between approaches more naturally formulated at the symbolic token level and approaches suitable to be used with acoustic tokens. As in other areas of structured audio processing, any attempt to split audio into the underlying units (e.g. phones in speech or chords in music) is a highly non-trivial endeavour requiring expert annotators to act under the uncertainty of determining boundaries among those units. In the case of music, this is further compounded by the need to decide on the level of some of those units (e.g. degree of loudness). Thus, approaches capable of automatically inferring segmentations or marginalising over all possible segmentations would need to be developed. 

\subsubsection{Long-Sequence Modelling}\label{subsec:long_sequence_modelling}
The length of a typical music recording (several minutes) and its resolution (48 kHz and above) firmly suggest that music belongs to a class of long-sequence modelling problems. For example, a 3-minute recording with a resolution of 48 kHz packs approximately 9 million samples. The need to handle long sequences has inspired a large body of approaches, which can be grouped into those that modify existing architectures and those that propose novel architectures that are better suited to handling long sequences. 

Approaches that modify the widely popular Transformer primarily focus on the two most important shortcomings of this architecture when handling long sequences: the complexity of attention and positional encoding. The quadratic complexity of the standard attention mechanism has been primarily tackled using sparsification in structured and unstructured forms leading to a drastic decrease in complexity with minimal or no loss in performance \cite{child2019generating, dai2019window, guo2019star, ye2019bptransformer, beltagy2020longformer, kitaev2020reformer, qiu2020blockwise, zaheer2020bigbird, xiong2022simple}. The standard (absolute) positional encoding seemingly appropriate for short sequences found in many speech and natural language processing tasks has been extended to various relative positional encoding schemes \cite{shaw2018self, dai2019window, t5, wu2022memformer} and hybrids of absolute and relative encodings \cite{su2024rope, sun2023length}. Although many of these approaches have been integrated into Transformer-based foundational music models \cite{huang2018musictransformer, yu2022museformer}, their ability to handle truly long music recordings remains to be seen. In this context, the recent work on long-context speech recognition \cite{flynn2023context}, where context length reached 1 hour of speech, suggests this to be a promising direction. 

Recently there has also been substantial interest in developing novel approaches with linear or near-linear complexity with respect to sequence length. Among them, structured state-space models \cite{gu2022efficiently} appear to yield promising results for sequences reaching 1 million tokens \cite{gu2023mamba}. These models can be interpreted as a combination of recurrent and convolutional architectures which additionally integrate principled mechanisms for modelling long-range dependencies. Another popular line of research explores hierarchical architectures where sequences are segmented into shorter units \cite{dai2019window, pappagari2019hierarchical, rae2020compressive}. This allows making use of separate mechanisms for modelling short-term (e.g. attention) dependencies within segments and long-term (e.g. implicit \cite{bulatov2024scaling} or explicit \cite{pappagari2019hierarchical} recurrence) dependencies across segments. Some of these approaches showed promising results on synthetic tasks with sequences reaching 2 million tokes.

\subsubsection{Causal Modelling}
The discussion about domain knowledge earlier in this section stopped short of mentioning perhaps one of the strongest kinds of domain knowledge - causal relationships. Such relationships we cultivate all of our lives. In childhood, many of us learned the impact of key pair variations on our perception of sound by tinkering with the piano keys. As we grow older and start watching movies we begin to judge choices made by sound directors in particularly dramatic scenes. The knowledge of these relationships and our ability to enact them enables us to conduct interventions (e.g. fix a key, increase tempo) and create counterfactual pieces (e.g. imagine a known piece in a new key). Do machine learning engineers need to know the causal relationships that the pianist Horowitz formed and his ability to manipulate them in order to automatically generate Horowitz-quality piano pieces?

Even in simpler scenarios than music, determining causal relationships is a non-trivial endeavour. In many cases this is hardly a possible task due to the lack of access (e.g. astronomy) or understanding (e.g. neurology). In other cases, the full set of causal relationships is not known or available. Latent variables (e.g. hidden Markov models \cite{rabiner1989hmm}, auto-encoders \cite{kramer1991autoencoder}) emerged as a powerful but implicit surrogate aimed at capturing some of these relationships. However, the lack of appropriate parameter estimation methodologies capable of learning causal rather than merely correlational relationships meant that latent variable models have so far achieved limited success in this area. Causal learning emerged in response to these deficiencies of modern machine learning. Starting from the work done by Pearl and colleagues \cite{pearl2009causal} it has now emerged into a new field full of exciting developments \cite{kaddour2022causal}. Despite much progress made in causal learning over the years, work done in music and related areas remains extremely limited. The key issue stems from the lack of data with labelled causal information and/or methods capable of automatically extracting causal relationships. Although some causal models have emerged for simple image generation tasks, where causal factors could be limited to shape, texture and background \cite{sauer2021counterfactual}, it is not clear how to extend them to other areas where causal relationships are much less trivial to specify, extract, and model. 

An interesting alternative trend recently emerged where large non-causal models are manipulated into exhibiting causal relationships \cite{lewis2020rag, meng2022locating, ouyang2022feedback, wei2022chain, li2023intervention}. One prominent example is the work of Wei and colleagues on chain-of-thought prompting for eliciting factually correct reasoning of large language models in question-answer type tasks \cite{wei2022chain}. Unlike standard prompting \cite{brown2020language}, where an example question is supplemented with a short factual answer (e.g. giving a numerical answer, such as 11, to the question about how many cans of food are left after a certain number of them go missing and some restocked), the chain-of-thought prompting supplies an answer that showcases how a human could arrive to that conclusion \cite{ling2017rationale}. This approach has been shown to be effective in arithmetic, common sense and reasoning tasks. Whether similar prompt manipulations can be extended to music foundation models remains to be seen.

\section{Datasets \& Evaluation}
\subsection{Datasets}
Data plays a central role in training the underlying model. The diversity, quantity and quality of the training data contribute greatly to the generalisability and robustness of the model\cite{xie2024doremi, soldaini2024dolma, zhang2024map}. For machine learning and computer music researchers, training music-based models may require hundreds of billions of tokens or even larger datasets, and selecting high-quality datasets of different music recordings is a great challenge. 
In the realm of computer music, datasets are categorised into composition-level, typically sheets represented in MusicXML and ABC notation, performance-level symbolic music exemplified by MIDI formats, and datasets comprising raw audio waveforms. This section explores extensive open-source music datasets that are beneficial for training LLMs or LDMs tailored for music applications. Additionally, we will discuss Python libraries that are pivotal for music processing. Subsequently, the focus will shift to multimodal music datasets. Finally, the evaluation methodologies for foundation models developed on these datasets will be examined.

\subsubsection{Composition-level Symbolic Music Datasets}\label{subsec:musicdatasets}
In this subsection, we focus on introducing symbolic music datasets that contain more than 1,000 music excerpts. We will exclude smaller datasets though they are potentially useful for evaluation or supervised fine-tuning in specific domains. For more information on music datasets, readers can refer to the datasets website page of the ISMIR community \footnote{\url{https://www.ismir.net/resources/datasets/}.}

\begin{table*}[!t]
    \caption{Open-source music dataset for pre-training.}
    \centering
    \begin{tabular}{c|ccc}
    \toprule
    \toprule
        Dataset & Modality & n files & Description\\
        \midrule
        \midrule
        Wikifonia& MusicXML & 2,252 CSV samples &CSV of MusicXML from Wikifonia.org\\
        MuseScore Lead Sheet Dataset& MusicXML, MIDI & 226 piece with 336k notes & Derived from MuseScore website\\
        Hooktheory Lead Sheet Dataset & MusicXML & 11,329 lead sheet samples & Derive from TheoryTab music theory forum \footnote{\url{https:// www.hooktheory.com/theorytab}}\\ 
        \midrule
       IrishMAN & ABC, MIDI, MusicXML& 216,284 & Scottish \&  Irish folk songs\\
        Nottingham Music Dataset& ABC notations& 1,200& Online corups of British \& American folk songs.\\
        ABC tune book of Henrik Notebook& ABC notations& 2,800 & Irish \&  Swedish folk songs\\
        \midrule
        Lakh MIDI Dataset &MIDI& 176,581 files& Mainly pop and rock music\\
        Yamaha Signature MIDI Collection &MusicXML, MIDI&1.4k& Piano performance, mainly Romantic pieces\\
        DoReMi & Image, MusicXML, MEI, MIDI&6k&Steinberg's Dorico \\
        ADL piano dataset& MIDI &11,086 & Pop, classical and jazz piano pieces\\
        Symphonies & MIDI & 46,359 files, 650 hours & Classical symphony, multi-instruments\\ 
        NES-MDB & MIDI & 5,278 & NES games BGM\\ 
        \midrule
        MAESTRO &MIDI, audio & 1.2k files & Classical Piano\\
        GiantMIDI-Piano & MIDI, audio & 10,855 pieces, 1237 hours& Machine transcribed classical piano\\
        Meta-MIDI & MIDI, audio & 436,631 MIDI files & 10M match to Spotify music tracks \\ 
        \midrule
       Free Music Archive (FMA)& audio & 106,574 tracks, 8.2k hours& Collected from FMA website\\
       MTG-Jamendo & audio& 55,701 tracks, 3.8k hours& Collected from Jamendo website\\
       Music4ALL & audio& 109,269 tracks, 911 hours& Collected from YouTube\\
       \midrule
       Million Song Dataset (MSD)& audio feature& 1,000,000& 1M pop song provided by Echo Nest\\
       AudioSet& URL of audio & 1,011,305 music clips &2,084,320 clips including general audio\\
       AcousticBrainz & audio feature& 2,524,739 &2M audio features with MusicBrainz metadata\\
       Disco-10M & feature \& URL of audio & 15,296,232 &10M features with diverse genres and artistis\\
       \bottomrule
       \bottomrule
        
    \end{tabular}
    \label{tab:datasets_analysis}
\end{table*}

\paragraph{MusicXML}
Besides representing traditional Western Classical music scores, MusicXML is widely utilised for storing lead sheets of popular music, encompassing the melody, lyrics, harmony, and various markings of a song. The melody is notated in modern Western music notation, with lyrics presented as text beneath the score, and harmony indicated by chord symbols positioned above the score. Notably, lead sheets provide little details regarding instrumentation or accompaniment.

The Wikifonia Lead Sheet Dataset \cite{lim2017chord}, originally from the now-defunct public repository Wikifonia.org, includes 5,533 MusicXML lead sheets of Western music across various genres. Before the service ceased in 2013, Lim et al. secured 2,252 lead sheets in major keys with typically one chord per bar. If a bar contained multiple chords, only the first was chosen. The dataset is converted and distributed to CSV format on the website\footnote{\url{http://marg.snu.ac.kr/chord_generation/}}. 
Besides ABC notation, it also provides a version of MusicXML and MIDI files.

To complement the Yamaha Signature MIDI Collection, Jeong et al. developed the MuseScore Lead Sheet Dataset \cite{jeong2019virtuosonet} by collecting MusicXML scores from MuseScore, a community-driven music score web platform, along with including their own transcriptions. These scores were transcribed voluntarily by community members. Although the dataset includes just 226 MusicXML pieces by 16 composers, it corresponds to 1,052 piano performances in MIDI format. Collectively, these pieces comprise a total of 666,918 notes in MusicXML and 3,547,683 notes in MIDI. 

Hooktheory Lead Sheet Dataset (HLSD) \cite{yeh2021automatic} include 11,329 music lead sheet samples collected from the TheoryTab forum on Hooktheory's website, a resource specializing in music education software. These samples, whose melodies are transcribed by users with corresponding chord progressions in high quality, are denoted by both chord symbols and functional labels. The dataset included an excessive 704 chord class. Due to copyright restrictions, only song snippets are shared on TheoryTab, with annotations such as structural and genre labels.

\paragraph{ABC Notation}
The ABC format is a plain text method for documenting music. Besides representing single-line melodies, it has the capability to fully represent polyphonic classical scores. 

The Irish Massive ABC Notation (IrishMAN) dataset \cite{wu2023tunesformer} is a comprehensive collection of 216,284 Irish tunes in ABC notation. It is primarily sourced from an online traditional Scottish and Irish collection \footnote{\url{https://thesession.org/}} and the official website of ABC notations\footnote{\url{abcnotation.com}}, 
To achieve uniform formatting, all tunes were converted from ABC to XML and back to ABC using automated scripts, with non-musical fields such as titles and lyrics removed to maintain focus on musical content. Additionally, each tune in the dataset is annotated with control codes for musical forms and structures.

Further, an open-source website has Western folk music in ABC format such as the Nottingham Music Dataset (NMD), which is comprised of 1,200 British and American folk songs. 
A recently refined version is available online \footnote{\url{https://ifdo.ca/seymour/nottingham/nottingham.html}}. Another example is the Henrik Norbeck collection that includes more than 2,800 ABC scores with lyrics, for Ireland and Sweden \footnote{\url{http://www.norbeck.nu/abc/}}. More information can be found in the dataset section of the music generation survey \cite{ji2020comprehensive}.

\subsubsection{Performance-level Symbolic Music Datasets}
\paragraph{MIDI}
There are many MIDI datasets and we only include the largest and the widely used portion.

The Lakh MIDI Dataset (LMD)\cite{raffel2016learning} is a large MIDI polyphonic music corpus with inconsistent expressive characteristics, consisting of various genres, instruments, and periods of time. The ``LMD full'' version includes the whole 176,581 MIDI files after removing duplication, and the ``LMD matched'' subset refers to the 45,129 files that are aligned with items in the Million Song Dataset (MSD).

The Yamaha Signature MIDI Collection\footnote{\url{http://www.yamahaden.com/midi-files}} comprises 1.4k high-quality solo piano MIDI performances recorded during an international competition for junior pianists. Predominantly featuring late Romantic pieces by Chopin and Liszt, along with Mozart sonatas, this dataset is essential for research in performance generation. A corresponding MusicXML version is available through the MuseScore Lead Sheet dataset, or the (note-)Aligned Scores And Performances dataset (nASAP) \cite{Peter-2023}.

DoReMi dataset \cite{shatri2021doremi} has around 6k pages of the score images, with the corresponding MusicXML, MEI, and MIDI scores. Its primary objective was optical music recognition (OMR) but can easily be used in other tasks. The MIDI is exported using Steinberg's Dorico \footnote{\url{https://github.com/steinbergmedia/DoReMi}} software. 

The Augmented Design Lab (ADL) Piano MIDI dataset \cite{ferreira2020computer} is derived from the LMD by retaining only one of the multiple versions of each song and only tracks featuring piano-family instruments identified by MIDI program index 1-8, resulting in 9,021 unique rock or classical piano MIDI files. 
To enhance genre diversity with additional styles like jazz, they incorporated 2,065 files sourced from Internet sources, culminating in a total of 11,086 pieces in the final dataset.

Symphonies dataset \cite{liu2022symphony} is a large corpus of symphonic music from various online sources. The collection consists of 46,359 MIDI files, primarily symphonic works, with a total duration of about 650 hours. 

The NES-MDB dataset \cite{donahue2018nes} encompasses 5,278 songs from the soundtracks of 397 NES games, involving compositions from 296 composers and containing over two million notes. This dataset is extracted from the assembly code of NES games, capturing precise timings and parameters for authentic chiptune renditions. The NES synthesiser includes five instrument voices—two pulse-wave, one triangle-wave, and one noise generator—with the complexity of its audio synthesis chip abstracted for researchers. NES-MDB is available in multiple formats to facilitate research in NES music, with additional details provided for those interested in deeper technical exploration.

Additionally, there are many online corpora for MIDI files such as 
BitMidi\footnote{\url{https://bitmidi.com/}} 
Classical Archives\footnote{\url{https://www.classicalarchives.com/}}, 
and FreeMidi\footnote{\url{https://freemidi.org/}}.

\paragraph{MIDI-audio dataset}
Apart from what has been mentioned above, there are multiple music transcription or generation datasets that not only include symbolic music but audio as well.

MIDI and Audio Edited for Synchronous TRacks and Organization (MAESTRO) dataset \cite{hawthorne2018enabling} consists of about 200 hours of professional piano performances including both MIDI annotations and audio recordings.

The GiantMIDI-Piano dataset \cite{kong2020giantmidi} is the largest classical piano dataset, encompassing 10,854 MIDI files by 2,784 composers, totalling 1,237 hours in duration. Such collection is transcribed using a high-quality, open-sourced piano transcription system \cite{kong2021high}. ATEPP \cite{Zhang2022ATEPPPerformance} is a similar transcribed classical piano dataset but with an emphasis on classical performance variety. It contains 11,674 performance tracks by 49 virtuoso pianists. 

MetaMIDI Dataset (MMD)\cite{ens2021building} is a large dataset of 436,631 MIDI files with metadata. It also includes a subset with 168,032 files that are matched to the audio clips in the MusicBrainz database.

\subsubsection{Acoustic Music Datasets}

The Free Music Archive (FMA) \cite{defferrard2016fma} is an expansive dataset that comprises 343 days of audio. It includes 106,574 tracks from 16,341 artists and 14,854 albums, organised into 161 genres. The dataset offers high-quality audio, audio features, and comprehensive metadata at the track and user levels, along with free-form text like artist biographies. Notably, a significant portion of the FMA consists of experimental music, which markedly differs from typical music used in downstream tasks. Therefore, it is advisable to exclude experimental tracks during pre-training to optimise relevance and applicability.

The MTG-Jamendo \cite{bogdanov2019mtg} Dataset is compiled from music available on Jamendo\footnote{\url{https://www.jamendo.com/}}, and features tags applied by the uploaders. This dataset contains 55,701 full audio tracks, each at least 30 seconds long and encoded at 320kbps MP3, totalling 509 GB of audio. The average track length is 224 seconds, summing up to 3,777 hours of audio. The tracks are annotated with 692 tags, derived from genres, instruments, and mood/theme categories. To streamline the dataset, variant spellings and semantically identical tags were consolidated, resulting in 99 remapped tags. 

Music4ALL\cite{santana2020music4all} database encompasses 109,269 songs from 15,602 anonymous users along with their listening histories, each represented by 30-second  audio clips, lyrics, and multiple metadata attributes like genres and tags. 

Aside from the aforementioned, there are many large-scale datasets which only provide the feature or URL of the audio, allowing users to collect the tracks from the internet. Some such examples include the Million Song Dataset (MSD) \cite{bertin2011million}, AudioSet \cite{gemmeke2017audio}, AcousticBrainz \cite{porter2015acousticbrainz}, and Disco-10M  \cite{lanzendorfer2024disco} etc.

\subsubsection{Software for Music \& Audio Processing}
Developing foundation models for music necessitates leveraging open-source software for tasks ranging from preprocessing and feature extraction to training the models and evaluation in both symbolic and audio domains. 
In symbolic music processing, libraries such as \texttt{mido} and \texttt{pretty\_midi} offer extensive capabilities for MIDI manipulation, whereas \texttt{note\_seq}, supported by Google's Magenta team, provides advanced functionalities for processing and training symbolic music. 
For audio processing, tools like \texttt{librosa}, \texttt{Essentia}, and \texttt{madmom} are crucial for their comprehensive audio analysis and feature extraction capabilities. 

Importantly, for audio I/O operations, the use of \texttt{torchaudio} with the \texttt{sox\_io} backend is advised due to its superior speed and performance compared to alternatives like the \texttt{soundfile} backend.

\begin{table*}[!ht]
    \caption{Open-source multimodal music dataset.}
    \centering
    \begin{tabular}{c|ccc}
    \toprule
    \toprule
        Dataset & Modality & n files & Tasks\\
        \midrule
        \midrule
        LP-MusicCaps-MSD & audio URL, text& 520k audio, 1.5M text& music captioning.\\
        Song Describer Dataset (SDD) & audio, text & 706 audio, 1.1k  &music captioning, text-to-music, retrieval\\
        MusicQA& audio, text& 12,542 clips, 112,878 Q\&A & acoustic music instruction following\\
        MusicInstruct& audio URL, text& 5.5k clips, 60,493  Q\&A& acoustic music instruction following\\
        MusicBench& audio, text& 52,768 text-audio pairs& text to music\\
        MARD& audio URL, text& 65,566 albums, 263,525 reviews& album music description\\
        MUEdit &audio pairs, text& 10815 text, 60.22 hours  &music editing with text prompt\\
        \midrule
        WikiMusicText (WikiMT) & ABC, text& 1,010& text to music, music captioning\\
        \midrule
        IMAC& audio URL, image URL& 85k images, 3,812 songs &affective music-image correspondences\\
        URMP &MIDI, Audio, Video& 44 pieces&audiovisual symphony separation\\
        URSing&audio, video& 65 pieces, 4 hours &audiovisual singing voice separation\\
        RAVDESS&audio, video&7356 piece& speech \& songs in different emotion and intense.\\
        EmoMV& audio, video& 5986 pairs& affective music-Video correspondencs\\
        SymMV& MIDI, audio, video& 1140 pairs, 76.5 hours& video background music generation\\
        MUImage &audio, image&9966 text, 27.72 hours&image to music\\
        MUVideo &audio, video&13203 text, 36.72 hours&video to music\\
        AnyInstruct &text, audio, images &108k instruction-following entries & instruction following w/ interleaved format. \\
        V2M &audio, video &190k pairs, 6403 hours & video to music \\        
        MMtrail &text, audio, video &20m pairs, 27.1k hours & text to music, video to music \\                
        \bottomrule
        \bottomrule
        
    \end{tabular}
    \label{tab:multimodal_datasets_analysis}
\end{table*}

\subsubsection{Multimodal Datasets: Present \& Challenges}

\paragraph{Text-Audio}
LP-MusicCaps-MSD\cite{doh2023lp} originates from the ECALS subset\cite{doh2023toward} of the Million Song Dataset\cite{bertin2011million} and includes 520k 30-second music clips. It features a diverse vocabulary of 1,054 labels spanning various music-related categories, such as genre, instrument, vocal attributes, mood, theme, and culture. On average, each clip is tagged with 10.2 labels. The 3 captions are generated by GPT-3.5 for each audio clip, involving one caption, one summary, and one rephrased version.

The Song Describer Dataset (SDD) \cite{manco2023song} introduces a curated dataset of 1.1k natural language descriptions paired with 706 music recordings, all licensed under a Creative Commons license, designed to provide a powerful tool for music and language models. This dataset helps evaluate three key tasks: music captioning, text-to-music generation, and music language retrieval, highlighting the critical role of cross-dataset evaluation. Each description in SDD contains rich musical characteristics such as genre, mood, and orchestration to provide semantically and syntactically comprehensive captions. SDD is unique in that it provides longer audio track segments and multiple descriptions of each track, thereby enhancing the robustness of the assessment. This dataset, combined with metadata from MTG-Jamendo, not only enriches the field of music research but also sets new standards for realistic and sustainable dataset practices in the field of music language.

MusicQA \cite{liu2024music} is a dataset of music question-answer pairs, generated by MPT-7B model \cite{MosaicML2023Introducing}. The dataset is created by using MusicCaps and MagnaTagATune that contain music captions or tags related to descriptive and inferential questions about the music. Each audio has five open-ended question-answer pairs focusing on aspects like music emotion, tempo, and genre. It include 12,542 music clips in the training set, 76.15 hours in total, with 112,878 question-answer pairs.

MusicInstruct \cite{deng2023musilingo}, also derived from the MusicCaps dataset's music-caption pairs, contains Q\&A pairs generated with few-shot learning techniques by GPT-4. It includes 5,521 YouTube ID of audio clips, a short version with 27,540 Q\&A pairs focused on detailed aspects such as emotion, instruments, vocals, tempo, and genre, typically eliciting concise one or two-sentence responses, as well as a long Version that contains 32,953 Q\&A pairs with broader questions about musical pieces, often resulting in more elaborate, paraphrased responses. 

MusicBench \cite{melechovsky2023mustango} is a text-to-music dataset, also based on the MusicCaps dataset, that %
consists of 5,479 samples. MusicBench is structured into various training and testing sets: the initial sets (TrainA and TestA) are further enhanced by splicing four control sentences that describe music features in chord, key, beat and tempo information into the original prompts to create TrainB and TestB. Subsequently, TrainB is rephrased using ChatGPT to generate TrainC. Additionally, an audio augmentation strategy that adjusts pitch, speed, and volume is implemented to improve data diversity. A comprehensive training set after augmentation consists of 52,768 text-audio pairs.

The Multimodal Album Reviews Dataset (MARD) \cite{oramas2016exploring} includes metadata and customer reviewers' comments elements from Amazon, enriched by metadata from MusicBrainz and descriptors from AcousticBrainz \cite{porter2015acousticbrainz}. This dataset includes 65,566 albums, accompanied by 263,525 customer reviews, supporting multimodal frameworks for analysing human preferences and behaviours to music across different genres. 

The MUEdit dataset \cite{hussain2023m} is curated to facilitate the development and training of models capable of understanding and editing music based on specific prompts. The MUEdit dataset includes 10-second music pairs, each selected based on rhythmic characteristics such as tempo and pitch to ensure consistency and relevancy, totalling 55.69 hours. The creation process involves generating descriptions for these music pairs using the MU-LLaMA model, followed by the generation of editing instructions via the MPT-7B model. These instructions are designed to mimic a conversational interface where the model not only receives detailed descriptions but also generates editing commands based on these descriptions, simulating a realistic user-model interaction for music editing tasks.

WikiMusicText (WikiMT) \cite{wu2023clamp} contains 1010 lead sheets in ABC notation format from Wikifonia.org, each with metadata including title and artist extracted directly from the scores. Additionally, it derives genre and description from Internet information. The genre labels are assigned by analysing related Wikipedia content and categorising it into 8 GTZAN taxonomy classes: Jazz, Country, Folk, R\&B, Pop, Rock, Dance, and Latin. Descriptions are synthesised from Wikipedia using the BART-large model. To emphasise musical content, extraneous text within the ABC notation has been removed.

\paragraph{Visual-Audio}

Image-Music Affective Correspondence (IMAC) database \cite{verma2019learning} is developed to advance research in crossmodal emotion analysis. This extensive dataset comprises the URL of over 85,000 images and 3,812 songs, totalling approximately 270 hours of audio. Each item in the database is categorised under one of three emotional labels: positive, neutral, or negative, facilitating detailed studies of emotional responses across different media.

The University of Rochester Music Performance (URMP) \cite{li2018creating} collection presents a comprehensive set of multimodal data for 44 distinct musical compositions. This dataset encompasses a wide range of resources, including synchronised audio and visual elements along with detailed annotations. For each musical piece, the dataset provides a dedicated folder containing various components: MIDI scores, visually enhanced PDF sheet music and audio recordings. These audio files are available in both isolated instrument tracks and mixed versions, stored in high-fidelity WAV format with 48 kHz sampling rate and 24-bit depth. Moreover, the dataset includes video recordings of performances, captured in 1080P resolution at 29.97 frames per second, with careful alignment to the musical score. To facilitate in-depth analysis, the URMP dataset also incorporates frame-by-frame pitch information and note-level transcriptions, both presented in ASCII text format.

Creating the University of Rochester Music Performance (URMP) dataset posed significant synchronisation challenges due to the need for both high-quality recordings and efficient dataset creation. To tackle this, various methods were tested to balance expressiveness and synchronisation without extensive joint rehearsals. Initial attempts included using a metronomic beat for synchronisation, which proved too rigid, and a pre-recorded piano guide, which failed to provide adequate synchronisation cues. Further approaches involved using rehearsal recordings with a conductor to enhance expressiveness and synchronisation, leading to better results but at the expense of scalability.
Subsequent strategies explored visual cues by having players follow video recordings of the first performer, which improved timing after long rests but still fell short of desired synchronisation levels. The most effective method involved a conductor directing the performance along with a pianist, recorded on video, which players followed during their individual recordings. This approach minimised the need for joint rehearsals while maintaining high synchronisation and expressiveness, proving to be efficient and scalable for creating the dataset. Each method's efficacy was carefully evaluated, considering both quantitative synchronisation measures and players' subjective feedback, ensuring a balanced and practical approach to dataset creation.

The University of Rochester MultiModal Singing Performance Dataset (URSing) \cite{li2021audiovisual} is designed to evaluate realistic singing performance models. It comprises 65 songs totalling four hours of audiovisual recordings from the University of Rochester students. The dataset creation involved singer recruitment with auditions for tuning accuracy, recordings using an AT2020 microphone and iPhone 11 within a semi-anechoic booth, and professional post-processing for audio quality. Annotations of mouth regions were added using the Dlib library, and the dataset includes solo and accompaniment audio tracks in WAV format, MP4 video recordings, and a benchmark set of 54 instrumental and 26 vocal-backed excerpts for detailed evaluation. URSing provides a rich resource for exploring audiovisual music information retrieval techniques. 

The Ryerson Audio-Visual Database of Emotional Speech and Song (RAVDESS) dataset \cite{livingstone2018ryerson} provides a validated collection of 7356 multimodal emotional voice and song recordings performed by 24 professional actors. 
This gender-balanced dataset includes multiple emotional expressions of varying intensities, provided in a combination of facial video and sound or in separate formats, focusing on multimodal sentiment recognition and analysis in speech and music.

The Lyra dataset \cite{papaioannou2022dataset} is a collection of traditional and folk music in Greek, containing 1,570 YouTube timestamp links of music videos and approximately 80 hours in total. 
Derived from a Greek documentary film series, the dataset provides a wealth of metadata on musical genres, instruments, and geographic origins. The dataset supports a wide range of computational musicological tasks, making it an invaluable resource for ethnomusicological research.

The EmoMV \cite{thao2023emomv} project is focused on assembling a trio of datasets aimed at facilitating the study of emotional correspondence between music and video modalities. The datasets, named EmoMV-A, EmoMV-B, and EmoMV-C, incorporate music video segments with EmoMV-A and EmoMV-B deriving content from existing datasets, while EmoMV-C features self-collected music videos from YouTube. Each pair within these datasets is annotated as either emotionally matched or mismatched.

Human experts conduct the annotation process for EmoMV-A, while a pre-trained deep neural network generates annotations for EmoMV-B and EmoMV-C. The reliability of these emotional labels is evaluated through a user study that examines their accuracy. In conjunction with the dataset development, researchers have created a benchmark deep learning model. This model is designed for binary classification of affective correspondence between music and video and has been optimised for affective music-video retrieval applications. This advancement provides a valuable resource for investigating how music and visual elements interact to express emotional content.

SymMV dataset \cite{zhuo2023video} is a larger video-music dataset that pairs music videos (MV) with piano MIDI and audio. It includes 1140 pop piano music pieces, along with their corresponding official MV, totalling 76.5 hours. Each MV pair is annotated for chord progression, tonality, and rhythm, enhancing its utility for research in MV emotional correspondence. 
The dataset functions as a valuable resource for developing algorithms that generate background music for videos. It enables the exploration of inherent relationships between visual and musical elements, leveraging the emotional and stylistic correspondence between video content and its soundtrack. The creation of this dataset involved a meticulous process: sourcing professional-grade piano renditions from select YouTube channels, aligning these with the original music videos, and utilising sophisticated transcription algorithms to convert the audio into MIDI format. To maintain high standards, professional musicians conducted a thorough review of the compiled dataset.

MUImage and MUVideo\cite{hussain2023m} are text/image/video-to-music datasets that facilitate the development of models for generating customised music to complement accompanying visual content. These datasets were constructed by carefully generating detailed descriptions of images, videos and music. The MU-LLaMA model is used to add captions to music files, while the BLIP and VideoMAE models provide captions for images and videos, respectively. These descriptions are then used as input to the MPT-7B model, which produces music-generated cues. This structured approach supports the creation of integrated arbitrary music datasets

The AnyInstruct\cite{zhan2024anygpt} dataset comprises 108k multimodal instruction-following entries, uniquely integrating text, speech, images, and music. Developed using GPT-4 for generating textual dialogues, and enhanced with images from DALL-E 3, music from MusicGen, and voices synthesised via the Azure Text-to-Speech API, this dataset features diverse modalities in an interleaved format. 
The dataset is organised into two parts, containing comprehensive multimodal dialogues along with approximately 205k images, 503k voice recordings, and 113k music tracks. 

The V2M~\cite{tian2024vidmuse} dataset comprises video and acoustic music pairs on a large scale. It includes three subsets: V2M-190K, which consists of approximately 190K video-music pairs totalling around 6,403 hours; V2M-20K, containing about 20K pairs with approximately 597 hours of content; and V2M-bench, which comprises 300 pairs totalling 9 hours in duration.

MMTrail~\cite{chi2024mmtrail} is a comprehensive multimodality dataset featuring over 20 million trailer clips accompanied by visual captions, as well as 2 million high-quality clips annotated with multimodal music captions.

\subsubsection{Future Work}
Advances in music modelling require more expansive and diverse datasets. Current resources, in spite of their extensive size, predominantly feature Western popular music, limiting model versatility. To develop robust models that grasp and generate music authentically across various cultures, it's crucial to integrate a broad spectrum of global musical styles and genres into these datasets.

The current Multimodal Music Instruct dataset needs substantial enhancement to support foundation models effectively. It should expand not only in volume but also in the complexity of its content, particularly in areas of sequential MIR downstream tasks like beat recognition and chord progression. Furthermore, introducing a chain-of-thought component is essential for enabling models to perform complex reasoning and explain the musical theories underpinning their responses, thus fostering deeper analytical capabilities in music understanding.

Additionally, datasets that simulate real interactions between users and music generation systems are necessary. These should include multi-round conversations with interleaved text-music data, including user queries about music recommendations or variations and the corresponding model responses. Such datasets would improve model responsiveness and adapt algorithms based on realistic user engagement, enhancing the overall user experience in music applications.

In addition to the data collection mentioned above, pre-training data cleaning strategies, mixing strategies of pre-training data from different sources, and curriculum learning with a specific priority of different types of datasets are still open issues to be explored.

Last but not least, there is a need to find reasonable ratios of balancing downstream tasks and pre-training data to develop effective tuning strategies for injecting new knowledge or expertise while avoiding catastrophic forgetting, e.g., allowing the GPT to learn more about music, adapting the model to new music genres, or make it more suitable to user preferences.

\subsection{Evaluation} 
Music understanding and generation are of increasing interest to both academic and industry researchers and have the potential to broadly impact the economics and culture of music. However, evaluation remains an open and difficult research problem, facing challenges like inconsistent downstream evaluation conditions, data leakage, model bias, etc. In the following subsection, we will review commonly used evaluation protocols and metrics for both music understanding and music generation.

\subsubsection{Evaluation for Music Understanding} 

\newcommand{\llark}{DBLP:journals/corr/abs-2310-07160}
\newcommand{\mugen}{DBLP:journals/corr/abs-2311-11255}
\newcommand{\mullama}{DBLP:journals/corr/abs-2308-11276}
\newcommand{\musilingo}{deng2023musilingo}

\begin{table*}[!htp]\centering
\caption{Evaluation of music understanding foundation models. * indicates human evaluation.}\label{tab:music_understanding_eval}
\scriptsize
\begin{tabular}{c|*{9}{c}|*{3}{c}}
\toprule
\toprule
\multirow{2}{*}{\textbf{Model}} & \multicolumn{9}{c|}{\textbf{Traditional MIR Tasks}} & \multicolumn{3}{c}{\textbf{Multimodal Tasks}} \\
& Beat & Chord & Pitch & Key & Tempo & Instrument & Genre & Emotion & Tags & Retrieval & Captioning & MQA 
\\\midrule \midrule
M2-Duo \cite{niizumiMaskedModelingDuo2024a} & &  &  \checkmark   &     &     &  \checkmark    &  \checkmark  &    &  &   &  & \\ 
Music2Vec \cite{li2022map} &  & &  &   \checkmark  &     &    &  \checkmark  & \checkmark   & \checkmark &  &   & \\ 
MusicHuBERT \cite{ma2023effectiveness} & & &  \checkmark   &   \checkmark  &    &   \checkmark   &  \checkmark  &   \checkmark  & \checkmark &   &  &  \\ 
MusicFM \cite{MusicFM} & \checkmark & \checkmark &     &  \checkmark   &     &      &    &    & \checkmark &   &  &  \\ 
MERT \cite{li2023mert} & & &  \checkmark   &   \checkmark  &   \checkmark  &   \checkmark   &  \checkmark  &   \checkmark  & \checkmark &   &  &  \\ 
COLA \cite{saeed2021contrastive} &  & &   &     &     &   \checkmark   &    &     &  &   & &  \\
CLAR \cite{al2021clar} &  & &  \checkmark &     &     &   \checkmark   &    &     &  &   &  & \\
MULE \cite{mccallum2022supervised} &  & & \checkmark  &   \checkmark  &     &   \checkmark   & \checkmark   &   \checkmark  & \checkmark &   & &  \\
JukeMIR \cite{castellon2021codified} &  & &   &  \checkmark   &     &      &  \checkmark  & \checkmark    & \checkmark &   & & \\
MuLaP \\ 
\midrule
MusCALL \cite{manco2022contrastive} &  & &   &     &     &      &   \checkmark &     & \checkmark & \checkmark &   &  \\
MuLan \cite{mulan} &  & &   &     &     &      &    &     & \checkmark & \checkmark &   &  \\
CLAMP \cite{wu2023clamp}  &  & &   &     &     &      &  \checkmark  &   \checkmark  & & \checkmark &   &  \\
\midrule
M\({}^{\mbox{2}}\)UGen \cite{\mugen} &  & &   &     &     &      &    &     & &  &   & \checkmark  \\ 
SALMONN \cite{tang2024salmonn} &   & &   &     &     &      &    &     & & &  \checkmark &   \\ 
MuLLaMa \cite{\mullama} &  & &  &     &     &     &      &    &   &  &  \checkmark & \checkmark  \\ 
MusiLingo \cite{\musilingo} &  & &  &     &     &     &      &    &   &  &  \checkmark & \checkmark  \\
Pengi \cite{deshmukh2023pengi} & & & \checkmark  &     &    &   \checkmark  & \checkmark    &    &  &   &  &   \\
LLark \cite{\llark} & & &     &   \checkmark  &   \checkmark  &   \checkmark   &  \checkmark  &     & & &  \checkmark* & \checkmark* \\ 
\bottomrule
\bottomrule
\end{tabular}
\end{table*}

\paragraph{Overview}
In this section, we elucidate the predominant evaluation protocols followed in assessing foundation models for music understanding. As seen in Section \ref{sec:music_understanding}, these types of models can either serve as common backbones for solving traditional MIR tasks or can engage multiple data modalities, most often visual or textual modalities, to perform a variety of novel tasks that require multimodal understanding. Evaluation, therefore, varies depending on model design, tasks supported, and input-output modalities. Here, we distinguish between the evaluation of two main types of tasks: traditional MIR tasks and multimodal tasks, providing an overview in Table \ref{tab:music_understanding_eval}.

\paragraph{Traditional MIR tasks}
We start by considering traditional \textbf{MIR tasks}. In this scenario, the goal is to measure music understanding by determining a model's ability to correctly recognise specific musical properties at different levels of granularity, from frame to track level, by adopting the canonical MIR evaluation approach for the corresponding task. For models operating in the audio domain, tasks typically included are a subset of key and mode detection \cite{\llark, li2023mert}, tempo detection \cite{\llark, li2023mert}, genre classification \cite{\llark, li2023mert, deshmukh2023pengi}, instrument recognition \cite{\llark, li2023mert, deshmukh2023pengi}, music tagging \cite{li2023mert}, emotion recognition \cite{li2023mert}, pitch detection \cite{li2023mert, deshmukh2023pengi}, beat tracking \cite{li2023mert}, and, in the less common performance domain, vocal technique detection \cite{li2023mert}, singer identification \cite{li2023mert}, and performance assessment \cite{Zhang2024FromJudges, Zhang2021LearnAssessment}, among others. MIR-based evaluation can be executed in two distinct ways: via probing \cite{li2023mert, deshmukh2023pengi, castellon2021codified} or via text-based instruction \cite{\llark}. Probing is used on models that produce audio representations: the audio backbone, which is either kept frozen or fine-tuned, is used as a feature extractor, and a lightweight probing network, typically a shallow MLP, is trained on top to solve the downstream task of interest.
Instead, in models equipped with a text interface, MIR-based evaluation can be realised by prompting the model with an adequate text input (e.g. \textit{``What is the key of this song?''}), and casting the output text (e.g \textit{``This song is in the key of F minor.''}) to the corresponding label in the dataset. 

Recently, there have been several efforts to unify this kind of evaluation methods across the full spectrum of music understanding tasks, by creating \textbf{benchmarks} that either include the music domain as a subset of interest \cite{turian2022hear} or fully specialise in it \cite{yuan2024marble, mir_ref}. The most extensive of these is MARBLE \cite{yuan2024marble}, which proposes a unified protocol for evaluating music audio representations across tasks at different levels of hierarchy, including high-level description (key detection, music tagging, genre classification, and emotion recognition), score-level (pitch tracking, beat tracking, melody extraction, chord estimation, lyrics transcription), performance-level (ornament and technique detection), and acoustic-level tasks (singer identification, instrument classification, source separation).

\paragraph{Multimodal tasks}
In the context of \textbf{multimodal} foundation models and beyond traditional MIR tasks, there is a similarly wide variety of tasks that are useful in evaluating music understanding. Among these, one that stands out as a common approach to evaluate fundamental capabilities in this domain is \textbf{cross-modal retrieval}, defined as the task of retrieving data samples of one modality (e.g. a music clip), based on a query sample of another modality (e.g. a textual description or a video). Models that are amenable to this type evaluation are most typically those trained to learn joint embedding spaces between two or more music-related data modalities, such as music-text contrastive models \cite{mulan, manco2022contrastive, wu2023clamp}, or those trained to learn correspondences between music and images or videos \cite{DBLP:conf/cvpr/McKeeSSR23, thao2023emomv, 10447276, DBLP:conf/cvpr/SurisVRS22, verma2019learning, 10.1145/3652583.3658045}. In the case of music-text retrieval, this evaluation approach is not only useful for directly measuring performance on the task itself, but it also offers a framework for evaluating models on music classification tasks by casting them as text-based retrieval. In this case, cross-modal retrieval results in a further variation of the MIR-based evaluation approach presented earlier \cite{doh2023toward}.

A further type of multimodal evaluation comprises tasks that are centred around language generation. As such, this is only suitable to music understanding models that take audio or (audio, text) pairs as inputs and generate text outputs. Under this setting, music understanding is measured by assessing language outputs that typically encompass several musical concepts.
In practice, this is articulated in two subtasks: \textbf{music captioning} \cite{tang2024salmonn, \mullama, \musilingo, \llark} and \textbf{music question answering} (MQA), which can be either open-ended \cite{tang2024salmonn, \mugen, \mullama, \musilingo} (sometimes also referred to as \textit{music reasoning} \cite{\llark}), or multiple-choice \cite{weck2024muchomusic}. In music captioning and open-ended MQA, the model is instructed to produce a language output that either describes the audio input (e.g. \textit{``Describe the contents of the provided audio in detail.''}) or answers a question about it (e.g. \textit{``What are some possible uses for this music in a film or TV show?''}). Performance is then measured in one of three ways: via automatic metrics, via human evaluation or via another LLM, using a strategy called LLM-as-a-judge \cite{zheng2023judging}. In the first case, automatic evaluation typically consists of computing text generation metrics such as BLEU \cite{papineni2002bleu}, METEOR \cite{banerjee2005meteor}, ROUGE \cite{lin-2004-rouge}, and BERT-score \cite{Zhang2020BERTScore}, which measure the linguistic or semantic similarity between the generated and reference text. We note that MQA evaluation bears similarities with the flavour of MIR-based evaluation that makes use of (question, answer) pairs. A key difference, however, is that the former is characterised by open-ended questions, while the latter focuses on a single musical aspect and assumes a one-to-one mapping between model output and ground truth.

While evaluation on traditional MIR tasks leverages standard metrics and benchmarks from the MIR literature and beyond, language-based evaluation follows less established protocols and is supported by only a handful of datasets. For the music captioning task, most works \cite{tang2024salmonn, \llark, \musilingo} make use of the MusicCaps dataset \cite{agostinelli2023musiclm}, while others \cite{\mullama} employ ad-hoc evaluation datasets created with the help of LLMs. For the MQA task, no standard dataset exists for the open-ended variant, and all evaluations are carried out on ad-hoc datasets \cite{\mullama, \mugen, \musilingo, \llark}. The MuChoMusic benchmark \cite{weck2024muchomusic} supports instead music understanding evaluation via multiple-choice MQA. In this setup, models are provided with an audio clip accompanied by a question and a set of answer options (A, B, C, or D) to choose from. The closed format of this task allows for more standardised evaluation, overcoming some of the limitations of match-based metrics.

\paragraph{Measuring music-domain knowledge in LLMs and Visual-Language Models}
Finally, we look at assessing music understanding abilities in non-music foundation models, through the lens of recent benchmarks that allow to evaluate music-domain knowledge encoded in LLMs \cite{yuan2024chatmusician, li2024music} and visual-language models \cite{yue2023mmmu}.

For LLM evaluation, we identify two key benchmarks, MusicTheoryBench \cite{yuan2024chatmusician} and ZIQI-Eval \cite{li2024music}, both employing ABC notation (see Section \ref{subsec:symbolic-acoustic}).
MusicTheoryBench \cite{yuan2024chatmusician} focuses on evaluating music knowledge and reasoning via a set of 372 multiple-choice questions designed to align with college-level music education standards. The questions span a variety of topics including notes, rhythm, chords, orchestration, and music-related cultural and historical knowledge and are aimed at assessing both the analytical and inferential skills of LLMs.
ZIQI-Eval \cite{li2024music} is instead designed to test LLMs' musical abilities in both comprehension (focusing on harmony, melody, and rhythm) and music generation or continuation. Extensive testing of 17 LLMs through this benchmark reveals that current models generally possess insufficient knowledge in the music domain and exhibit poor compositional understanding and generation abilities, underlining the need for targeted improvements in musical training for LLMs.

Considering instead visual-language models, the Massive Multi-discipline Multimodal
Understanding and Reasoning Benchmark (MMMU) \cite{yue2023mmmu}, among its broad evaluation of multimodal models across various disciplines, includes a dedicated subset focused on music sheet understanding. This features 369 music sheet images accompanied by targeted questions which probe knowledge of musical theory, including harmonic intervals, instrument tuning, and chords within specific bars.

\paragraph{Challenges and open questions}
Despite the flourishing of research in this field, evaluating music understanding models in a consistent and reliable fashion remains a challenge due to limited benchmarks dedicated to the music domain. As a consequence, existing works follow inconsistent evaluation practices which make use of different datasets, metrics, experimental settings, and even task formulations. This becomes particularly prominent in language-based evaluation, where, as highlighted earlier in this section, there is little in the way of standardisation, and automatic metrics become unsuitable due to the open-ended nature of the tasks \cite{\llark}. Text generation metrics are also known to deviate from human judgement when there isn't a sufficient number of reference sentences or, similarly, when there are several potentially correct outputs, as is often the case in music understanding. This has prompted some to conduct human studies to obtain more reliable evaluations on open-ended tasks \cite{\llark}.

Finally, comprehensive evaluation should go beyond assessing task-specific performance and include other key factors like robustness, safety, efficiency, and alignment with human preferences. While this is becoming commonplace in other domains, it remains unexplored in the context of music understanding.

\subsubsection{Objective Evaluation for Music Generation}\label{subsubsec:obj-eval}

\paragraph{Overview}
Similar to the field of music generation itself, research on objective evaluation for music generation models is in a nascent stage--The field has yet to agree on a set of standard metrics and/or benchmarks, and new metrics have been constantly proposed in an ad-hoc manner with new methods.
Objective evaluation of machine-generated music is challenging for several main reasons:
(i)~The aesthetics of music is multifaceted, e.g., harmony, melody, instrumentation, and how they are assembled together, which are both hard to quantify individually and highly interrelated with one another.
(ii)~Music generation is, in fact, a family of tasks that can take a variety of user inputs, ranging from a short text description~\cite{agostinelli2023musiclm, copet2024simple} to fine-grained musical concepts like melody~\cite{copet2024simple}, chords~\cite{von2022figaro}, beats~\cite{musiccontrolnet}, and even music composition with rendering (performance) objective \cite{Peter2023SoundingPerformance, app14156543}, hence necessitating task-specific metrics.

We provide an overview of the landscape of objective evaluation metrics for music generation
according to three axes:
(i)~\textbf{functional type},
(ii)~\textbf{purpose},
and (iii)~\textbf{input domain}.

A \textbf{functional type} (or \textbf{type} in short) defines the high-level workflow to map a set of generated (and/or reference) samples to a metric value.
Let one sample of \textit{reference} (i.e., human-composed) music be denoted by $\bm{x}$, and one sample of \textit{generated} music by $\bm{\Tilde{x}}$.
Also, let $X = \{\bm{x}^{(i)}\}_{i=1}^N$ denote the set of $N$ reference samples, and $\Tilde{X} = \{\bm{\Tilde{x}}^{(i)}\}_{i=1}^{\Tilde{N}}$ the set of $\Tilde{N}$ generated samples.
We further define $f(\bm{x})$ or $f(\bm{\Tilde{x}})$ to be any function that extracts an \textit{attribute} or some \textit{features} from music, either global ones (e.g., a text description for the music) or local ones (e.g., a chord sequence), which may sometimes also be independently given by users as conditions, and $\mathrm{score}(\cdot, \cdot)$, or $\mathrm{score}(\cdot)$ to be a deterministic function that computes the reported metric value.
Commonly adopted metrics may be categorised as follows:
\begin{itemize}
    \item \textbf{Pair-wise aggregate} (\textit{P-A}) -- $\mathrm{score}(\{f(\bm{x}^{(i)})\}_{i \in [N]} \, , \, \{f(\bm{\Tilde{x}}^{(i)})\}_{i \in [\Tilde{N}]})$.
    \textit{P-A} metrics seek to compare the distribution of outputs induced by music generation models to a sample of the distribution of human-created music.
    To compute \textit{P-A} metrics, two sets of samples (generated and reference) with or without direct correspondence are needed.
    Attributes/features are extracted from each sample, and the $\mathrm{score}(\cdot, \cdot)$ function would then calculate some aggregate statistics of the attributes/features in each set, between which some notion of similarity or distance is computed.
    Frech{\'e}t audio distance (FAD) \cite{kilgour2019frechet} is a representative example of a \textit{P-A} metric.
    \item \textbf{Pair-wise individual} (\textit{P-I}) -- 
    $\mathrm{mean}_{i \in [\Tilde{N}]}(\mathrm{score}(f(\bm{x}^{(i)}), f(\bm{\Tilde{x}}^{(i)}))$, or $\mathrm{mean}_{i \in [\Tilde{N}]}(\mathrm{score}(f_1(\bm{x}^{(i)}), f_2(\bm{\Tilde{x}}^{(i)}))$, where $f_1$, $f_2$ are different attribute, or feature, extraction functions.
    \textit{P-I} metrics are used when there is a clear \textit{target} (e.g., a piece of human-composed music, some musical features, or semantic attributes) that each \textit{individual} machine-generated piece is expected to resemble or reflect.
    \textit{P-I} metrics require sample-wise direct correspondence between reference and generation, and hence assert $N = \Tilde{N}$.
    CLAP score \cite{wu2023large} is a widely adopted \textit{P-I} metric.
    \item \textbf{Single individual} (\textit{S-I}) -- $\mathrm{mean}_{i \in [\Tilde{N}]}(f(\bm{\Tilde{x}}^{(i)}))$.
    \textit{S-I} metrics typically serve as proxies for musical aspects that are hard to quantify, e.g., how ``good'' the harmony, rhythm, or repetitive structure is.
    \textit{S-I} metrics are computed without a reference.
    Therefore, unlike \textit{P-A} or \textit{{P-I}} metrics, where we may specify higher/lower ($\uparrow$/$\downarrow$) is better, \textit{S-I} metrics are often specified as the closer to reference samples, i.e., $\mathrm{mean}_{i \in [N]}(f(\bm{x}^{(i)}))$, the better.
    For example, pitch-class histogram entropy \cite{wu2020jazz} is an \textit{S-I} metric.
    We note that, while it is more desirable to compare the distribution of $f(\bm{\Tilde{x}}^{(i)})$ induced by machine-generated music to that of human-composed music, i.e., $f(\bm{x}^{(i)})$, effectively turning \textit{S-I} metrics into \textit{Single aggregate (S-A)} metrics following our taxonomy, researchers have gravitated to comparing simple $\mathrm{mean}$s of features $f(\bm{\Tilde{x}}^{(i)})$ and $f(\bm{x}^{(i)})$.
\end{itemize}

The \textbf{purpose} of a metric is associated with either of the two following major focuses:
\begin{itemize}
    \item \textbf{Quality}-focused: Such metrics try to characterise how close generated samples are to human-composed music, and can focus on a variety of aspects, like audio quality, musical diversity, rhythm, melody, harmony, repetitive structures, and so on.
    \item \textbf{Control}-focused: Such metrics examine how well a generative model adheres to the conditions given by users.
    Frequently used conditions are text descriptions, musical genres, emotions, instrumentation, tempos, chord sequences, melodies, etc.
\end{itemize}

The \textbf{input domain} of an evaluation metric largely depends on the domain the evaluated model/system targets(cf.~\ref{subsec:symbolic-acoustic}).
Audio-domain metrics accept audible waveforms as inputs, while symbolic-domain metrics apply to note/event sequences.
Audio-domain metrics may be used on symbolic-domain samples with the help of audio synthesisers, but not vice-versa in general as music transcription \cite{benetos2019transcription} remains a challenging task.
We note that, compared to audio-domain models, evaluation of symbolic-domain models is relatively less standardised -- While some works follow the mechanisms proposed in evaluation-focused papers \cite{wu2020jazz, yang2020evaluation}, the majority define their own metrics for evaluation.

\paragraph{Audio-domain, quality-focused metrics}
\textbf{Frech{\'e}t audio distance} (\textbf{FAD}; type: \textit{P-A}, $\downarrow$) \cite{kilgour2019frechet} is the most widely adopted metric to assess the quality of generated audios, which has been shown to correlate well with human perception.
FAD employs an audio encoder pre-trained on, e.g., multilabel audio classification \cite{gemmeke2017audio}, as $f(\bm{x})$ to map an input audio to a feature vector.
Then, the $\mathrm{score}(\cdot,\cdot)$ function first estimates a Gaussian covariance matrix from all feature vectors obtained from generated audios, and likewise for reference audios.
Finally, the Frech{\'e}t distance between the two estimated Gaussian distributions \cite{dowson1982frechet} is computed.
As the procedure suggests, FAD conflates audio quality, musical quality, cross-sample diversity, and potentially many more aspects together, and measures a notion of \textit{overall realisticness} of audios.

As for metrics that focus on one specific aspect, the \textbf{Structureness indicator} (type: \textit{S-I}) \cite{wu2020jazz} is a representative example, which examines the presence of \textit{repetitive structures} across all time granularities.
The Structureness indicator leverages Fitness Scape-plots \cite{muller2012scape} as $f(\bm{x})$, and then applies a $\max$ operation on the scape-plot as the $\mathrm{score(\cdot)}$ function, which is meant to describe how much the most-repeated excerpt within a specified time granularity (e.g., 10$\sim$20 seconds) is repeated throughout the entire audio.

\paragraph{Audio-domain, control-focused metrics}
With the recent rise of text-to-audio music generation models \cite{liu2023audioldm, agostinelli2023musiclm, copet2024simple}, the most popular control-focused metrics currently are concerned with \textit{adherence to text} (or genre/emotion/instrument tag-based) inputs:
\begin{itemize}
    \item \textbf{CLAP score} (\textbf{CLAP}, type: \textit{P-I}, $\uparrow$) \cite{wu2023large} depends on a text encoder and an audio encoder as feature extractors $f_1, f_2$.
    The two encoders should have been jointly pre-trained via contrastive learning \cite{oord2018representation} to form an aligned latent feature space.
    Then, CLAP score is simply the cosine similarity between the encoded features of the input text and those of the generated audio.
    MuLan score \cite{mulan} is an almost identical metric, only with a different bi-encoder backbone.
    \item \textbf{KL divergence} (\textbf{KL}, type: \textit{P-I}, $\downarrow$), as opposed to CLAP score, measures text input adherence indirectly.
    It requires a pre-trained audio classifier \cite{zeghidour2021leaf, koutini2021efficient} as $f(\bm{x})$.
    The KL divergence is computed between the output class distribution of the reference audio (for the text input) and that of the generated audio.
\end{itemize}

To evaluate adherence to more fine-grained and musically-centred types of controls, the metrics are often proposed in individual works and hence not yet standardised.
Commonly considered musical controls and their corresponding metrics are as follows:
\begin{itemize}
    \item \textbf{Melody}: \textbf{Chroma cosine similarity} (type: \textit{P-I}, $\uparrow$) \cite{copet2024simple} leverages the chromagram (or pitch-class profile) \cite{takuya1999realtime} as $f(\bm{x})$ to extract the relative strength of 12 semitones from reference and generated audios, from which the frame-wise (or chunk-wise) cosine similarity is computed.
    If the chromagram is frame-wise binarised to be one-hot, the metric is equivalent to frame-wise \textbf{chroma accuracy} \cite{musiccontrolnet}.
    \item \textbf{Chords}: \textbf{Chord recognition accuracy} (type: \textit{P-I}, $\uparrow$) \cite{han2023instructme, melechovsky2023mustango} use a trained chord recognition model (e.g., \cite{cheuk2022jointist}) to predict the chord sequence from reference and generated audios.
    Then, an alignment (with various levels of tolerance \cite{melechovsky2023mustango}) between the two sequences is performed to compute the score.
    \item \textbf{Tempo}: \textbf{Tempo bin accuracy} (type: \textit{P-I}, $\uparrow$) \cite{melechovsky2023mustango} relies on a beat detection model \cite{bock2016madmom, heydari2021beatnet}, whose predictions can be converted to beats per minute (BPM).
    The metric then checks whether the BPMs of the reference and generated audios fall in the same bin, e.g., Adagio (60$\sim$70 BPM).
    \item \textbf{Rhythm}: \textbf{Beat F1 score} (type: \textit{P-I}, $\uparrow$) \cite{musiccontrolnet} also employs a beat detector and follows the evaluation for beat detection models \cite{davies2009evaluation}.
    It aligns the timestamps of beats predicted from reference and generated audios with a 70ms tolerance.
    \textbf{Beat count} (type: \textit{S-I}) \cite{melechovsky2023mustango} is a simpler version that only counts the number of detected beats.
    \item \textbf{Intensity}: \textbf{Dynamics correlation} (type: \textit{P-I} or \textit{P-A}, $\uparrow$) \cite{musiccontrolnet} computes the smoothed frame-wise loudness in decibels, and then calculates the Pearson's correlation between the frame-wise values from reference and generation.
\end{itemize}

\paragraph{Symbolic-domain, quality-focused metrics}
This family of metrics often fall under the \textit{S-I} type, and use the corresponding $f(\bm{x})$'s to aggregate/summarise some feature (concerning aspects like pitch, rhythm, and tonality) from the token sequence representing $\bm{x}$, and then compute the score on the feature.
\begin{itemize}
    \item \textbf{Pitch-class histogram entropy} (type: \textit{S-I}) \cite{wu2020jazz} collect the counts of 12 pitch classes (i.e., C, C\#, \dots, B) over a certain period (e.g., 4 bars), and computes the entropy on the resulting histogram as a measure of the diversity of pitch classes used.
    \item \textbf{Groove consistency} (type: \textit{S-I}) \cite{wu2020jazz} divides each bar into segments (e.g., in 32th note increments), and construct a binary \textit{groove vector} for each bar by checking whether each segment has note onsets.
    Then, some similarity (e.g., Hamming similarity) between groove vectors of neighbouring (or all) bar pairs is computed.
    \item \textbf{Scale consistency} (type: \textit{S-I}) \cite{mogren2016c} computes the fraction of pitch classes that belong to the 7 pitch classes of each of the 24 major/minor scales, and takes the maximum over the 24 scales as the scale consistency.
    \item \textbf{Similarity error} (type: \textit{P-A}, $\downarrow$): was proposed by \cite{yu2022museformer} and encompasses both pitch and rhythm to examine whether generations are structurally similar to human-composed music.
    It first constructs a \textit{note set} for each bar, whose elements describe the (pitch, duration, onset time relative to downbeat) of each note, and then computes the mean intersection-over-union (IoU) similarity of all bar pairs that are $t$ bars apart (where $t$ is preset).
    Finally, the difference between the mean IoUs between human-composed and generated pieces is taken.
\end{itemize}
\paragraph{Symbolic-domain, control-focused metrics}
Unlike audio-domain models, where text descriptions are the prevailing control form, control inputs for symbolic models are much less standardised, and hence so are the evaluation metrics.
Therefore, we group the metrics roughly according to controls that targeted aspect, and refer readers to relevant papers for further details.
\begin{itemize}
    \item \textbf{Pitch/melody:} melody matchness (\textit{P-I}, $\uparrow$)~\cite{hsiao2021compound}, pitch distribution L2 distance (\textit{P-I}, $\downarrow$)~\cite{zhang2023sdmuse}.
    \item \textbf{Chords:} chord matchness~\cite{hsiao2021compound}, chord F1 (between controls and generated music; \textit{P-I}, $\uparrow$)~\cite{von2022figaro}, and chord progression L2 distance (\textit{P-I}, $\downarrow$)~\cite{zhang2023sdmuse}.
    \item \textbf{Time signature:} time-signature accuracy (\textit{P-I}, $\uparrow$)~\cite{von2022figaro}.
    \item \textbf{Instrumentation:} instrument F1 (\textit{P-I}, $\uparrow$)~\cite{von2022figaro}.
    \item \textbf{Intensity:} Pearson's correlation of rhythmic intensity or polyphony (\textit{P-A}, $\uparrow$)~\cite{wu2023musemorphose}, and note density L2 distance (\textit{P-I}, $\downarrow$)~\cite{zhang2023sdmuse}.
\end{itemize}

\subsubsection{Subjective Evaluation for Music Generation}
Due to the highly artistic and subjective nature of music, subjective studies, i.e., listening tests with human participants, still remain the gold standard of evaluation music generation models \cite{yang2020evaluation, ji2023survey}.
Regardless of whether the model operates in the audio or symbolic domain, the samples presented to participants are typically (synthesised) audio waveforms.
Below we explain two common protocols for subjective studies.

\paragraph{Pair-wise Comparisons}
In each set of the study, a participant is provided with two samples $\bm{\Tilde{x}}_1, \bm{\Tilde{x}}_2$ (without being able to infer the underlying generative models), and asked to choose from the three options: (i)~$\bm{\Tilde{x}}_1$ is preferred over $\bm{\Tilde{x}}_2$, (ii)~$\bm{\Tilde{x}}_2$ is preferred over $\bm{\Tilde{x}}_1$, or (iii)~no preference among $\bm{\Tilde{x}}_1, \bm{\Tilde{x}}_2$.
Associating the responses with the underlying models known only to the study setters, the \textit{win-loss-tie} counts for every pair of the involved models can be obtained, on which the Wilcoxon signed-rank test \cite{wilcoxon1992individual} is performed to check whether one model is superior to another with statistical significance.
Works that adopt this protocol include \cite{huang2018music, von2022figaro, 
agostinelli2023musiclm,
thickstun2023anticipatory}.

\paragraph{Absolute-scale Ratings}
Under this protocol,
participants are asked to independently rate each sample $\bm{\Tilde{x}}$, possibly with an extra given condition $\bm{c}$, e.g., a melody, or a short prompt.
The ratings are typically in 5-, 10-, or 100-point scales.
The most frequently evaluated aspects are:
\begin{itemize}
\item \textbf{Overall quality (OVL)}
\item \textbf{Relevance to input text (REL)},
\end{itemize}
which are seen in \cite{liu2023audioldm, ghosal2023text, copet2024simple, liu2023audioldm2, han2023instructme, melechovsky2023mustango} and focus on quality and control respectively.

On the other hand, compared to pair-wise comparisons, more fine-grained aspects are often considered.
Although various terms and questions have been used, the fine-grained aspects may roughly be grouped into several categories:
\begin{itemize}
    \item \textbf{Audio quality}: Whether the audio is clear, without noises \cite{schneider2023mo, melechovsky2023mustango}.
    \item \textbf{Melody quality}:
    Whether the melody is prominent, pleasant-sounding, and leaves a strong impression \cite{ zhang2022structure, schneider2023mo}.
    \item \textbf{Rhythm quality}:
    Whether the rhythm is clear, stable, and reasonable \cite{zhang2022structure, melechovsky2023mustango}.
    \item \textbf{Harmony}:
    Whether the melody, accompaniment, and chord sound pleasant together \cite{schneider2023mo, melechovsky2023mustango, han2023instructme, zhang2022structure}.
    \item \textbf{Orchestration}: For multi-instrument models, whether the instruments are used and arranged properly \cite{dong2023multitrack, liu2022symphony}.
    \item \textbf{Coherence}: Whether the transitions between musical phrases are natural and well-connected \cite{hsiao2021compound, zhang2022structure, liu2022symphony, wu2023compose, dong2023multitrack}.
    \item \textbf{Structure}: Whether the music exhibits clear repetitive structures and reasonable development \cite{wu2020jazz, hsiao2021compound, zhang2022structure, liu2022symphony, yu2022museformer, wu2023compose}.
    \item \textbf{Richness}: Whether the music is rich, creative, and interesting \cite{hsiao2021compound, liu2022symphony, dong2023multitrack, wu2023compose, melechovsky2023mustango}.
\end{itemize}
To examine statistical significance with absolute ratings, a Student's $t$-test \cite{student1908probable} is usually performed.

\section{Ethics \& Social Impact}\label{sec:ethic}
The development and deployment of FMs for music applications raise many ethical and cultural issues. This section delves into four main aspects identified through related work aimed at uncovering potential issues within the broader space of Music Information Retrieval (MIR). While little of this work specifically addresses FMs, most of these aspects are critical to understanding the ethical landscape surrounding FMs in music applications. The first section focuses on ethical and social impact issues. The first subsection highlights how the lack of diversity within the space where FMs are developed can negatively impact various ``ologies'': ontology, epistemology, and axiology. The second subsection addresses specific issues related to fairness, bias, transparency, and explainability. The third subsection examines the potential adverse consequences of FMs on humans. Finally, the fourth subsection provides suggestions on taking responsibility for the work done in this field. Then, the second section will discuss music copyright issues.

\subsection{Ethical issues}
\subsubsection{The Sectarian -ologies} 
\begin{quote}
Might a more diverse population of MIR practitioners favour awareness of and sensitivity to a wider spectrum of the world’s musics? \cite{born_diversifying_2020}
\end{quote}

A major concern is the threat to diversity embedded in the sociotechnical systems that form the backdrop of FMs. ML models often embed specific and narrow values and cultural backgrounds, which fail to capture and support diverse musical provenance and heritages. There is a fundamental need to account for cultural pluralism in MIR and to include contributions from non-Western traditions \cite{clancy_reflections_2021,huang_-centering_2021,morreale_where_2021,berkowitz_artificial_2024}. While FMs should be able to handle all kinds of music, they indeed often fail to capture or support cultural heritage and diversity. The consequences are tangible: biases toward Western popular music, fostering of commonalities rather than distinctiveness, and threats to regional creative work. A significant concern is the prevalence of Western popular music in training FMs, which exacerbates the historical trend of Westernisation and displacement of local content. This issue mainly arises from the narrow socio-cultural-economic background of FM creators: MIR practitioners and researchers come predominantly from WEIRD (Western, Educated, Industrialised, Rich, Democratic) demographics \cite{holzapfel_ethical_2018}, which leads to cultural biases reproduced in their work. Against this backdrop of cultural homogeneity, \cite{huang_beyond_2023} suggests that an ethical turn in MIR should involve rethinking the four ologies—ontology, epistemology, methodology, and axiology. We borrow this framing from them to structure our discussions around what we consider the sectarian -ologies: the narrow ontologies, epistemologies,\footnote{Different from Huang, this essay combines epistemology and methodology, viewing methodology as a way to reach knowledge.} and axiologies that currently affect the FM space.

\textbf{Ontology:}
When reporting on the issues around recommender systems, \cite{born_artificial_2021} highlights that research practices are shaped by those who create them. Thus, the ontological question ``what music is'' whose engagement is fundamental for the development of FMs, should account for various music forms and ideas (ibid). The theory of ``what music is'' that MIR embodies, which \cite{born_for_2010} called analytical ontology, includes ontological assumptions about what counts in music. The prioritised ontologies are not only those musical aspects that are important for Western tonal music (harmony, melody, rhythm), but also ``those sounds that, through recording, have been disembedded from originating bodies, socialities and locales'' (ibid). 
\cite{born_diversifying_2020} invites MIR researchers to question, "Whose music and which music, among the vast ocean of sounds in the world, gets to be the focus of MIR’s influential scientific practices?". While, in recent years, one may notice a rise in the use and focus on non-Western music, especially from India, there is a need to diversify our understanding of music further. Music is not just a sequence of notes; it carries significant cultural value, including the use of specific instruments and scales. The lyrics may convey the feelings or opinions of a specific group of people.

Notably, ontological endeavours are not limited to unpacking ``what music is,'' but also need to account for ``what \textit{good} music is.'' \cite{newman_human-ai_2023} discusses how the Eurocentric Westernisation of aesthetics influences training data and is misaligned with much of the world. This aspect is particularly critical in FMs, as exemplified by the reliance on LAION-Aesthetics for image FMs like Midjourney and Stable Diffusion, which were rated visually appealing by WEIRD individuals and AI-art developers \cite{knowing_machines_models_2024}\footnote{\url{https://knowingmachines.org/models-all-the-way}}.
There is a need to dissuade the idea of ``universal'' foundation models in favour of culture-specific and culture-preserving models. It remains to be seen whether supervised fine-tuning techniques can build rich culture-specific models or merely project non-Western musical languages into a space of Western musical concepts. This comment is particularly relevant when considered against the growing body of literature that criticised FMs for the lack of universalism. For instance, as discussed by \cite{bender2021dangers}, GPT-2 was trained on sources that mostly represent males in their twenties. Thus, music FM developers should avoid perpetuating hegemonic ontologies or at least acknowledge that their system is not representative of all music. The risk is that, in the case of music generation, representation will be further skewed as music libraries are inundated with synthetic music influenced by inequitably designed FMs.

\textbf{Epistemology}
The second ``ology'' affected by the lack of diversity among FM researchers is epistemology, which deals with what constitutes knowledge and how to reach it. Cultural homogeneity in the field results in similar methodologies and methods being employed across the field, which result in a limited palette of epistemological tools that are considered ``pertinent.'' \cite{born_diversifying_2020} asks ``How could MIR equip itself with tools suited to analysing and modelling a greater diversity of musics?'' \cite{morreale_where_2021} proposes that MIR research ``is not merely a matter of mathematical models and computational optimisation,'' and that the field should embrace an ``epistemological turn'' to broaden the skill sets of MIR researchers, in order to allow them to take ownership around discussions on effects of AI on music making and consumption. Similarly, \cite{chen_data_2019} proposes that MIR researchers should be educated on an array of non-technical matters (i.e. ethical, cultural, and financial issues) related to use of music data. \cite{holzapfel_ethical_2018} discusses the adaptation of MIR developments, emphasising that they must be possible without requiring extensive engineering expertise.
\cite{born_diversifying_2020} highlights the numerous issues with universalising non-universal techniques, noting that ``the techniques and parameters employed in MIR tend to derive from, and reflect, commercially dominant areas of global popular music. Yet those techniques and parameters come to be applied in powerful technologies as though they were universal, with inevitable `de-pluralizing' effects.''

Notably, calls for more plurality and diversity in AI and MIR research \cite{serra2011multicultural, holzapfel_ethical_2018, born_diversifying_2020, morreale_where_2021, huang_beyond_2023} align with the core tenets of feminist epistemology \cite{stinson2024feeling}, particularly with the concepts of \textit{situated knowledge} \cite{haraway1988situated} and \textit{strong objectivity} \cite{harding1995strong, harding2015objectivity}. These concepts are intended to surface how systemically marginalised groups may possess greater epistemic insight due to their different experiences and expertises compared to dominant groups. Although incorporating all perspectives is essential for a comprehensive understanding, placing a focus on marginalised standpoints helps to ensure that the influence of structures of oppression on scientific research is accounted for. Such attention towards diversity should not only be reflected and fostered in the composition of teams of researchers, both in terms of demographics and fields of expertise, but also fundamentally embedded in study design \cite{tannenbaum2019sex, schiebinger2011gendered}.

\textbf{Axiology:}
Axiology, the study of value, is proposed by \cite{huang_beyond_2023} as a critical consideration in MIR: the demographic and cultural narrowness of MIR researchers can impact what is deemed good or valuable research. This section focuses on the values and agendas embedded in FM research and who benefits from it \cite{morreale_data_2023,moore2023serge}. \cite{born_artificial_2021} indicates that ontological assumptions extend beyond music itself to include specific theories of human subjectivity embedded in MIR. For instance, recommender systems frame users as individuals overwhelmed by choice, and seek to maximise the time they spend on the platform rather than, for instance, on delivering diverse music. The human subjectivity that underlines this representation opposes the view that users are sovereign individuals who are fully aware of the specific music they want to listen to. The ideology of recommender systems is thus based on serving something palatable to users while denying them the ability to make those choices.
 \cite{born_artificial_2021} notes naturally competing interests behind AI technologies, as corporate private interests do not necessarily match public interests. This comment applies to FM-related work. \cite{morreale_data_2023} found that creators of datasets used in MIR tend not to prioritise musicians' rights and demands, often ignoring the extent to which ML models are fair for musicians. Similarly, \cite{born_artificial_2021} invites us to question "what masters/mistresses does MIR serve?"

Analysing the values embedded in FMs is particularly important as the field is now inundated with venture capital \cite{morreale_where_2021}. The music industry has a large interest in AI and FMs, but understanding the industry's effect on culture, society, musicians, listeners, and researchers remains challenging. New technologies created opportunities for new industry classes, such as streaming services and generative AI. However, \cite{born_artificial_2021} suggests that when MIR research is tied to the drive for economic growth, issues of sustainability \cite{devine_decomposed_2019} and human values (e.g., transparency, accountability, privacy, and security) are at stake. This research should seek to identify ``other goals and values to guide future science and engineering'' and consider how it would look under a different set of assumptions and incentives linked to human flourishing. \cite{born_artificial_2021} advocates for a logic of diversity (rather than similarity) and collectivism.
An important aspect for FM developers to consider is whether non-Western music should be represented and made globally available through digital data in the first place. \cite{born_diversifying_2020} suggests that non-Some forms of Western music have not been incorporated into global digital music archives. These types of music have diverse ontologies that are often deeply embedded in social, religious, or cosmological traditions, and they differ greatly from the universal music ontologies assumed by MIR and related disciplines. \cite{morreale_unwitting_2023} indicates that ML applications should resist misappropriating original intents, and \cite{huang_beyond_2023} explains how Western-centric theories might be insensitive to local practices, traditions, knowledge, and values that form distinct music ecosystems.
\cite{huang_-centering_2021} adds several relevant topics to this discussion. By turning to Chinese and Indigenous philosophies, the authors urge MIR developers to broaden their scope and consider both human and non-human rights when discussing AI ethics\footnote{Notably, this move towards the non-human or more-than-human has become increasingly popular even within Western schools of thought like posthumanism \cite{braidotti_posthuman_2016} and even in scientific discourse \cite{barad2007meeting}}. For instance, AI systems that produce billions of songs ``just because they can'' clash with Mohist's condemnations of ``wasteful productions and performances of music.'' This more-than-human ecomusicological perspective is particularly relevant to FMs, given the exponential environmental costs of energy-consuming neural networks \cite{bender2021dangers,crawford_generative_2024}. Furthermore, following \cite{huang_-centering_2021} invites inquiry into the effect of excessive productions and appoints music AI developers the responsibility to consider how their products impact our soundscape's health amidst environmental crises. 

\subsubsection{Fairness, Transparency, and Bias}\label{subsec:bias}
Since there can be no accountability without transparency, and without accountability fairness in unachievable, here we approach an understanding of how the themes of fairness and transparency are operationalised and proceduralised in AI and MIR. We discuss how achieving fairness in AI requires rejecting the myth of value-free science and  value-neutrality \cite{zhao2024measuring,miller2021technology}, which can perpetuate inequalities by ignoring the underlying social structures that AI systems can replicate. By acknowledging the inherent values and impacts of technological artefacts, designers and developers must accept that their creations are not morally neutral and instead should consider integrating explicit ethical and moral considerations early in the design process to tackle concerns upfront rather than shifting blame to users later \cite{miller2021technology}. 

Critics have long argued that technological development must consider cultural-historical dimensions to avoid unintentional consequences and ensure ethical development. As a matter of fact, AI has a long history of contentious interactions with its critics. As early as three decades ago, \cite{agre1998toward} suggested that these conflicts might be due to the AI field's limited engagement with cultural-historical explanations that scholars from social studies instead take for granted:
\begin{quote}
Without the idea that ideologies and social structures can be reproduced through a myriad of unconscious mechanisms such as linguistic forms and bodily habits, all critical analysis may seem like accusations of conscious malfeasance. Even sociological descriptions that seem perfectly neutral to their authors [i.e., the critics] can seem like personal insults to their subjects [i.e., the AI practitioners] if they presuppose forms of social order that exist below the level of conscious strategy and choice.
\end{quote} 

This misunderstanding and lack of involvement with social studies mostly stems from the longstanding myth that science is value-free. \cite[p. 7-11]{harding2015objectivity} argues that this myth, which became the hegemonic understanding of ``objectivity'' since the end of World War II, has contributed to establishing a hierarchy of knowledge, with STEM at the top and humanities at the bottom, and brought about a sense of autonomy of science from politics\footnote{Although certainly not from corporate interests \cite[p. 7-11]{harding2015objectivity}.}. This divide, at present, is causing a lot of confusion within the AI community about what terms like ``fairness'', ``diversity'', ``transparency'', and ``bias'' really mean \cite{zhao2024measuring,hesmondhalgh2023impact,batlle2023transparency, mulligan2019thing}. In addition, several scholars outside \cite{harding2015objectivity} and within  \cite{holzapfel_ethical_2018,huang_beyond_2023,morreale_where_2021} MIR argue that value-free/neutral assumptions hinder a truly ethical development of science and technology, as these assumptions can at times be unintentionally \cite{holzapfel_ethical_2018} instrumental in preserving inequalities in the distribution of (algorithmic) power and resources. In other instances, it can even be deliberately directed towards ``attempts to isolate science from sensitive questions'' \cite[p. 267]{proctor1991value}. 
 
\textbf{Fairness:} Understanding the concepts of ``fairness'' and ``diversity'' \cite{porcaro2023fairness} requires first restricting their scope within a precise subset of tasks that are relevant in MIR research: fairness in automated decision-making (ADM) and diversity in dataset creation and curation. According to \cite{watch2019taking} ADM refers to the delegation of decision execution to systems that use algorithmic models to carry out actions based on data. Additionally, the term ADM encompasses not only complex machine learning applications but also simpler yet impactful rule-based systems like risk assessment scores. To date, the most predominant MIR application pertaining to ADM are music recommendation systems \cite{hesmondhalgh2023impact,dinnissen2022fairness}. 

Researchers should reject value-free assumptions whilst implementing operationalisations (e.g., measurements) of ``fairness'' in ADM and dataset creation, and instead align with well-established theories of social justice\footnote{Such as diversity and inclusion \cite{mitchell2020diversity}, intersectional approaches to political economy  \cite{kasy2021fairness}, and critical race theory  \cite{hanna2020towards}. All of these examples address intersectionality \cite[p. 57]{harding2015objectivity}, which, in addition to being itself a theory of justice, also serves as an analytical framework \cite[p. 2]{collins2020intersectionality}.} all the while adhering to principles of empirical verifiability, as for example in \cite{kasy2021fairness, mitchell2020diversity, hanna2020towards}. A concrete example of value-free assumption would be adopting the notion of ``merit'' in ADM fairness \cite{kasy2021fairness}.\footnote{See \cite{trevisan2022psychologising} for how meritocracy developed into a value-free construct.} Whereas, in MIR the most prevalent manifestation of the myth of value freedom is currently the lack of ethical training, resulting for example in critically overlooked aspects during dataset creation \cite{morreale_data_2023, scheuerman2021datasets}. In fact, Zhao et al. \cite{zhao2024measuring} specifically identify value neutrality as one of the main issues in dataset curation in machine learning research at large, pointing out how practitioners often use value-laden terms such as ``diversity'' and ``bias'' without first giving a clear definition. In their research, Zhao and colleagues directly address this gap by applying principles from measurement theory to identify key considerations and provide recommendations based on social science insights. To the best of our knowledge our field has yet to produce precise guidelines for operationalisations of diversity for music datasets. In this regard, \cite{zhao2024measuring} and \cite{mitchell2022measuring} offer clear and potentially transferable insights for future MIR research.

Finally, it is important to highlight that (un)fairness in dataset creation and curation begins at data collection and annotation. In one respect, data is being collected to build (music) datasets without obtaining consent from content creators \cite{morreale_data_2023,morreale_unwitting_2023} raising clear ethical concerns related to intellectual property and labour exploitation. Moreover, in terms of employment conditions, \cite{miceli2022data} argue that a combination of precarity and financial dependency leads to a situation where millions of \textit{data workers} from the Global South \cite{kassi2021many, williams2022exploited} are alienated and made compliant \cite{miceli2022data}. This issue is exacerbated by the way historical \textit{power relations} are embedded in artefacts like interfaces and performance metrics that restrict workers' autonomy \cite{whittaker2023origin} and promote the standardisation of specific interpretations of data \cite{miceli2022data}, which as a side-effect, also introduces bias. We thus believe that researchers, developers, and companies should avoid supporting platforms for large-scale data annotation that are well-known places of exploitation, such as \textit{Amazon Mechanical Turk} \cite{pittman2016amazon, adda2011amazon}, where half of the workers in 2018 earned around or less than $~\$2/h$, and only one in 25 workers earned more than $~\$7.25/h$ \cite{hara2018data}.

\textbf{Transparency:}
In their systematic review of transparency in music Gen AI, \cite{batlle2023transparency} identified a lack of precise terminology as one of the main challenges in compiling their collection, so they adopted the policy-oriented definition of algorithmic transparency proposed by the IEEE \cite[p. 82]{ieee2017glossary}:
\begin{quote}
Transparency is a characteristic which describes a process whereby information is requested and then disclosed completely within the limits of public law, without distortion, and with respect to the computational and cognitive capacities of the information recipient in order to enable those recipients to interpret the information so that they are able to make rational, informed decisions. 
\end{quote}
The authors propose that, based on the current literature in AI and MIR, transparency can itself be proceduralised in terms of explainability, interpretability and documentation.

According to \cite{arrieta2020explainable}, explainability methods are categorised into transparent models, which are inherently interpretable, and post-hoc explainability methods that use external techniques (model-agnostic or specific). Explanation methods and interpretable models can enhance audio applications throughout most MIR tasks, from music generation, to genre classification and recommendation systems. For a systematic review of explainability and interpretability methods in music (Gen) AI see \cite[p.13]{batlle2023transparency}.\footnote{Another valuable overview of XAI in MIR and related fields is offered by the first ICASSP workshop on Explainable Machine Learning for Speech and Audio \url{https://xai-sa-workshop.github.io}.} 

While in terms of documentation, \cite[p.15]{batlle2023transparency}, different methodologies are emerging in both research and industry. Proposed approaches involve both documenting datasets and models while others focus on AI services, reporting on purpose, performance, safety, and security. These documentation strategies improve data practices, facilitate optimal dataset selection, and enhance algorithm quality and trust in AI services. In European law, the AI Act addresses transparency by requiring all providers of general-purpose AI models to adhere to documentation obligations. These include:
\begin{quote} 
Information on the data used for training, testing and validation, where applicable, including the type and provenance of data and curation methodologies (e.g. cleaning, filtering etc), the number of data points, their scope and main characteristics; how the data was obtained and selected as well as all other measures to detect the unsuitability of data sources and methods to detect identifiable biases.\footnote{Annex XI, Point (a), as referred to in Article 53(1), as approved by the European Council on May 14, 2024.}
\end{quote}

Outside academia and policy-making, several initiatives are promoting transparency with regard to AI applications in the music industry. \textit{Fairly Trained}, a non-profit organisation, aims to ensure fair treatment for human creators by distinguishing AI providers who obtain consent from data owners from those who do not, using a certification system based on data handling practices. \textit{AI:OK} provides similar certifications for music products and services within the generative AI and emerging technologies landscape. Additionally, \textit{Water\&Music}, a collaborative platform, has developed the \textit{Music AI Ethics Tracker}\footnote{Available for consultation at: \url{https://www.waterandmusic.com/data/ai-ethics-tracker}}, a living document that examines and analyzes ethical statements on AI from the music industry.

\textbf{Bias:} As previously discussed also the term ``bias'' is surrounded by confusion about its meaning. \cite{bommasani2021opportunities} explain that, in order to understand the relationship between inequity and foundation models, we must recognise that FMs serve as intermediary assets adapted for user-impacting applications. They propose a coarse dichotomy distinguishing between intrinsic biases (inherent properties of foundation models that could cause harm) and extrinsic harms (harms arising from specific applications using these models). Whereas, in a seminal work from the early era of \textit{value-sensitive design}, \cite{friedman1996bias} provide a more nuanced taxonomy of bias. They use the term bias to refer to ``computer systems that systematically and unfairly discriminate against certain individuals or groups of individuals in favor of others,'' and they categorise this into three types: technical, emergent, and preexisting bias.

\textit{Technical bias} is due to limitations in technology, such as hardware constraints, flawed algorithms, issues with pseudorandom number generation, and issues with the formalisation of human constructs. \textit{Emergent bias} in computer systems arises after the system's design is complete and is used in real-world contexts. This type of bias occurs because the system's original design assumptions no longer match the evolving conditions and characteristics of its user base, both at an individual and societal level, such as issues with user expertise\footnote{Friedman and Nissenbaum give the example of an ATM designed with extensive written instructions being used in a community with low literacy rates.}, or cultural values. 
Building on Friedman and Nissenbaum's taxonomy, \cite{dobbe2018broader} interpret technical bias as an epistemological problem, as the methods AI practitioners use to analyze and formalise problems are themselves sources of bias\footnote{Thus, bias can be even introduced in operationalisations of fairness, especially in the presence of value-free assumptions, such as "merit" in \cite{kasy2021fairness}.}; and emergent bias as a dynamical feedback phenomenon, whereby it occurs within evolving societal changes and user interactions. Feedback loops in decision-making systems can reinforce and amplify biases over time, necessitating a dynamic perspective to understand and mitigate these effects. For example, as also discussed in the following section, recommender systems are notoriously affected by a feedback loop on the emergent ``popularity bias'' \cite{mansoury2020feedback}.

Finally, \textit{preexisting bias} arises from social attitudes and institutions, either through explicit intentions or implicit norms. Notably, it can cascade into every aspect of AI research and production, from the lack of diversity in developers' teams to sampling biases in datasets and model or system outputs. The previously-discussed Western-centric tradition in MIR risks to marginalise or incorrectly appropriate non-Western music, all the while a lack of engagement with historical ethnic and gender-based power relations in education and the music industry at large \cite{topaz2022race, wang2019gender, betti2023large} can result in them being mirrored and amplified through the use of FMs in ADM, as for example in recommendation systems. Particularly to music recommender systems, \cite{hesmondhalgh2023impact} found that there is minimal collaboration between the fields of academic computer science and critical social sciences and humanities. This, they argue, leads to a situation where specific preexisting biases, such as popularity and demographic biases (e.g., gender), are commonly identified in computational studies \cite{dinnissen2022fairness}, whilst other biases like ethnicity and socioeconomic status are less studied. For this reason, research in recommender systems---of which music systems are a subset---is increasingly considering multidisciplinary and intersectional approaches \cite{deldjoo2023fairness}. In this context, increasing the presence of underrepresented demographics in STEM\footnote{Which includes addressing early on widespread gender bias in (music) technology education \cite{armstrong2008hard}} is a crucial first step. However, as also discussed in the previous section (under ``Epistemology''), true diversity goes beyond mere physical presence in research teams, as ``what is desired is the kind of diversity that fully respects the values and interests of all citisens while protecting those of the most economically and politically vulnerable groups'' \cite[p. xi]{harding2015objectivity}. This, when developing FMs for music, first and foremost entails addressing the needs of (long tail) musicians and listeners.

\subsubsection{Adverse Effects on Humans} 
FMs raise several questions on how they can affect the relationship between music and society. Music creation is not only a technical process but also, and mostly, a creative practice that binds itself to social and cultural contexts \cite{barton2018relationship}. Thus, the adversarial impact of FMs on humans and society cannot be ignored. Current literature in MIR and related fields recognises the harmful impact of not centring design on humans (and, by extension, living beings) \cite{friedman_value_2019,born_artificial_2021, huang_-centering_2021, holzapfel_ethical_2018}. In this subsection, we call for a deeper exploration into the potential threats posed by FMs, extending a call made by \cite{born_artificial_2021} for music recommendation systems. We also explore how FMs that do not consider ethical issues at their design stage can amplify known effects such as isolation, replacement, reduced agency and taste manipulation. 

\begin{quote}
    As platforms like Spotify become the primary means of distribution and circulation of music and other content, they begin to shape the ways in which music is produced and readied for them, either through explicit policies, rules, and guidelines they impose, or through more hidden acts of infrastructural and algorithmic politics \cite{born_artificial_2021}
\end{quote}
As from the above quote about music platforms re-shaping the future of how we consume music, questions arise on more fundamental tools: the data-driven machine learning models that support such platforms. As applications of FMs are developed for such platforms, they too can shape the ways in which music is created, curated, and distributed. This shaping process can happen through the known assumptions, rules, and limitations surrounding the models, or through, what \cite{born_artificial_2021} describes as ``infrastructural and algorithmic politics.'' Indeed, the opacity of such algorithmic decisions is evident not only at the deployment stage but also in understanding their long-term impact on individuals, the music industry, and society as a whole.
Although \cite{holzapfel_ethical_2018} stated that MIR algorithms are not fatal to individuals, we argue that ill-defined FMs can bring about the erasure of cultures, sub-cultures and identities. This may occur in the hands of premature deployment of FMs without careful reflection on the scope of the models. \cite{holzapfel_ethical_2018} presents a grounding example where a community has to adapt its music style because existing software can only cater to a dominant culture's style, thereby threatening local music traditions and manipulating the future of the local music. Design decisions and constraints play an important role in evaluating the impact of FMs. It is necessary to not only express the scope of the model, but also evaluate how it might affect communities beyond it. For this, \cite{born_artificial_2021} asks of recommendation algorithms: ``How will detaching people’s taste-forming musical habits from their wider ecologies disrupt ordinary forms of social and cultural relationality, and the linking of music to other forms of experience?'' Short-term goals such as performance gain are only as valuable as their term in the long run, it is indeed necessary to push for more human-centered long-term evaluation that can help understand the unknown effects that may persist and amplify without proper intervention. It should be noted that although not all unknown effects are adversarial, such as what is deemed as ``emergent properties'' in FMs, their exploration is necessary to not only avoid ill effects on humans, but also to promote the transparency of the algorithms. 

One such reason for these adverse effects might be the disjoint relationship between MIR researchers and the end-users \cite{holzapfel_ethical_2018}. Lack of communication can drive a wall of isolation between the communities researchers take from versus give to. \cite{holzapfel_ethical_2018, newman_human-ai_2023} push for reducing this gap between MIR researchers and musician communities by engaging in conversation, implementing feedback, and connecting stakeholders to researchers. \cite{holzapfel_ethical_2018} provide another grounding example where a fictional company, Adaetal, scrapes data from an online community to generate music in the specific style of the community. Although the company is successful in the task, the contribution of Adaetal in preserving the traditional music remains in question. \cite{holzapfel_ethical_2018} asks ``Is this research only benefiting Adaetal, to the detriment of the tradition they are using? How is the work of Adaetal contributing to this tradition?'' The implications of Adaetal's work remain adversarial to the preservation of the tradition; whilst Adaetal benefits from the recognition and commercial success of developing algorithms that mimic a traditional style, human artists from the community may face possible loss of livelihood. 

Human replacement remains a major threat to human livelihood when it comes to the deployment of FMs. How do music FMs affect the role of musicians, producers, musicologists, or even the listener? \cite{morreale_where_2021} suggests that the long-tail problem\footnote{The long tail is a common problem, found especially in recommendation, where the distribution of musicians' popularity is largely skewed. Few musicians gain immense popularity, while the rest remain to have incomparable popularity (the long tail) \cite{morreale_where_2021}.} might become more competitive for human musicians when the space is infiltrated by AI-musicians. This ideology is not limited to Western circles. As mentioned above, \cite{huang_-centering_2021} finds that the school of Mohism, an eastern philosophy, believes such excessive production of music is wasteful. It is evident that across the world, the idea of producing music for the sake of producing music could not only be meaningless, but also unfavourable to humans. Despite obvious ethical dilemmas, generative FMs are attracting increasing venture capital, given their potential to generate millions of tracks every day. In fact, \cite{morreale_where_2021, sturm_artificial_2019} recognise that the reduced labour cost is a strong incentive for music platforms to adopt such generative models. It is necessary for researchers developing such FMs to reflect on these potential issues before they deploy their models. 

While FM companies demand an act of faith that their models will create new roles and skills for musicians rather than replace them, a callback to the transparency of such models is important. As \cite{born_artificial_2021} points out, the lack of transparency regarding the fundamental decision-making done by algorithms might lead to a lack of agency for its users: the musicians its meant to support. This lack of control is an important issue that resonates in conversations with musicians \cite{newman_human-ai_2023}. Musicians recommend more open-source code for greater transparency of their controls as well as to enable changes as required for their task (ibid.). ``Control'' is not only limited to the capabilities of the FMs, but also speaks to the human intervention that can be supported. Research suggests that musicians do not find that creative pursuits with AI are meaningful without evidence of human labour in the production process (ibid.). ``Intention and choice'' are as valuable as the output as well, signifying the human touch and agency is necessary for musicians to connect with the music (ibid.). As one participant in a study on perception of AI music says, ``For me, creativity also involves the decision-making in a big way. And then to determine where to end things. It doesn't seem that my experience with AI so far affords these possibilities'' \cite{newman_human-ai_2023}. 

Lack of transparency and control are not only detrimental to musicians, but can also unknowingly impact the community of listeners. In its most obvious form, taste manipulation can adversely affect the perception and reception of music. As seen in another example developed by \cite{holzapfel_ethical_2018}, a fictional Digital Audio Workstation (DAW) company adopts technology developed to find similar music. However, the limitation of the training data disallows any music from a time signature apart from 4/4 to work well with their model. This manipulates how music is produced, at the detriment of the local music in a region. Such design decisions can reshape what is acceptable and not acceptable. However, this raises questions on who has the right to determine what is acceptable music or not. Hence, it is not only necessary to carefully reflect on design decisions, but also to document them indiscriminately to allow understanding of any changes which may reflect certain biases and affect the taste of listeners \cite{holzapfel_ethical_2018}. The historical prevalence of western music training datasets can also promote culture homogeneity because FMs trained on such data can be deployed as ``universal'' models, algorithmically enforcing western styles to become the dominant, and the only, style of music available as AI-music. This might promote an individualistic perception of music listening, which stands in contrast to the real world where our music perception is not only a product of our individual characteristics but also of the culture and environment we are a part of \cite{born_artificial_2021}. At its worst, this hyper-focused attention on individuals and neutralisation of culture might increase the echo-chamber effect highlighted by \cite{morreale_where_2021}. This radicalisation of ``personal attention'' has been discussed in other fields where urgent calls have been made to redefine and shift our motivations to develop technology that feeds on user engagement \cite{de2024attention, hermann2022artificial}.

    \subsubsection{Take Responsibility}
In this subsection, we will discuss the responsibilities of researchers, music educators, policymakers and companies to address ethical concerns.

\textbf{Researchers in MIR} must recognise their ethical responsibilities regarding music foundation model development. Unfortunately,
genuine investigations into these matters are unlikely from large corporations primarily driven by capital interests since corporate ethical inquiries often serve as strategic manoeuvres to buffer against criticism~\cite{bietti2020ethics}. 
Thus, many researchers suggest the research community should self-appoint as ethical regulators to face the potential social impact and ethical concern~\cite{morreale_where_2021, terzis2020onward}.

In ML and MIR communities, scholars have begun addressing these ethical considerations. The data-centric attribute of ML can somehow perpetuate current social biases embedded in training data \cite{barocas2016big}. Researchers must critically evaluate and document the capability, bias and limitations of their algorithms, ensuring transparency for users and stakeholders. The IEEE community advocates processes guided by ethical principles to mitigate bias and increase transparency \cite{hand2018aspects, chatila2019ieee}. 
Besides, there are many ethical discussions in MIR. Some have focused on recommendation bias and fairness \cite{holzapfel2018ethical, gomez2019fairness}, gender representation in music streaming \cite{epps2020artist}, and user data concerns \cite{saurel2014changing, chen_data_2019}. Legal aspects of AI-generated music, such as the use of copyright-protected training sets and transparency regarding AI involvement in music creation, have also been examined \cite{sturm2019artificial}. 

Researchers can enhance the ethical standing of music AI by improving the interpretability of the foundation model. Interpretability \cite{doshi2017towards} also known as explainability \cite{mittelstadt2019explaining}, refers to making AI behavior and outcomes understandable. This improvement helps in correcting errors, resolving biases, and providing causal inferences \cite{zeng2017interpretable}. Additionally, it addresses music copyright issues by clarifying how the system generates specific content, which aids in attributing credits appropriately to users, programmers, or musicians involved in the training set \cite{sturm2019artificial}.

Another measure researchers can take in the context of foundation models for music is to prioritise fairness and transparency discussed in previous subsections. 

\textbf{Music education.} Given the capabilities of foundation models in music creation and production, and considering the significant implications for career trajectories and opportunities in the music ecosystem, music educators are likely to incorporate AI-assisted composition and the use of foundation models and prompt engineering into their curricula.
However, given the rapid pace of technological advancement, AI tools are expected to replace entry-level jobs in small media companies or recording studios. So such a measure is only a temporary measure. Developers and administrators must consider the long-term implications of deploying music AI and explore ways to mitigate adverse consequences. Additionally, the current legal framework might be necessary to be adapted to better support artistic innovation and creativity in this evolving environment \cite{sturm2019artificial}.

\textbf{Policymakers} can take multiple measures to address the ethical concerns of foundation models to the music industry while boosting creativity, such as introducing rights and obligations on ``AI-generated music'' labels, training set record keeping for copyright supervision as well as enhancing personality rights to prevent DeepFake \footnote{Artificial Intelligence and the Music Industry – Master or Servant? A report from British All-party Parliamentary Group (APPG) \url{https://www.ukmusic.org/research-reports/appg-on-music-report-on-ai-and-music-2024/}\label{APPG}}. 

The first measure is to label AI-generated songs on streaming platforms. 
AI-driven tools, especially foundation models, have been increasingly used in music creation with better and better capabilities, from composing and mixing to streaming. These models shape users' music consumption behaviours and can foster unethical practices, such as promoting AI-generated ``artists'' to boost revenue at the expense of human artists \cite{eriksson2019spotify}. Transparency about AI's role in music creation and recommendation can empower both artists and listeners. For instance, flagging AI-generated music or offering options to avoid such music could be beneficial \cite{aguiar2018platforms}. This measure is all compatible with consumer rights such as the UK Consumer Rights Act 2015, leaving enough information for users to make decisions on their own.
However, defining AI involvement levels is challenging. Decomposing music creation into stages and accurately documenting AI's contribution is complex and varies individually. This process may stifle creativity and create barriers for artists, especially those lacking institutional support who rely on AI-assisted compositional tools. Moreover, it is unclear whether consumers understand or care about AI's extensive involvement in music creation. Despite these challenges, further research is needed to assess the potential harms of not informing consumers about AI involvement and to develop effective transparency measures. Researchers must balance transparency with practical considerations to ensure ethical and fair practices in AI-driven music creation and consumption \cite{sturm2019artificial}.

Besides, policymakers should require full documentation of all data used by developers.
For commercial companies, explicit permission from musicians or music companies is required to use copyrighted datasets. Governments should consider market access regulations and include models developed in countries or regions where copyright protection is relatively lax especially when copyright compliance can hardly be demonstrated for model training data. Furthermore, policymakers shall introduce a specific human right to protect musicians from being affected by DeepFake including misappropriation and false endorsement of their voice, image, name, and likeness (VINL) \footnote{\ref{APPG}}. 

\textbf{Streaming platforms}, though, can leverage AI-generated music to potentially reduce payouts to musicians, such practices could stifle creativity and long-term music diversity. In response, leading companies like Spotify and Deezer have implemented measures to reportedly better reward professional musicians \footnote{Modernizing Our Royalty System to Drive an Additional \$1 Billion toward Emerging and Professional Artists
 \hyperlink{https://artists.spotify.com/blog/modernizing-our-royalty-system}{Website.}+} \footnote{Universal Music Group and Deezer to launch the first comprehensive artist-centric music streaming model. \hyperlink{https://newsroom-deezer.com/2023/09/universal-music-group-and-deezer-to-launch-the-first-comprehensive-artist-centric-music-streaming-model/}{Website.}}
In 2023, Deezer introduced an ``Artist-Centric Model'' that prioritises streams from established artists, actively searched-for songs, and combats fraud. This resulted in the removal of millions of low-quality tracks generated by AI or music amateurs and a projected 7-10\% royalty shift towards real artists \footnote{Music in the Air -- Focus on monetisation, Emerging Markets and AI; updating global music industry forecasts \hyperlink{https://www.goldmansachs.com/intelligence/pages/music-in-the-air-focus-on-monetisation-emerging-markets-and-ai.html}{Website.}}.

\subsection{Copyright}

Intellectual Property (IP) is a type of property that includes intangible creations.
Intellectual creations such as names of products or brands, designs or looks of products, inventions, literary works, films, art, photographs or music compositions are examples of forms of intellectual property. Depending on the nature of the IP, different types of protection apply, such as designs and trademarks, patents or copyright. 
By preventing third parties from using or copying the IP without authorisation from the IP owner for a limited period of time, the intent of IP protection is to encourage creativity and innovation. 
Producing the initial creation requires an investment and the IP protection allows the creators to derive a benefit from their original creations, therefore providing an incentive to put in the effort in the first place. 
Analogous to a patent protecting an invention, copyright is an instrument that allows protection of creative works. 
The Statute of Anne (1709) \footnote{Statute of Anne, London (1710), Primary Sources on Copyright (1450-1900), eds L. Bently \& M. Kretschmer, \url{www.copyrighthistory.org}} is often regarded as the first act of law to provide copyright regulation. 
Since then copyright protection has been introduced in most jurisdictions and copyright has been considered a fundamental human right since the Universal Declaration of Human Rights (1948). 
In the music domain, copyright typically applies to compositions and sound recordings.

Copyright is sometimes described as a negative right because it only limits the rights of others. Copyright law disenfranchises a community of derivative creativity. Longen argues that ``term limit extensions, increased protectionist treatment of secondary works online, and the functional lack of access to proper licensing mechanisms have rendered users’ rights impotent.'' \cite{longen2023} Finding a balance between encouraging creative work through copyright and fostering innovation through a more open creative commons has been a long-running controversy. \cite{lessig2001}

Composers, songwriters, performing musicians and record producers have benefited from the protection of their work and a way of earning income from it. 
Recent progress in generative AI saw this technology acquiring the ability to produce outputs of comparable quality to human creators. It, therefore, becomes a potential form of competition with human creative output and a threat to the related sources of income. 
This brings forth copyright concerns mostly in two areas: usage of copyrighted material for training AI systems and the eligibility of the output of Gen AI systems for copyright protection. 

In the former case, the concern consists in determining whether or not training an AI system on copyrighted material without obtaining a license from the right holders is a breach of copyright. Artists and the wider creative community typically argue that it constitutes a violation of copyright, while some commercial organisations offering generative AI services argue it is not. Multiple lawsuits have already been filed for alleged copyright violations \cite{samuelson2023legal}, and more may follow as new companies and applications make AI-generated music widely available, and legal precedent has not been set yet.  
Interestingly, independently of legal obligations, some commercial providers of GenAI services have voluntarily adopted the position that they will only train on licensed content and offer compensation to the creators whose content is used as part of the training set. 
This is motivated by various factors: the recognition that training data is a major contributor to the value created by machine learning systems, the belief that artists whose creative works are used for training should be compensated for their contribution to the value creation, pressure from investors who cannot risk the legal exposure of copyright cases, and the high value of corporate reputation.

There is also the concern of determining whether content generated by Gen AI system can be protected by copyright. Given that copyright is traditionally aimed at protecting human creativity, there is usually a requirement that sufficient human creative input is part of the creation process for the work to qualify for copyright protection. 
It therefore appears unlikely that content fully machine-generated may be eligible for copyright protection. However, copyright institutions need to determine how much creative human input is enough to qualify. 
As an example, the US Copyright office in its AI policy guidance recalls its view that ``copyright can protect only material that is the product of human creativity''\footnote{Copyright Registration Guidance: Works Containing Material Generated
by Artificial Intelligence \url{https://copyright.gov/ai/ai_policy_guidance.pdf}}. As a result, ``If a work’s traditional elements of authorship were produced by a machine, the work lacks human authorship and the Office will not register it.'' Based on this rationale, the US Copyright Office determined that in the case where a user only provides a prompt to a Generative AI system, the ``traditional elements of authorship'' are provided by the machine, not the user, and therefore that the resulting output cannot be registered for copyright protection. 
Although this determination may be easy to make in simple scenarios, it will be vastly more difficult in cases where the interplay between a user/creator and the GenAI system is more intricate. Auditing creations that potentially contain elements of GenAI in order to determine their eligibility for copyright protection will also constitute a monumental challenge, partly because of the scale at which it will need to be done and partly because technical tools to rely on something else than an honest disclosure from the applicant do not currently exist. 

Copyright infringement in the form of plagiarism is not a risk that is new or specific to Gen AI systems, but the broad availability of this technology makes it possible on a scale much larger than before. Empirical evidence shows that music’s distinctive pitch and rhythm sequences can be memorised by current Gen AI models and are recognizable when re-expressed in new contexts or outputs. In other domains, AI systems tend to paraphrase or re-express learned information from the text and images, providing a defence against infringement claims. The more abstract character of music perception poses a challenge for the makers of Gen AI services, whose original intent is not to infringe copyright, as well as for the rights holders. Gen AI providers may need to establish appropriate guardrails to prevent their systems from plagiarising copyrighted material.

From the rights holders perspective, since Gen AI may be susceptible to produce outputs that infringe copyright in large numbers of small transactions, it may be impractical to discover and take legal action on a case-by-case basis.
In response, copyright owners across the space are “opting out” (denying AI companies the right to use their copyrighted material for training). While this is entirely understandable, it could have severe adverse effects in the long term. It may be in the music industry’s own interest to avoid another Naptser-like situation, with aggressive use of legal action against any and all training of AI models on copyrighted material \cite{mccourt2003napster}. An alternative strategy is embracing AI while trying to infuse the space with desired values and offering attractive solutions in terms of technology and access to desirable data. A failure to do so could yield a competitive advantage to actors who, by ignoring rights, would thus be able to train better models at a significantly lower cost. Over time, this could effectively dilute the concept of copyright and take away the abilities of industry, citizens and artists to influence and control the Gen AI space.

Large language models for music, through the application of “disentanglement” of rhythm, style, emotion, timbre, and other musical elements, may lead to an expansion of current concepts of copyrightable content beyond existing copyright coverage of composition, lyrics, and recordings. An important question for Gen AI and for music copyright is: to what extent are generated music elements derivatives of original copyrighted music recordings and compositions vs. original expressions of learned musical knowledge?

Because of the scale and new nature of copyright challenges raised by Gen AI, the development of appropriate and scalable technical tools and solutions may be required to address them. Improved technology could be used to avoid copyright infringement as well as to make copyright infringement detection practical in the new era of mass production by AI.
For example, technology performing the following tasks in a tractable, robust and scalable way: identification of GenAI output, training set membership inference, content provenance, authenticity and authorship tracking. 
As of today, most of these tools either do not exist or are not mature enough for deployment at scale, which undoubtedly warrants more research. 

Even though Gen AI raises some profound copyright-related questions, it is undeniable that this technology possesses enormous potential to engender a new generation of creative tools that will empower artists and creatives across domains, including music. 
The desirable outcome of copyright debates is one where all stakeholders (artists, right holders, Gen AI companies, legislators etc.) establish a model for usage and management of copyrighted material that allows the adoption and continued development of Gen AI technology while preserving the ability of the creative sectors to thrive.

\subsection{Personality Rights}

In addition to the copyright-related challenges discussed in the previous section, Gen AI systems bring additional risks and challenges related to a different type of rights: Personality Rights. Personality Rights, which are sometimes also referred to as right of publicity, give an individual control over the commercial use of their name, image, likeness, and any traits that may allow to identify them. This includes an individual's voice, which is of particular interest and relevance in the music domain. 

Voice, image and likeness cloning technology is particularly problematic in the music domain. 
Owing to recent progress in the development of Gen AI technology, voice cloning systems are now capable of cloning both spoken and singing voice with a level of realism that makes a synthetic output virtually indistinguishable from an authentic recording. Given that these systems are automated, broadly available and that music and sound recordings can be accessed (perhaps illegally) on the internet to train them, this means that voice cloning is already being performed at scale and in a variety of scenarios. 
Doing so without the authorisation of the individual (artist) being cloned constitutes a profoundly unethical behaviour. 
Some actors are totally oblivious to ethical considerations. For instance, there exists today a number of commercial organisations that sell services to clone known artists voices, without having received prior authorisation, or proposing any form of compensation. 
The risks associated with such a questionable use of this technology in the music domain include reputational harm to the artist, misinformation, and deceit of the artist's audience or the general public, to name only a few. 
Because the likelihood that these risks materialise is significant, and the consequences can be devastating for the artist and for society at large, it is extremely important that they are acknowledged, considered and addressed jointly by the Gen AI and music communities, and, possibly, also by the legislators. 

There is very little debate that using someone's likeness without their prior consent cannot be deemed as an ethical or acceptable behaviour. 
In most jurisdictions, a legal framework establishing personality rights and their protection already exists (though it may vary depending on the jurisdiction). 
However, because of the specificity of the threats resulting from the broad availability of likeness cloning technology powered by Gen AI, certain jurisdictions recognise the need to introduce AI-specific regulation to protect personality rights. 
In the state of Tennessee (USA), the ELVIS act is generally acknowledged as the first piece of legislation voted into law specifically designed to protect musicians against unauthorised use of their vocal likeness in scenarios such as deep fakes, voice cloning, etc. 
This law has been criticised for its broad reach and potential conflict with other legal rights\footnote{
ELVIS Act Needs to Be 'Returned to Sender'
\url{https://www.realclearpolicy.com/articles/2024/02/29/elvis_act_needs_to_be_returned_to_sender_1015123.html}}. 
With the introduction of any new legal protections comes a risk is that may be used or abused beyond the original intent of the legislation. 
 Hypothetically, a famous person could assert personality rights claims to attempt to block the career of another artist, regardless of how they came to sound alike \cite[p.~368]{stamets1994}. 
 In the context of AI and voice-cloning, similarity to human artists may become a tunable parameter that intensifies the already existing gray area defining the boundaries of identity.
Nevertheless, one may anticipate that other jurisdiction may also consider the introduction of similar legislation, given the growing prominence of voice cloning technology. 

While the law provides legal constraints, voluntary initiatives can raise ethical standards beyond (or in spite of) legal requirements. 
Some providers of music AI technology are already adopting an ethical position that goes beyond strict legal requirements \footnote{\url{https://aiformusic.info}}. 
This type of approach is broadly in line with an open call from musicians and a vast number of music industry organisations for more consideration being given to musicians' rights, livelihood and perspective in the development of Gen AI technology \footnote{\url{https://www.humanartistrycampaign.com}}. 
The development of technical tools may be helpful to not only combat harm when done but also prevent harm being done in the first place. 
In particular, tools such as the following are identified as possible aids: Automatic identification of a likeness (e.g. image or vocal), detection of cloned vs. authentic voice \cite{desblancs2024realclonedsingeridentification}, content provenance and authenticity tracking \cite{c2pawebsite}, and perhaps attribution models to fairly compensate authorised use of a likeness. 
For the most part, these tools are in their infancy and will require further research and development before they can be effective guardrails at scale.

Another use case that raises arduous ethical questions  consists in using likeness cloning technology for deceased artists. 
There is already evidence that the public is receptive to this type of use \cite{Chen2020ArtificialCI, cnnsouthkoreavoice}, which means we can most likely expect that it is a commercial offering that may experience some growth in the future. 
Personality rights are fundamentally designed to give a (living) individual rights over their likeness in the general sense of the term. After death, it is often the deceased's estate that is in a position to make decision regarding the potential use of the deceased's likeness. 
Should there be limits to what can be deemed ethical or acceptable to do with the deceased's likeness? Especially if they did not explicitly express their will during their lifetime? How should those limits be established? Should there be time limits for protection as in copyright? Should artificial personalities be protected, and could, say, 1 billion personalities be detected or distinguished?

Despite a number of risks and pending ethical questions, and much like the case of copyright, provided it is done in an ethical and responsible way, likeness cloning technology can be an enabler of valuable and innovative products or experiences. For example voice cloning alone could be used to: create personalised artist messages to fans, games, localise content (i.e. adapt to local languages), produce avatars to reach audiences that are not able to attend concerts etc. 
The artist community has already started embracing the technology, showing the exciting path of new creative possibilities for the use of technology that is respectful of artists' personality rights.

\section{Conclusion \& Discussion}
In conclusion, this survey has articulated the transformative potential of foundation models (FMs), particularly large language models (LLMs) and latent diffusion models (LDMs), in the realm of music. These models exploit self-supervised learning to process complex musical data, significantly enhancing music information retrieval, generation, and multimodal interactions. The integration of these technologies into the music industry promises substantial benefits across various sectors, including entertainment, education, therapy, and cultural preservation. The survey has delineated several core areas where FMs are making an impact: representation of music, downstream tasks, pre-training strategies, adaptation techniques, and the pressing issues of ethics and copyright in music AI applications.
The outlined advancements underscore the need for advanced LLMs, visual-language models, and other interdisciplinary approaches to address the challenges of long-sequence modelling and domain-specific knowledge architecture.
In addition, we collect the available corpus for music FM development, including a number of large, high-quality and diverse datasets in multiple modalities and evaluation metrics for music understanding and generation. 
Besides, we suggest that due to the strong socio-cultural and music industry connections, ethical issues and data copyrights need to be taken into account.

Targeted at computer music researchers unfamiliar with LLMs and LDMs and ML researchers interested in music applications yet inexperienced with traditional music processing techniques, this survey serves as a foundational reference. It aims to guide future research directions, encouraging a deeper engagement with the ethical implications and societal impacts of deploying foundation models in music, thereby shaping a responsible trajectory for the evolution of music technologies.

Our exploration of FMs' \textbf{representation in music} has revealed that current models predominantly focus on a limited spectrum of musical representations such as spectrum, waveform, MIDI, and ABC notation. Besides, we observe a considerable gap in fully exploiting multimodal representations that unify symbolic music, audio, text, and music score images within FM development.

On the \textbf{music applications} front, music pre-trained models demonstrated robust capabilities across a wide array of music-related tasks—ranging from tonality and harmony analysis to sophisticated music generation and medical applications. These include chord and key detection, melody extraction, pitch detection, and more dynamic tasks like music source separation and emotion recognition. 
The potency of foundation models in NLP and CV communities, exemplified by systems like chatGPT and stable diffusion, in addressing diverse and previously unseen natural language or image generation tasks suggests a promising future. Products such as Qwen-audio and SunoAI, which offer extensive music understanding and generation capabilities, indicate that multimodal LLMs combined with LDMs or employed as Music Agents could substantially impact music technology.
This burgeoning field necessitates exploration of the advanced LLM and LDM techniques in music to enable models' versatility, along with ethical dimensions ensuring that advancements enhance.

Moreover, the survey highlights several significant technical advances and remaining challenges in leveraging LLMs and LDMs for music, including pre-training paradigms, domain adaptation, in-context learning, model architectures, controllability, music agents and scaling law.
\textbf{Training paradigms}, including contrastive learning, generative pretraining, and mask language modelling, have shown potential in learning from large-scale, unsupervised music data. However, all of these models require fine-tuning on downstream tasks, and models developed with the combination of pre-training and instruction tuning are yet to be fully realised. Besides, combining the music domain to develop pre-training targets has not been fully discovered.

\textbf{Domain adaptation techniques} such as adaptor and prefix tuning have proven effective for fine-tuning pre-trained language models on downstream music tasks, enhancing the multimodal capabilities of these models without extensive retraining. These methods facilitate the integration of music-specific knowledge into LLMs, allowing for nuanced understanding and generation of musical content. However, the research field still lacks a comprehensive understanding of the potential of MLP, Q-former or other architecture on music FMs.

\textbf{In-context learning (ICL)} techniques have enabled foundation models to perform tasks with little to no task-specific training, leveraging the model's ability to generate contextually appropriate responses based on provided examples. This approach is particularly promising for music applications where training data may be scarce or highly diverse. However, the development of supervised fine-tuning data, which gives model music ICL capabilities, is still rare. And how music capabilities at different levels (e.g., acoustic information, performance-level information, high-level text description, etc.) can be improved by which type of advanced prompt techniques, such as chain-of-thought (CoT) remain underexplored.

\textbf{The architectural} of most current foundation models, primarily based on the Transformer architecture, supports extensive scalability but faces challenges with long-sequence music data. Innovations in model architecture are required to better handle the complexities of musical structure and longer context windows. Although the GPT-4 and Claude-3  can handle more than 100k tokens, there is still much room for improvement in the ability to handle long sequences. Specific architectural adjustments or algorithms may be needed to enhance the modelling and exploitation of long contextual information. Another issue is that current foundation models are mainly decoder-only architectures, and training such models places high demands on the quality of the music-text tokeniser. Whether the Encoder-decoder architecture is an alternative is also worth exploring.

\textbf{controllability} remains a critical aspect, with current models needing better mechanisms to ensure music generation adheres to specific stylistic or structural constraints. The integration of explicit control mechanisms and the exploration of interpretable models will enhance the utility and applicability of foundation models in music, making them more transparent and aligned with users' creative intentions.

\textbf{Scaling laws }in LLMs have underscored the importance of model size and training data scale for achieving superior performance. Training good foundation models is very challenging due to huge computational consumption and sensitivity to data quality and training techniques. The scaling law solves this problem to some extent. However, the abilities of musical FMs can be inferred from small models, which are emergent abilities that need further exploration to fully leverage their potential in creative and analytical tasks. Further, the tokenisation of the music may greatly influence the ratio of training resources to training data and model parameters. The exact link between the two needs to be explored.

Lastly, \textbf{music agents }represent a promising area of development, transitioning from semi-autonomous to fully autonomous systems that can interact with users, manage complex workflows, and integrate various musical tasks and modalities seamlessly. These agents are also poised to revolutionise how we interact with and create music, offering personalised and adaptive music experiences.

Apart from the training methodology mentioned above, the efficacy of foundation models for music heavily depends on \textbf{the quality and diversity of the datasets} used for training and evaluation. Challenges related to the size, diversity, and copyright restrictions of music datasets necessitate rigorous efforts in data collection, cleaning, mixing, and curriculum learning to ensure that models can generalise effectively across varied musical backgrounds and cultures. Despite the availability of extensive, open-source, high-quality music audio and scores encompassing various epochs, instruments, and genres, there remains a notable gap in the strategies for data preparation, such as cleaning and instruction tuning for multimodal data involving diverse musical tasks. Additionally, the deployment of these models requires the development of strategies that appropriately balance data to avoid catastrophic forgetting and continuously adapt to new musical trends and user preferences.

Additionally, the security aspects of models capable of distilling training data require rigorous examination. This scrutiny enables music publishers to assess the legality of training datasets and simultaneously exposes models to the risk of revealing proprietary training data. An illustrative case is a lawsuit initiated by Universal Music Group Publishing, Concord Music Group, and ABKCO against Anthropic, which involved the unauthorized use of copyrighted material in 500 instances for training AI models \footnote{{Universal Music Sues AI Company Anthropic for Copyright Infringement - Levi's Sues Coperni for Trade mark Infringement.} \url{https://intellectual-property-helpdesk.ec.europa.eu/news-events/news/universal-music-sues-ai-company-anthropic-copyright-infringement-levis-sues-coperni-trade-mark-2023-10-26_en}}.

On the \textbf{evaluation} front, the generalisation of these models, whether in understanding or generation tasks, requires more uniform and comprehensive metrics. Often, models are only tested on a subset of downstream tasks, and many benchmarks use music audio from public datasets that the models might have already been exposed to, leading to potential data contamination and leakage. Furthermore, there is an urgent need to address biases in models that have predominantly been trained on Western music, neglecting the vast array of world music, which raises significant ethical concerns.

Ethically, the use of FMs in music necessitates careful consideration of \textbf{ethics issues and copyright issues}, including transparency, accountability, and bias. By fostering collaboration among researchers, practitioners, and policymakers, we hope to better utilise the potential of AI in enriching musical experiences while ensuring its benefits are widely accessible to everyone ethically.

This survey underscores the importance of interdisciplinary collaboration in advancing music foundation models. We call upon researchers in computer music and machine learning to engage with ethical research and development practices rigorously. By doing so, we aim to bring transformative changes to the music industry and human musical culture through the responsible advancement of music foundation models.

\section{Acknowledgement}
Yinghao Ma is a research student at the
UKRI Centre for Doctoral Training in Artificial Intelligence and Music, supported by
UK Research and Innovation [grant number
EP/S022694/1]. 
Emmanouil Benetos is supported
by a RAEng/Leverhulme Trust Research Fellowship [grant number LTRF2223-19-106].

We thank Dr. Zhiyao Duan's suggestions on the introduction, presentation sections and multimodal dataset subsection. We thank Dr Jie Fu's suggestions on the multimodal music understanding subsection.
We thank Pedro Sarmento for his help documenting initiatives towards AI transparency in the music industry.
We also thank Andrew Zigerelli, Qixiao Zhu, and Rikki Hung for their help in evaluation methods of music generation.

Last but not least, we acknowledge Junhong Li's kind help with illustrations.

\appendix
\subsection{Division of Work in Section Order}
\begin{itemize}
\item Introduction: Simon Dixon, Yinghao Ma
\item II Representations of Music: 
\begin{itemize}
    \item II-A Music Perception \& Notation: Charalampos Saitis
    \item II-B Computer Representations of Music
    \begin{itemize}
        \item II-B1 Acoustic-level Representations of Music: Huan Zhang, Yinghao Ma
        \item II-B2 Symbolic Music Format \& Their Content: Elona Shatri, Gy\"orgy Fazekas
        \item II-B3 Transformed Symbolic Music Representations for Foundational Models: Gy\"orgy Fazekas, Huan Zhang
    \end{itemize}
    \item II-C Multimodal Music Representations: Shangda Wu, Yinghao Ma 
\end{itemize}
        
\item III Applications
\begin{itemize}
    \item III-A Music Understanding
    \begin{itemize}
        \item III-A1 Traditional Music Information Retrieval Tasks: Emmanouil Benetos
        \item III-A2 Multimodal Music Understanding Tasks: Shangda Wu, Xingjian Du, Yinghao Ma
        \item III-A3 Vocal Music Understanding: Jiawen Huang, Zhizheng Wu
    \end{itemize}
    \item III-B Music Generation
    \begin{itemize}
        \item III-B1 Symbolic Music Generation: Zeyue Tian
        \item III-B2 Language Model for Acoustic Music Generation: Shun Lei, Zhiyong Wu
        \item III-B3 Audio Diffusion Model: Max W. Y. Lam, Shiyin Kang
        \item III-B4 Vocal Music Generation: Jiawen Huang, Zhizheng Wu
    \end{itemize}
    \item III-C Music Therapy \& Medical Applications: Bleiz MacSen Del Sette, Chenghua Lin, Yizhi Li
\end{itemize}
        
\item IV Technical Details of Foundation Models
\begin{itemize}
    \item IV-A Model Design \& Pre-training Strategies: Christos Plachouras, Julien Guinot, Ruibin Yuan, Ziyang Ma, Wenhao Huang, 
    \item IV-B Music Domain Adaptation for Foundation Models: Ge Zhang, Wenhu Chen, Xingwei Qu, Yinghao Ma
    \item IV-C Audio Tokenisers: Qiuqiang Kong
    \item IV-D Model Architectures: Qiuqiang Kong
    \item IV-E Interpretability \& Controllability on Music Generation: Gus Xia, Li-wei Lin, Ziyu Wang
    \item IV-F Foundation Models as Music Agents: Xu Tan
    \item IV-G Scaling Laws for Music: Xu Tan
    \item IV-H Additional Future Improvements: Anton Ragni
\end{itemize}

\item V Datasets \& Evaluation
\begin{itemize}
    \item V-A1-4 Music Datasets: Yinghao Ma
    \item v-A5-6 Multimodal datasets: present and challenges: Xingjian Du, Xingjian Du, Yinghao Ma%
    \item V-B1 Evaluation for music understanding: Ilaria Manco
    \item V-B2-3 Evaluation for Music Generation: Chris Donahue, Shih-Lun Wu
\end{itemize}

\item VI Ethics \& Social Impact
\begin{itemize}
    \item VI-A Ethical issues
    \begin{itemize}
        \item VI-A1 The Sectarian -ologies: Fabio Morreale
        \item VI-A2 Fairness, Transparency, and Bias: Luca Marinelli
        \item VI-A3 Adverse Effects on Humans: Fabio Morreale, Megha Sharma
        \item VI-A4 Take Responsibility: Yinghao Ma
    \end{itemize}
    \item VI-B Copyright: Anders Øland, Roger B. Dannenberg, Shuqi Dai 
    \item VI-C Personality Rights: Fabio Morreale, Roger B. Dannenberg
\end{itemize}

\item VII Conclusion \& Discussion: Yinghao Ma
\end{itemize}

\subsection{Author List}
\subsubsection{Main Coordinators}
\begin{itemize}
    \item Yinghao Ma, ~\IEEEmembership{Student Member, IEEE}
\end{itemize}

\subsubsection{Coordinators}
\begin{itemize}
    \item Emmanouil Benetos, ~\IEEEmembership{Senior Member,~IEEE,} 
    \item Fabio Morreale
    \item Shiyin Kang, ~\IEEEmembership{Member, IEEE}
\end{itemize}

\subsubsection{Main Contributors}
\begin{itemize}
\item Anton Ragni
\item Bleiz MacSen Del Sette
\item Charalampos Saitis
\item Chris Donahue
\item Christos Plachouras
\item Elona Shatri
\item Huan Zhang
\item Ilaria Manco
\item Jiawen Huang
\item Julien Guinot
\item Liwei Lin
\item Luca Marinelli
\item Max W. Y. Lam
\item Megha Sharma
\item Qiuqiang Kong, ~\IEEEmembership{Member, IEEE}
\item Roger B. Dannenberg, ~\IEEEmembership{Member, IEEE,}
\item Shangda Wu
\item Shih-Lun Wu
\item Shun Lei, ~\IEEEmembership{Student Member, IEEE},
\item Simon Dixon, ~\IEEEmembership{Senior Member,~IEEE} 
\item Wenhu Chen
\item Xu Tan, ~\IEEEmembership{Senior Member, IEEE}
\item Yizhi Li
\item Zeyue Tian
\item Zhiyong Wu, ~\IEEEmembership{Member, IEEE}
\end{itemize}

\subsubsection{Contributors}
\begin{itemize}
\item Anders Øland
\item Chenghua Lin
\item Ge Zhang
\item Gy\"orgy Fazekas
\item Gus Xia 
\item Ruibin Yuan
\item Shuqi Dai
\item Wenhao Huang 
\item Xingjian Du
\item Xingwei Qu
\item Zhizheng Wu
\item Ziyang Ma, ~\IEEEmembership{Student Member, IEEE}
\item Ziyu Wang
\end{itemize}
\newpage

\bibliographystyle{alpha}
\bibliography{IEEEabrv}

\end{document}